\documentclass{article}
\usepackage[utf8]{inputenc}
\usepackage[left=0.8in,top=0.8in,right=0.8in,bottom=0.8in]{geometry}

\newcommand{\be}{\begin{equation}}
\newcommand{\ee}{\end{equation}}
\newcommand{\bea}{\begin{eqnarray}}
\newcommand{\eea}{\end{eqnarray}}
\newcommand{\ba}{\begin{array}}
\newcommand{\ea}{\end{array}}
\newcommand{\rf}[1] {(\ref{#1})}

\newcommand{\eps}{\epsilon}
\newcommand{\cn}{\mbox{cn}}
\newcommand{\sn}{\mbox{sn}}
\newcommand{\dn}{\mbox{dn}}

\usepackage{Sam}	% Change to match style file

\usepackage{enumitem}
\usepackage{authblk}
\usepackage{bm}

\title{Soliton-like Rogue Wave Dynamics in Dissipative Higher-Order NLS  Models: A Floquet Spectral Perspective}
\author{C.M. Schober\thanks{Corresponding author: cschober@ucf.edu} \hspace{6pt} and A. Islas \\
Department of Mathematics, University of Central Florida}

\date{}

\begin{document}
\def\ri{\mathrm{i}}
\def\rd{\mathrm{d}}
\def\e{\mathrm{e}}

\maketitle

\begin{abstract}
 We investigate rogue wave formation and spectral downshifting in the higher-order nonlinear Schrö\-dinger (HONLS) equation and its dissipative extensions: the nonlinear mean-flow damping model (NLD-HONLS) and the viscous damping model (V-HONLS). 
 Using Floquet spectral analysis, we characterize 
 i) the organization of the underlying dynamical background and ii) the
 structure of the emergent rogue waves,
 distinguishing soliton-like rogue waves (SRWs) -- which are sharply localized and spectrally
   coherent --  from more diffuse rogue-wave events that lack coherent
   localization. In the conservative HONLS, SRWs arise only for
 sufficiently steep initial data, with the evolution
 intermittently switching between SRW formation and  disordered multi-mode
 dynamics, while
 moderately steep initial data produce only broader, less coherent rogue waves.

 Nonlinear mean-flow damping in the NLD-HONLS model suppresses
 disorder and
promotes a sustained, well-organized Floquet spectrum
that  supports  persistent soliton-like states from which SRWs emerge.
In contrast, viscous damping in the V-HONLS model  generates a disordered  Floquet spectral evolution,  broader rogue wave events,  and 
enhanced phase variability.
Furthermore, the NLD-HONLS exhibits a close link
between rogue wave events and the time of permanent spectral downshift, whereas  these phenomena appear
decoupled in the V-HONLS model.

These results clarify how dissipation type and wave steepness interact to regulate coherence,  rogue wave formation, and spectral downshifting in near-integrable wave systems. Moreover, the persistent coherent states identified in the NLD-HONLS align with the focused, highly organized wave groups observed  immediately prior to  wave breaking in laboratory experiments, providing a Floquet spectral framework that is conceptually relevant to pre-breaking wave dynamics.
 \end{abstract}

\section{Introduction}
\label{Introduction}

In studying the long-time evolution of modulated Stokes
waves, a foundational series of laboratory experiments by
Lake {\it et. al.} revealed that wave trains subject to modulational
instability (MI) also exhibited an
asymmetric spectral evolution,
leading to a permanent downward shift in the dominant frequency
\cite{LYRF1977,LYF1978}. This phenomenon, now known as frequency
downshifting, revealed limitations of the nonlinear Schr\"odinger (NLS) equation which, despite effectively modeling
the onset of MI and recurrence phenomena, cannot capture
irreversible processes due to its conservation of energy and momentum.

To address these limitations, several higher-order extensions
of the NLS equation were developed. Among them, the Dysthe equation and its Hamiltonian variant formulated by Gramstad and Trulsen (here
referred to as HONLS) include additional nonlinear
and dispersive terms that break certain symmetries of the standard NLS equation \cite{D1979,GT2011}.
These models provide better agreement with laboratory observations and a 
more realistic modeling of wave group dynamics \cite{LM1985,DTKS2003}. 
However,
neither the Dysthe nor conservative HONLS equations capture permanent frequency downshifting \cite{LM1985,SS2015}.

Recognizing that damping plays a critical role in driving frequency downshifting in deep water wave trains \cite{M1982,T1999,HM1991,TD1990}, several dissipative extensions of the HONLS equation
have been proposed.
The HONLS equation with nonlinear mean-flow damping  (NLD-HONLS) qualitatively reproduces key features of spectral downshifting observed in laboratory experiments \cite{KO1995,SS2015,CHB2019}.
Separately, a viscous extension of the HONLS equation  (V-HONLS)
derived from a weakly viscous modification of the Euler equations  \cite{CG2016,DDZ2008},  captures frequency downshifting
without relying on wind forcing or wave breaking \cite{ZC2021}. The V-HONLS equation has recently been used as the base model in a blended machine learning framework for predicting
wave breaking \cite{YETBYAB2023}.

 Modulational instability is widely recognized as playing a key
 role in permanent frequency downshift and in the development of
 rogue waves, which are rare,  very steep waves that
 can spontaneously arise in deep water \cite{PK2016}.
 Together, the HONLS, V-HONLS, and NLD-HONLS models
 provide a family of physically motivated frameworks for studying the effects of higher-order nonlinearity, dispersion, and dissipation in deep-water wave trains. They are particularly relevant for
 examining how these effects shape rogue wave formation,
 coherence, and frequency downshifting, offering complementary perspectives beyond the scope of the NLS equation.

 In prior work, we compared
 numerical simulations using the HONLS equation with
  laboratory experiments to investigate the long-time evolution
 of modulated periodic Stokes waves \cite{AHHS2001}. Both the experiments and
 HONLS simulations revealed chaotic behavior. Even so,
 the dynamics can be interpreted as near-integrable and  key aspects of the wave evolution -- such as the stability and spatial-temporal structure of the underlying nonlinear modes -- can be effectively analyzed using the Floquet spectral theory of the NLS equation.
 By applying a Floquet   spectral
 decomposition directly to both the  laboratory data and the HONLS simulations,
 we identified complex critical point crossings and 
 left-right transitions of the spectral  bands, which correspond to 
 bifurcations 
 between standing waves and left- and right-traveling
 modulated  wave trains  \cite{AHHS2001}.
 Notably, although  damping in the experiments
 reduced the number of unstable modes at later times, the
 fundamental left-right switching mechanism driving chaos
 persisted. This suggests  that while damping affects
 the amplitude, the underlying
 chaotic dynamics still has the opportunity to manifest itself.
 Although the qualitative agreement between
 laboratory observations and the HONLS simulations supports the
 conclusion that the HONLS equation captures the essential
 macroscopic features of the chaotic long-time evolution,
  incorporating damping would improve quantitative
 predictions of amplitude decay.

Certain exact solutions of the NLS equation are
 regarded as prototypes for rogue waves.
 Among the more analytically tractable examples are spatially periodic breathers (SPBs), which can be
 interpreted as heteroclinic orbits of  modulationally unstable Stokes waves \cite{OOS2000,CS2002,IS2005}.
In a recent study \cite{IS2022}, we investigated the stabilization of SPBs within the framework of the  NLD-HONLS equation. In that work, Floquet spectral  analysis revealed  a novel feature: the  emergence of rogue waves
was consistently accompanied by the formation of tiny bands of complex spectrum in the Floquet decomposition of the NLD-HONLS data.
This  spectral
signature 
indicates the emergence  of a highly localized coherent structure
that closely resembles a 
one- or two-soliton state near its peak (see Figure~\ref{NLD_SPB_PVD}(c)), while still  developing on a non-uniform, periodically
modulated background \cite{IS2022}.

We refer to such events as soliton-like rogue waves
(SRWs) to distinguish them  from generic, more diffuse rogue waves
that may satisfy a strength-based criterion but lack coherent localization.
By definition, SRWs form a subset of rogue waves that satisfy
both (i) a standard strength criterion and (ii) a spectral criterion indicating
soliton-like localization. Importantly, SRWs do not develop from soliton initial conditions; rather, they arise dynamically from modulationally unstable wave fields and represent coherent, pre–wave-breaking states within the broader rogue-wave context.

The earlier study \cite{IS2022} focused on a fixed damping strength and a specific class of initial conditions.  As such, it did not address the
robustness of SRWs,  whether they persist across varying damping strengths, broader classes of initial data, or under different model formulations.
It also left open two key  questions: (i)  whether SRWs can arise in the conservative HONLS without damping, and (ii) how  SRWs influence, or are influenced
by, the onset and timing of frequency downshifting. 

In \cite{CESS2023}, we examined the linear stability of damped Stokes waves within the framework of the V-HONLS equation, and 
provided an analytical
 description of the transition from an initial phase of exponential growth to a later regime characterized by predominantly oscillatory behavior on a slowly growing background.
 Viscosity was found to play a destabilizing role, amplifying the initial instability and delaying the onset of oscillations.
 Furthermore, for perturbed Stokes wave initial data, we
 proposed new criteria to identify the onset of permanent
 frequency
 downshift in the spectral peak and determined its relationship with
 the time of the global minimum in wave momentum.

 In this paper, we unify and extend several related lines of inquiry initiated in  our previous  studies  on dissipative extensions of the  HONLS equation.
 In earlier work, each model was  analyzed in isolation in connection with a single phenomenon
 -- for example,  chaotic dynamics  in the conservative HONLS,
 the emergence  of localized SRWs in the NLD-HONLS, and the
mechanism for spectral downshifting in the V-HONLS. 
Moreover, these studies were restricted to  specific initial conditions,
leaving open the broader question of how 
each dissipative mechanism
influences both SRW formation and frequency downshifting across 
 wider parameter regimes.
 
 Here, we address this gap by conducting a systematic comparative study across the HONLS, V-HONLS, and NLD-HONLS models.
 Our aim is to determine
 how viscous dissipation and nonlinear mean-flow damping  separately affect the emergence, persistence,  and coherence of SRWs as well as their relationship to
 spectral downshifting, over  a broader range of initial conditions and damping strengths than previously considered.

 To isolate the role of dissipation 
we begin by examining the   undamped HONLS. 
 Establishing whether SRWs arise in the HONLS is essential both for
 identifying phenomena intrinsic to the HONLS dynamics and for distinguishing them from effects that arise specifically due to dissipative mechanisms.

 The Floquet spectrum serves as our key diagnostic tool. It provides a
 framework  to 
(i) characterize the structural organization of the dynamical background and ii) classify rogue waves by  distinguishing sharply localized, soliton-like structures from  broader, less coherent events.
 Importantly, the Floquet spectrum not only identifies
 the presence of SRWs but also captures transitions between coherent and disordered regimes, revealing the dynamical context in which SRWs arise.

Our analysis considers two 
classes of initial data distinguished  by wave steepness: steep
conditions, generated from SPB profiles, and moderately steep
conditions, arising from generic Stokes wave perturbations.
 Across these initial conditions and a range of damping strengths, 
we address the following  key questions:
{\bf(i)} How should  rogue waves be classified? Do they consistently appear as
SRWs, or do more diffuse, nonlocalized events also occur?
{\bf(ii)} What background conditions favor  the formation of SRWs
as opposed to generic rogue waves?
{\bf (iii)} Do the SRWs emerge from transient or long-lived, persistent
soliton-like regimes?

In addition, we investigate  frequency downshifting, examining whether the onset of permanent spectral downshift is dynamically linked to SRW formation or whether the two phenomena are largely independent.
Taken together, these questions clarify how the type of dissipation, wave steepness, and initial conditions  shape
 the broader dynamical background that either supports or inhibits SRWs within the HONLS model and its dissipative extensions.

The remainder of this paper is organized as follows. Section 2
introduces the governing equations for the HONLS, NLD-HONLS, and V-HONLS models, along with  two complementary diagnostics:  Floquet spectral analysis and phase variance. This combined approach
provides a framework to distinguish localized, coherent SRWs from broader,
less coherent events and for characterizing the underlying dynamical
background. 
Section 3 presents numerical experiments grouped by initial data type (first, steep initial data  and second, moderately steep conditions), comparing results across all three models.
Section 4  summarizes the  key findings and their implications for rogue waves in dissipative, near-integrable wave systems.

\section{Background}
\subsection{Governing equations}
The comparative study of the formation of SRWs and frequency downshifting mechanisms, as discussed in the Introduction, is carried out within the framework of two distinct dissipative extensions of the HONLS equation. These models incorporate dissipation either through nonlinear mean-flow damping or via weak viscous effects, thereby enabling an investigation of the influence of different physical damping mechanisms on
wave dynamics:

\begin{equation}
\label{DHONLS}
\ri u_t + u_{xx} + 2u|u|^2 + \ri \Gamma u + \epsilon\left[ 2u (1 + \ri\beta)\mathscr{H} \left(|u|^2_x\right) - 8\ri|u|^2 u_x + \frac{\ri}{2} u_{xxx} + 2\Gamma  u_x \right] = 0,
\end{equation}
with $\epsilon >0$ and $0 < \Gamma, \beta <<1$.
The complex envelope of the wave train, $u(x,t)$, is assumed to be periodic in space with period $L$ and
 $\mathscr{H}$ denotes the Hilbert transform, defined by
 $\displaystyle \mathscr{H}(f)(x) := \frac{1}{\pi} \int_{-\infty}^{\infty} \frac{f(\xi)}{x - \xi} d\xi$.

For $\epsilon = \Gamma = 0$, equation~\rf{DHONLS} reduces to the integrable NLS equation.
Aspects of the associated Floquet spectral theory, which serves as a diagnostic tool in this study, are discussed in Subsection \ref{Subsection 2.2}.
Analytical expressions for the relevant NLS solutions are provided in the Appendix.

The numerical experiments in this study are based on the
following models: 

\vspace{6pt}
\noindent {\bf The HONLS with $\bm \epsilon \neq 0$, $\bm \beta = \bm \Gamma = 0$:} Equation~\rf{DHONLS} simplifies to the conservative HONLS equation, a Hamiltonian variant of the Dysthe equation as derived in \cite{GT2011,FD2011}. This conservative version captures higher-order nonlinear and dispersive effects beyond the standard NLS approximation while preserving fundamental invariants such as energy and momentum, and thus cannot account for irreversible frequency downshifting.

\vspace{6pt}
\noindent {\bf The viscous HONLS (V-HONLS) with $\bm \beta = 0$, $\bm \Gamma \neq 0$:}
Carter and Govan derived the V-HONLS equation from a weakly viscous extension of the Euler equations \cite{CG2016}. This model accounts for energy loss due to background fluid viscosity. Viscous effects enter through two  terms: a leading-order linear damping term, $\Gamma u$
and a higher-order dispersion correction term, 
$\epsilon\Gamma u_x$, both of which act directly on the entire wave envelope.
The combined influence  of these terms has been shown to correlate with
experimentally observed
spectral downshifting in nonlinear waves \cite{ZC2021,CESS2023}.

A diffusive term of the form $-\ri \epsilon^2\delta\Gamma u_{xx}$ is sometimes
included in the literature to suppress a weak long-time high-frequency instability associated with the $2\Gamma u_x$ term (see \cite{WLY2006,CHB2019}).
However, our previous analysis \cite{CESS2023}, together with supplementary
simulations conducted for the present study, 
confirms that for the  range of damping strengths considered here ($\Gamma =\mathcal{O}(10^{-3})$), this instability has only a minimal effect on the 
dynamical behavior of interest and doesn't affect the key results of this paper.
We therefore restrict our study to the V-HONLS formulation without diffusion.

\vspace{6pt}
\noindent {\bf The mean-flow damped HONLS (NLD-HONLS) with $\bm\beta \neq 0$, $\bm \Gamma = 0$:}  
Damping of the induced mean flow was incorporated in an ad hoc manner into Dysthe-type equations to account for frequency downshifting
\cite{UK1994,KO1995,IS2011},  and this approach has proven effective in  modeling and reproducing results from
laboratory experiments \cite{CHB2019,GMAECKB2023}.

In the derivation of Dysthe's equation under inviscid, conservative assumptions, it was  shown that modulation of the wave envelope $u(x,t)$ generates  a 
slowly varying  mean flow.
This arises because the wave envelope carries not only energy but also momentum. As it evolves and becomes spatially modulated (due to nonlinearity and dispersion), a momentum imbalance develops in the fluid. Since mass and momentum are conserved, this imbalance drives the formation of a mean flow which then feeds back into the wave dynamics. 
This feedback is conservative and enters  the HONLS as a
nonlocal dispersive effect represented by the term 
$u \mathscr{H} \left(|u|^2_x\right)$, which in turn  modifies the envelope evolution.
 
In realistic dissipative environments, e.g. due to viscosity,  wave breaking, or turbulence, the mean flow response  may
% exhibit a phase lag relative to the wave envelope. This introduces
introduce a complex coefficient, i.e. the $1 + \ri \beta$ term  into the feedback term.
The $\ri \beta$ term  in the NLD-HONLS modifies the mean flow’s influence on the wave envelope by introducing dissipation which breaks energy and momentum
conservation. As evident from
equation \rf{derivE&derivP}, the wave energy is damped  when  $\beta >0$.

\begin{figure}[htp]
  \centerline{
  %\includegraphics[width=0.3\textwidth]{Fig1a_new}
  %\hspace{6pt}
  \includegraphics[width=0.25\textwidth]{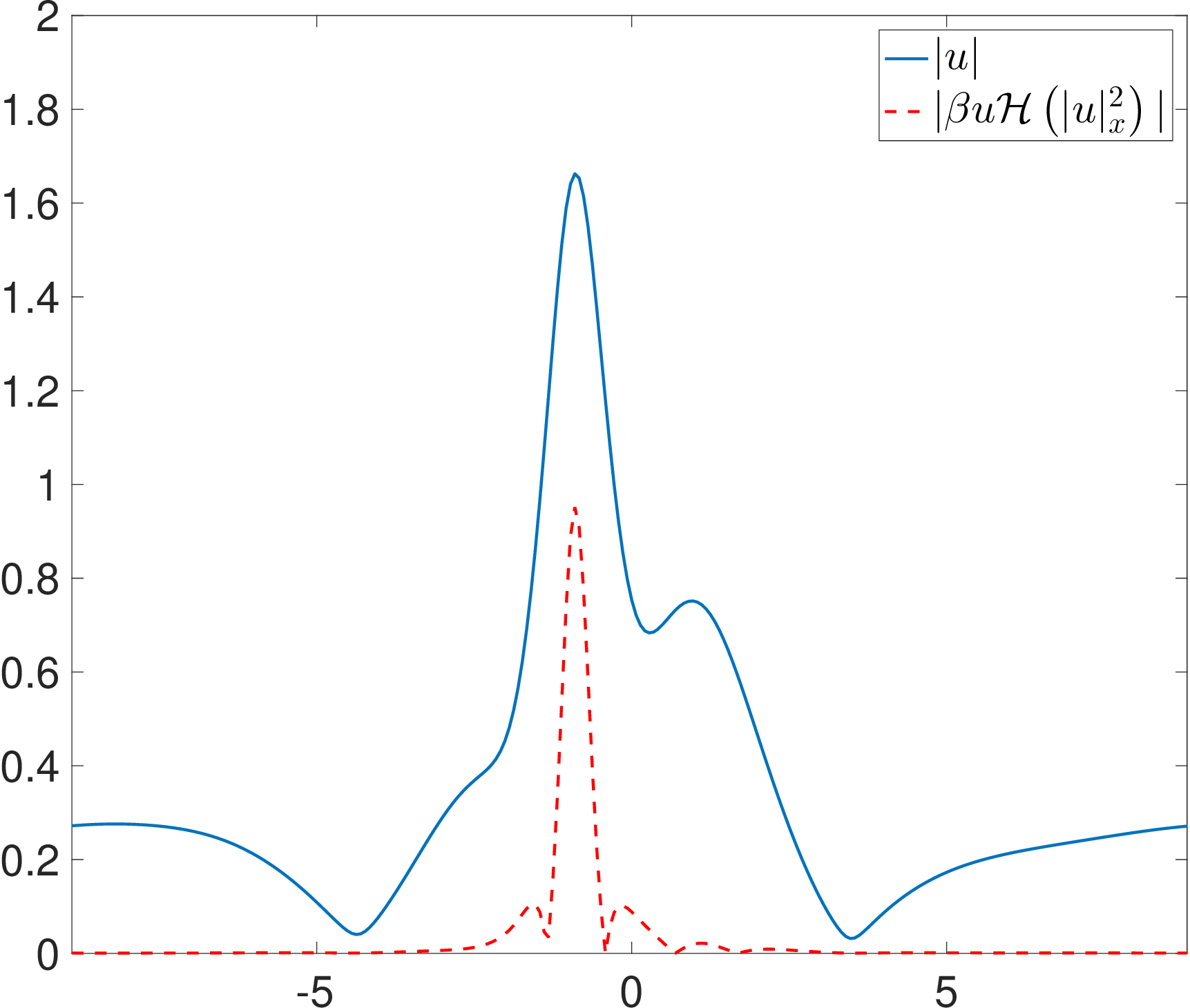}
  }
  \caption{$|u(x,t)|$ (solid line) versus
   $\left|\beta u {\cal H}\left(|u|^2_x\right)\right|$ (dashed red line) at $t = t^*$.
The nonlinear damping term is steepness dependent and is 
effective only near the crest of the envelope.}
  \label{Beta_term}
  \end{figure}

%While the NLD-HONS equation is not derived from first principles, it serves as a proxy model for wave breaking, capturing its core feature: the irreversible transfer of energy  from the wave envelope to smaller-scale structures.
%The damping it introduces
Mean-flow damping is proportional to both wave amplitude and its steepness.
This relationship is illustrated in Figure~\ref{Beta_term}, which  compares
the wave envelope $|u(x,t)|$ (solid blue line)  with the
nonlinear damping term 
$\left|\beta u {\cal H}\left(|u|^2_x\right)\right|$ (dashed red line)
 at the time $t = t^*$ when the wave train is strongly modulated. Here $u(x,t)$ is obtained from the NLD-HONLS model with $\beta = 0.1$, initialized using the 2-mode SPB data
\rf{SPB_ic}.  The plot shows  that the nonlinear damping term is
significant only near the wave crest and  is nearly negligible elsewhere. As a result, the nonlinear damping acts in a  highly localized manner in space and time, acting most strongly when the wave is steepest.

\subsection{Identifying Soliton-like Rogue Waves}
\label{Subsection 2.2}
%We appeal to the Floquet spectral theory of the NLS equation  to interpret and characterize the perturbed dynamics in terms of nearby integrable solutions of the NLS.
The motivation for using the Floquet spectrum as a diagnostic lies in its ability to quantify the number, spatial and temporal structure, and stability of the nonlinear modes present in the wave field. This enables a clearer classification of rogue waves, distinguishing sharply localized, soliton-like structures from more diffuse, spatially extended waveforms arising from the interaction of multiple modes.

To complement and corroborate the spectral classification, we introduce an additional diagnostic -- phase variance -- which quantifies the coherence of the wave field in Fourier space. Low phase variance is indicative of strongly localized, phase-coherent structures, while higher values reflect increased spectral disorder and multimodal interactions. This scalar measure serves as a complementary tool for distinguishing soliton-like rogue waves
from more disordered events.

Below, we present a brief overview of the Floquet spectral theory for the NLS equation, followed by the criteria employed to identify soliton-like rogue waves in the numerical experiments, incorporating both the Floquet spectral and phase coherence diagnostics.

\vspace{6pt}
\noindent {\bf Floquet Spectral theory for the NLS equation:}
The NLS equation
(when $\epsilon = \Gamma = 0$ in Equation~\rf{DHONLS}),  is a completely integrable system arising as the compatibility condition of  the  Zakharov-Shabat (Z-S) linear system \cite{ZS1972}:

\bea
\label{Lax1}
\mathcal{L}^{(x)} \mathbf v& =&\begin{pmatrix} \partial/\partial x + \ri\lambda & -u \\ u^* & \partial/\partial x -\ri\lambda \end{pmatrix} \mathbf v = 0,
 \\
\label{Lax2}
  {\cal L}^{(t)} \mathbf v & = &\begin{pmatrix} \partial/\partial t -\ri\left(|u|^2 - 2 \lambda^2\right) &  -\ri u_x - 2\lambda u \\ -\ri u^*_x +2 \lambda u^* & \partial/\partial t +\ri\left(|u|^2 -  2\lambda^2\right) \end{pmatrix} \mathbf v = 0,
\eea
where $\lambda$ is the spectral parameter and $u(x,t)$ is a solution of the NLS equation itself.

%\be
%\sigma(u) 
%%:= \left\{\lambda \in \C \, | \mathcal{L}^{(x)} \mathbf v = 0, |\mathbf v|
%\mbox{ bounded } \forall x\right\}.				
%\ee

Given a fundamental solution matrix of the Z-S system,$\Psi$, the Floquet discriminant is  defined as 
$\Delta(u,\lambda) = \mbox{Trace}\left(\Psi(x + L; \lambda)\Psi^{-1}(x;\lambda)\right)$,
which determines how  the eigenfunctions $\mathbf v$ evolve over one spatial period $L$.
The characterization of the  Floquet spectrum in terms of the
 discriminant is:
\be
\sigma(u) := \left\{ \lambda \in \C \, | \, \Delta(u,\lambda)\in\R,
 -2 \leq \Delta(u,\lambda) \leq 2 \right\}.
\ee 
Notably, $\Delta(\lambda)$ is conserved under the NLS flow and encodes the infinite family of NLS constants of motion. Correspondingly, $\sigma(u)$ is invariant in time.

The spectrum for  an NLS solution consists of the entire real axis and curves or ``bands of spectrum'' in the complex $\lambda$ plane due to the non self-adjoint nature of $\mathcal{L}^{(x)}$.
Special values of $\lambda$ for which $\Delta = \pm 2$ are referred to as
periodic/antiperiodic points (abbreviated here as  periodic points). 
Among these, the  simple periodic points form the set
\be
\sigma^s(u) = \{\lambda_j^s\,|\, \Delta(\lambda_j^s) = \pm 2, \partial \Delta/\partial \lambda \neq 0,\}
\ee
which occur in complex conjugate pairs off the real axis and determine the endpoints of the bands of spectrum.

\vspace{6pt}

\noindent {\bf Nonlinear modes:}  A general  $L-$periodic solution of the NLS equation admits a  representation  in terms of a set of
nonlinear modes whose structure and stability characteristics are
determined by its Floquet spectrum.
Key elements of the spectrum are the periodic  points.
For generic initial conditions, the Floquet spectrum typically contains an infinite set of simple points $\lambda_j^s$. Each pair $(\lambda_{2j}^s, \lambda_{2j+1}^s)$ defines a stable, dynamically active nonlinear mode of the system.

In many cases, NLS dynamics are 
well approximated by  $N$-phase quasiperiodic solutions  of the form
$u(x,t) = u_N(\theta_1, ...,\theta_N)$  where $N$ is finite. These $N$-phase solutions $u_N$ can be  constructed explicitly using Riemann theta functions associated with the hyperelliptic Riemann surface $\cal R$ of
$\sqrt{\Delta^2(u,\lambda) - 4}$, whose branch points are located at the $2N$ simple periodic points $\lambda_j^s  \in \sigma^s(u)$. The evolution of each phase $\theta_j$ is governed by the operator $\mathcal{L}^{(t)}$ and evolves linearly as $\theta_j = \kappa_j x + \omega_j t + \theta_j^{(0)}$, where the wave numbers $\kappa_j$ and frequencies $\omega_j$ depend on the associated spectral data $\lambda_j \in \sigma^s$.

\vspace{6pt}

 Two important elements are identified within the spectral bands:
\begin{enumerate}
\item Critical points, $\lambda_j^c$, where the derivative of the discriminant vanishes:
$\frac{\partial \Delta}{\partial \lambda} |_{\lambda_j^c} = 0$. 
\item Double points
  $\lambda_j^d $,  corresponding to degenerate periodic points satisfying:
  
$ \Delta(\lambda_j^d) = \pm 2, \partial \Delta/\partial \lambda = 0,\;
\partial^2 \Delta/\partial \lambda^2 \neq 0\}$.
\label{critical}
\end{enumerate}
While  double points are technically a subset of the critical points, 
 in this paper we reserve the term ``double points'' to refer to those degenerate periodic spectrum
with $\Delta = \pm 2$ . The term
``critical points'' is used exclusively for non-periodic degeneracies where
$\Delta \neq \pm 2$.

\vspace{6pt}
\noindent {\bf Floquet spectral characterization of instabilities:}
Critical points and double points play an important role when identifying instabilities of an $N$-phase solution. In particular, a real double point, 
$\lambda_j^d \in \mathbb R$,  corresponds to
a  stable inactive mode.
In contrast,  complex double points  $\lambda_j^d \in \mathbb \C$, which can represent either an active or inactive mode,
are typically associated with exponential instabilities
\cite{EFM1990}.

More recently, complex critical points -- arising from transverse intersections of spectral bands --  were  shown to be
linked to weaker instabilities
and to play an important  role in shaping the dynamics \cite{SI2021}.

\vspace{6pt}
\noindent {\bf Critical points as bifurcation points:} Complex
critical  points occur far more frequently than  complex
double points in the 
HONLS and its dissipative generalizations, largely  due to the
system's inherent asymmetric evolutions.
In \cite{AHS1996} we showed that for a standing
wave state, a complex critical point -- formed  when two
spectral bands intersect -- acts as a bifurcation point.
When the initial data for the standing wave state is slightly
perturbed, the spectrum breaks at the critical point and undergoes 
an asymmetric deformation into one of 
two possible configurations:  either  a left- or
right-spectral  configuration which corresponds to 
a left- or right-traveling
modulated  wave train.

In our current numerical simulations, frequent crossings of
complex critical points are observed in
the nonlinear mode decompositions of
both HONLS and V-HONLS flows,
leading to  a disordered  Floquet spectral evolution. In some
regimes, real degenerate
critical points also arise as points of transition. In both cases, critical point crossings  indicate non-localized dynamics characterized by multimodal mixing.

%A more detailed discussion of the differing impacts of real versus complex critical point crossings on the waveform is provided in the Appendix.

\vspace{6pt}
\noindent {\bf Classification as ``Soliton-like'' structures:}
Unlike in the integrable case, the spectrum is not conserved under the
HONLS flow or its damped extensions. In the numerical experiments, the spectrum is computed at each time step to track its evolution. As time progresses, one or two
spectral bands noticeably contract, with their lengths becoming extremely small (see, e.g., Figure~\ref{HONLS_SPB_spec}(b) and  Figure~\ref{HONLS_SPB_spec}(c), respectively). For a qualitative understanding  of the associated waveforms, these shrinking bands are viewed as nonlinear modes that have transitioned into a localized, ``soliton-like'' structure.

This terminology is motivated by considering the limiting behavior of the following family of 3-phase solutions to the NLS equation \cite{AEK1987}:
 \be 
  u_0(x,t) = \frac{\kappa}{\sqrt{2}}\e^{\ri t}\frac{\cn\left(\sqrt{\frac{1+\kappa}{2}}x, k\right)\,
    \cn(t,\kappa)+
    \ri\sqrt{1+\kappa}\,\dn\left(\sqrt{\frac{1+\kappa}{2}}x,k\right)\,\sn(t,\kappa)}
  {\sqrt{1+\kappa}\,\dn\left(\sqrt{\frac{1+\kappa}{2}}x,k\right) -     \cn\left(\sqrt{\frac{1+\kappa}{2}}x,k\right)\,
    \dn(t,\kappa)},
  \label{3phase}
  \ee
where $0 < \kappa <1$,
and $k = \sqrt{\frac{1-\kappa}{1+\kappa}}$.
  The solutions are
periodic in space  and quasi-periodic in time; the spatial period, $L_x $, and the temporal period of the modulated phase, $L_t $,
    are functions of the complete elliptic integrals of
  the first kind,   ${{\cal K}_x(k)}$ and ${{\cal K}_t(\kappa)}$ respectively.

  The  Floquet spectrum of $u_0(x,t)$ has two gaps in the spectrum along the imaginary axis.
  As $\kappa \rightarrow 1$ in \rf{3phase}, the gaps close  to complex double points, and $u_0(x,t)$ limits to a scaling of a one-mode SPB.
  On the other
  hand as  $\kappa \rightarrow 0$, the bands shrink to  discrete  points and
  $u_0(x,t)$ limits to a scaling of the  NLS soliton
 solution.

\vspace{6pt}
\noindent {\bf Rogue Wave Identification (Strength-based criterion):}
In the numerical experiments, we  adopt a widely used threshold based criterion to detect rogue waves. Specifically, the wave strength is defined as 
\begin{equation}
  S(t) = \frac{U_{max}(t)}{H_s(t)},
  \label{rwcond}
\end{equation}
where $U_{max}(t) = \mbox{max}_{x\in [0,L]} |u(x,t)|$ and $H_s(t)$ is the significant
wave amplitude.  A rogue wave occurs at time $t^*$ if $S(t^*) \ge 2.2$.
This criterion classifies an event as a rogue wave based solely on its strength.

\vspace{6pt}
\noindent {\bf Soliton-like Structure (Spectral Criterion):}
In addition to strength, we monitor the Floquet spectrum to
determine whether an event exhibits localized, soliton-like behavior.
When one or more spectral bands contract below a prescribed threshold, the corresponding mode of the $N$-phase solution is interpreted as
exhibiting soliton-like behavior  within the evolving multi-mode background.

{\sl Spectral condition for a soliton-like state:} Let  $\gamma(t;\lambda_m,\lambda_n)$
 denote the band with endpoints $\lambda_m$ and $\lambda_n$. 
If one or two of the band lengths satisfy  
 \be
|\gamma(t;\lambda_m,\lambda_n)| = |\lambda_m(t) - \lambda_n(t)| < 0.025,
 \label{s_state}
 \ee
 then the 
 spectrum is said to be in a one- or two-mode soliton-like configuration. In this case, the waveform is described as having a one- or two-mode soliton-like structure. These soliton-like structures (also
 called states) may be transient    or they may be persistent.

 \noindent {\bf Definition of Soliton-like Rogue Wave (SRW):}  An event is identified as a {\sl soliton-like rogue wave (SRW)}  if and only if it
 satisfies both:
 \begin{itemize}
   \item The strength based rogue wave condition $S(t) \ge 2.2$
     \rf{rwcond},  and
     \item The soliton-like spectral localization condition \rf{s_state}.
\end{itemize}
 Thus, SRWs form a distinct subset of rogue waves characterized not only by large strength, but also by  spectral signatures consistent with strong spatial localization and a coherent soliton-like structure.

 Two issues need to be addressed with respect to the terminology used:
 \begin{itemize}
   \item   In the numerical study  the background Stokes wave supports two unstable modes. As such,  the one-mode soliton-like configurations are not genuinely single-mode in a strict sense. Rather, the solutions
   are hybrid structures, characterized by partial localization: one spectral component enters a soliton-like regime, while the other remains extended and aligned with the  background.

   For brevity, we retain the use of the term
   "one-mode soliton-like structure" and "one-mode SRW" throughout, but the reader should interpret these as hybrid structures involving a mixed spectral signature.

   \item The bandlength criterion for identifying an SRW was developed by comparing the numerical solution, at times when it exhibits two tiny Floquet  spectral bands,
 with an exact two-soliton solution of the NLS, constructed using the
 spectral data obtained from the numerical solution at that time. This comparison was carried out across a set of representative
 experiments to provide a practical benchmark for distinguishing soliton-like  waveforms in our simulations.
When the bandlengths satisfy this criterion, the numerical waveform near its peak closely matches the structure of the two-soliton solution (see Figure~\ref{NLD_SPB_PVD}(c)).
 \end{itemize}
 
\vspace{6pt}
\noindent {\bf Phase Variance:}
In our numerical simulations, phase variance is used as a diagnostic that complements the more detailed modal information provided by Floquet spectral analysis. The phase variance captures  the spectral phase alignment across the
dominant Fourier modes,  providing a statistical indicator of coherence -- 
 of whether the structures are organized and localized or  instead are more  diffuse and disordered.
 Biondini and Mantzavinos \cite{BM2016} linked phase coherence to transitions between breather-like and incoherent states during modulational instability. Agafontsev and Zakharov \cite{AZ2015} employed phase statistics to characterize integrable turbulence.

Given the Fourier representation for $u(x,t)$ each Fourier mode can be expressed as $$\hat u_k(t) = |\hat u_k(t)| \e^{\ri\phi_k(t)}$$ with $\phi_k(t)\in [-\pi,\pi)$. Let $A_k = \phi_k\left|u_k\right|^2/\sum_{k\in{\cal K} } |\hat{u}_k|^2$. Then the weighted phase variance is defined by
  \be
  \label{pvd}
  \mbox{PVD}(t) = \frac{1}{N-1}\sum_{k\in{\cal K}}\left(A_k(t) - \langle A\rangle \right)^2,\qquad \langle A\rangle = \frac{1}{N}\sum_{k\in{\cal K}} A_k(t),
  \ee
  where $\langle A\rangle$ is the mean weighted phase and $N$ is the number of modes in ${\cal K}$.

Low PVD values indicates a strong phase alignment among dominant modes. This is characteristic of coherent, soliton-like structures, where the nonlinear interactions reinforce a collective phase evolution.
On the other hand, high PVD values suggest a broader spread in phase, typically associated with a more disordered, multi-mode dynamics.

Simulations with NLD-HONLS consistently maintain low peak phase variance (typically PVD$ \le 0.08$), aligning with the formation of sharply localized, SRWs. In contrast, the HONLS and V-HONLS models show elevated phase variance (often PVD$ > 0.2$), reflecting less coherent rogue events arising from a more disordered or chaotic background  with the interaction of many modes. 
Together, these diagnostics clarify the distinct coherence-preserving versus decohering dynamics underlying SRW formation in different dissipative regimes.

\section{Numerical Experiments: Rogue Wave Formation}
In this section, we numerically investigate how different forms of dissipation,
nonlinear mean-flow damping in the NLD-HONLS and linear viscosity in the V-HONLS, shape the broader dynamical background and influence the nature of 
rogue waves that emerge 
in deep-water wave trains.

Soliton-like rogue waves (SRWs) are localized, high-amplitude wave events
that emerge dynamically on a modulated, non-uniform background \cite{SI2021}.
Although
they closely resemble one- or two-soliton states near their peak, SRWs arise
spontaneously during evolution and do not originate from soliton initial conditions.

To evaluate whether the rogue waves consistently take the form of SRWs, or instead manifest as 
broader less coherent structures, we consider two types of initial conditions that represent different wave steepness regimes: (i)  exact SPB  profiles which serve as idealized representatives of steep waves, and (ii) generic perturbed 
Stokes wave profiles that represent moderately steep conditions, more typical of realistic ocean wave fields. This contrast allows us to determine how initial
wave steepness interacts with damping to influence both SRW formation and the evolution of the surrounding wave environment.

To establish a baseline and isolate the role  of dissipation, we first examine the undamped HONLS, whose ability to support SRW formation has not been previously studied.

\vspace{6pt}
\noindent {\bf  Numerical Method:}
The NLD-HONLS and V-HONLS equations are numerically integrated using a highly accurate fourth-order exponential time-differencing Runge-Kutta method (ETD4RK). This scheme employs a Fourier spectral decomposition in space and utilizes Padé approximations for the matrix exponential terms in time stepping \cite{KMWY2009,LKX2015}.
%Details of the ETF4RK algorithm and its convergence properties are provided in the Appendix.
The number of Fourier modes and the time step are chosen based on the complexity of the solution. For example, for perturbed Stokes wave initial data in the two unstable-mode (UM) regime, we set the domain size to $L = 4\sqrt{2} \pi$, use $N = 256$  Fourier modes, and select a time step of
$\Delta t = 10^{-3}$.
With this mesh, the conserved quantities of the HONLS equation, i.e  the total energy $E(t)$, the  momentum $P(t)$ (see equations \rf{EandPdefn} for their definition), and Hamiltonian
\be\label{hamiltonian}
H(t) =  \int_0^L \left\{ | u_x |^2 - | u |^4 - \ri\epsilon\left[
1/4\left( u_x u_{xx}^* - u_x^* u_{xx} \right) 
+ 2 | u |^2\left(u^* u_x - u u_x^*\right) - |u|^2
\mathcal{H} \left( | u |^2_x \right)\right]\right\}\,dx
\ee
are preserved to within $\mathcal{O} (10^{-11})$ over the time interval $0 \le t \le 100$.
Moreover, in previous studies of the NLS equation, this numerical scheme produced solutions accurate enough that the Floquet spectrum -- computed independently -- was shown to be preserved with high fidelity, making it a reliable tool for analyzing the dynamics of the NLD-HONLS and V-HONLS equations

\vspace{6pt}
\noindent {\bf Nonlinear mode decomposition of the dissipative flows:}
The Floquet spectral decomposition of the HONLS, V-HONLS, and NLD-HONLS data is typically computed for $0 \le t \le 100$ with output every $\delta t = 0.1$
using the numerical procedure developed by
Overman et. al. \cite{OMB2086}.  After solving 
system \rf{Lax1}, the  discriminant $\Delta$ is constructed.
The zeros of $\Delta \pm 2$ are determined using a root solver 
based on M\"uller's method and then the curves of spectrum are filled in.
The spectrum is calculated with an accuracy of ${\mathcal O}(10^{-6})$.

The numerically observed short-time spectral evolution of SPB initial data,
under both the HONLS and NLD-HONLS dynamics, is consistent
with a perturbation analysis,  which predicts that
the complex double points undergo asymmetric splitting
at order $\mathcal{O}(\epsilon)$  \cite{SI2021}. 

\vspace{6pt}
\noindent {\bf  Spectral plot notation:} In the spectral plots, periodic/antiperiodic spectrum are marked with 
a large ``$\bigtimes$'' when $\Delta = -2$ and a large box when $\Delta = 2$.
The continuous spectrum is represented by smaller symbols: a small ``$\bigtimes$'' denotes a region where $\Delta <0$  and a small box indicates where $\Delta >0$.
Due to symmetry of the NLS spectrum with respect to  complex conjugation,
only the upper half of the $\lambda$-plane is shown.

\vspace{6pt}
\noindent {\bf Setup of the numerical experiments:}
We consider two families of initial conditions that represent different steepness levels and are known to be conducive to  both rogue wave formation and
frequency downshifting (discussed in Section 4). 
\begin{itemize}
\item {\bf Initial data for steep waves:} 
 We initialize the simulations using exact 
two-mode spatially periodic breathers (SPBs) of the integrable NLS equation:
\be
\label{SPB_ic}
  u(x,0) = U(x,t_0,\rho,\tau),
\ee
where $U(x,t,\rho,\tau)$ is given by \rf{comboSPB} in the  Appendix. This initial data is for highly
structured, localized and steep waves, making it well suited for examining coherent SRW formation in the HONLS and dissipative HONLS models.

\item{\bf Initial data for moderately steep  waves:}
  For moderately steep waves we initialize the simulations using a small perturbation of the Stokes wave:
  \be
\label{Stokes_ic}
  u(x,0) = a(1+ \alpha \cos{\mu x}),
%\qquad \alpha  10^{-2}, \, \mu = \frac{2\pi}{L},
\ee
where $a = 0.45$, $\alpha = 10^{-2}$, $\mu = \frac{2\pi}{L}$, and $L = 4\sqrt{2} \pi$.
These parameter values  ensure the background Stokes wave is modulationally unstable, with two unstable modes   at $t = 0$. In contrast to the SPB initial data, the initial waveforms are less structured and require time to develop coherent features. However, the perturbed Stokes initial data eventually evolves into a modulated state, although
less steep and localized than the SPB case, providing a useful testbed for investigating SRW formation under more generic initial conditions.

\item{\bf Parameter values:} The HONLS parameter is fixed at $\epsilon = 0.05$
  in all the simulations.  For the NLD-HONLS model, we  
systematically vary  $ \beta \in [0.1,0.8]$  
in increments of  $\Delta \beta = 0.1$, while for the V-HONLS model, we 
vary $ \Gamma \in [0.001,0.008]$ with increment $\Delta \Gamma = 0.001$.
All simulations are conducted with $\Gamma \le 0.008$, a range previously
shown in \cite{CESS2023} to lie below the threshold at which the asymptotic
instability associated with the $2\Gamma u_x$ term affects
the key dynamical outcomes. 

For each model and each type of initial condition, we provide a detailed analysis of the system's behavior at a representative 
value of  $\Gamma$ and $\beta$, specifically $\Gamma = 0.002$ and $\beta = 0.1$
 We then present summary plots showing how increases in  the viscous 
 damping strength $\Gamma$ or  mean-flow damping parameter $\beta$  affects 
 the frequency and structure of SRW events, as well as the organization of
 the underlying wave field.

\item{\bf Floquet spectral evolution up to the first rogue wave event:}
As demonstrated  in the numerical experiments, for small values of $\Gamma$ and $\beta$, for each type of initial data, the spectral evolution is qualitatively similar across all three models, up to the onset of the first rogue wave. However,  this rogue wave marks a turning point,
 after which the dynamics diverge significantly between the models.
 Note that the spectral evolution itself differs between the two types of initial data from the outset.

%We restrict our choice of damping strengths to these ranges because, for larger values, rogue waves are not observed in the Stokes wave case, and in the SPB case, the initial steep structure remains the only rogue wave—already evident within the range considered.

 \end{itemize}

\subsection{Steep Wave Initial Data and the Formation of Rogue Waves}

Before analyzing the effects of dissipation, we first examine the undamped HONLS model to establish a baseline for the background dynamics and rogue wave formation, and to determine whether damping is necessary for  the formation of SRWs.

In the following numerical simulations, we observe that for each type of initial data, the spectral evolution remains qualitatively similar across all three models -- HONLS, V-HONLS, and NLD-HONLS -- up to the onset of the first rogue wave. This first rogue wave then marks a turning point, after which the dynamics diverge significantly between the models. As a result, we first present in detail the short-time Floquet spectral evolution under the HONLS model. We then use this as a reference point when discussing the corresponding results for the V-HONLS and NLD-HONLS models and focus on how the dynamics evolve after the first rogue wave.

 \subsubsection{Steep Wave Initial Data: HONLS equation}  
\begin{figure}[htp!]
\centerline{
\includegraphics[width=0.3\textwidth]{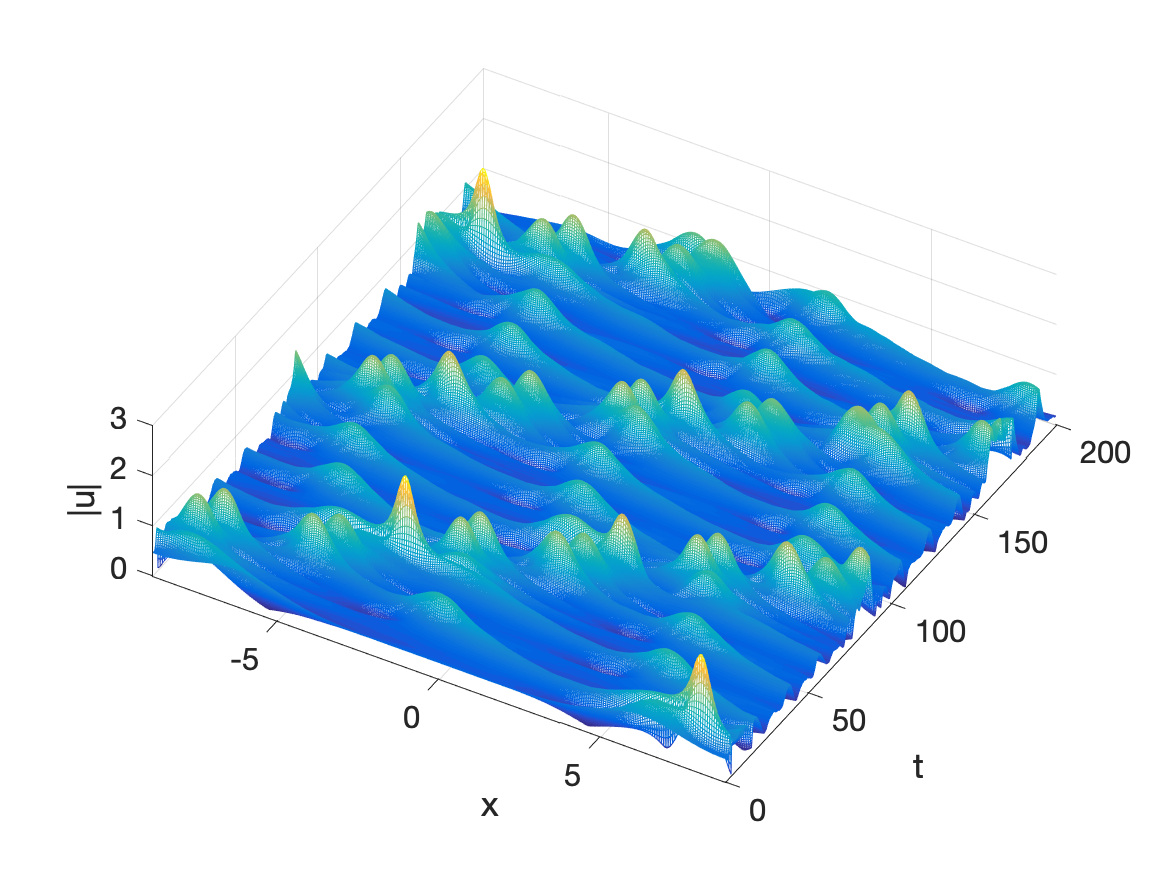}
\hspace{6pt}\includegraphics[width=0.25\textwidth]{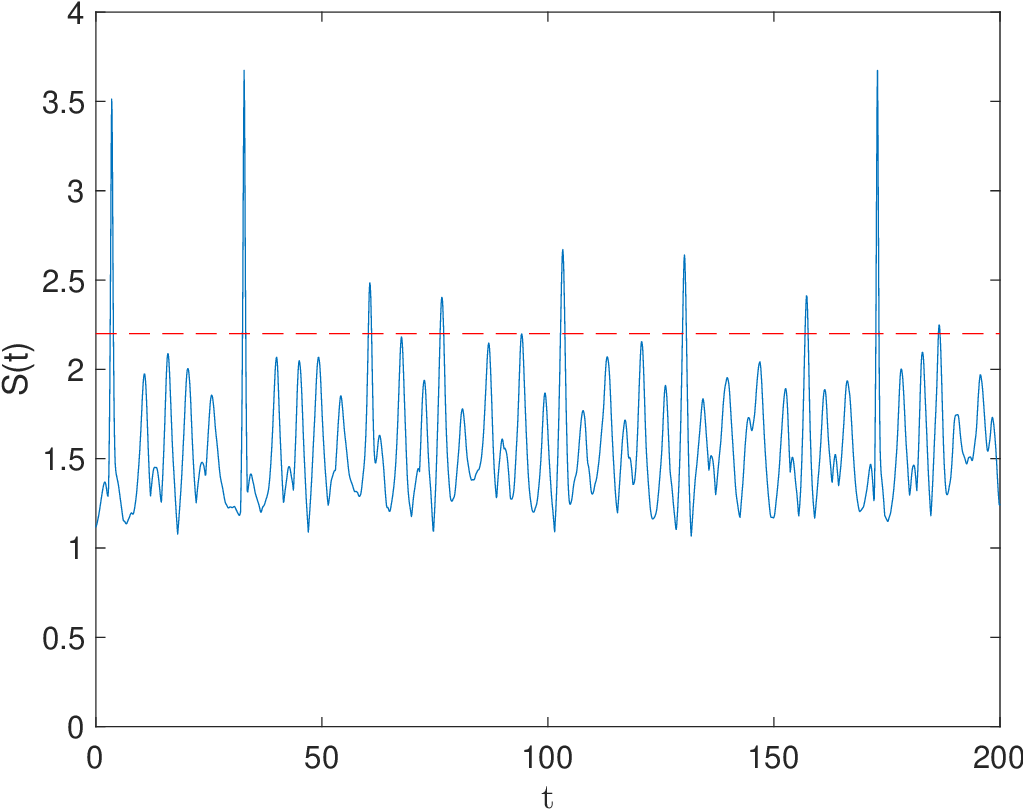}
\hspace{6pt}\includegraphics[width=0.25\textwidth]{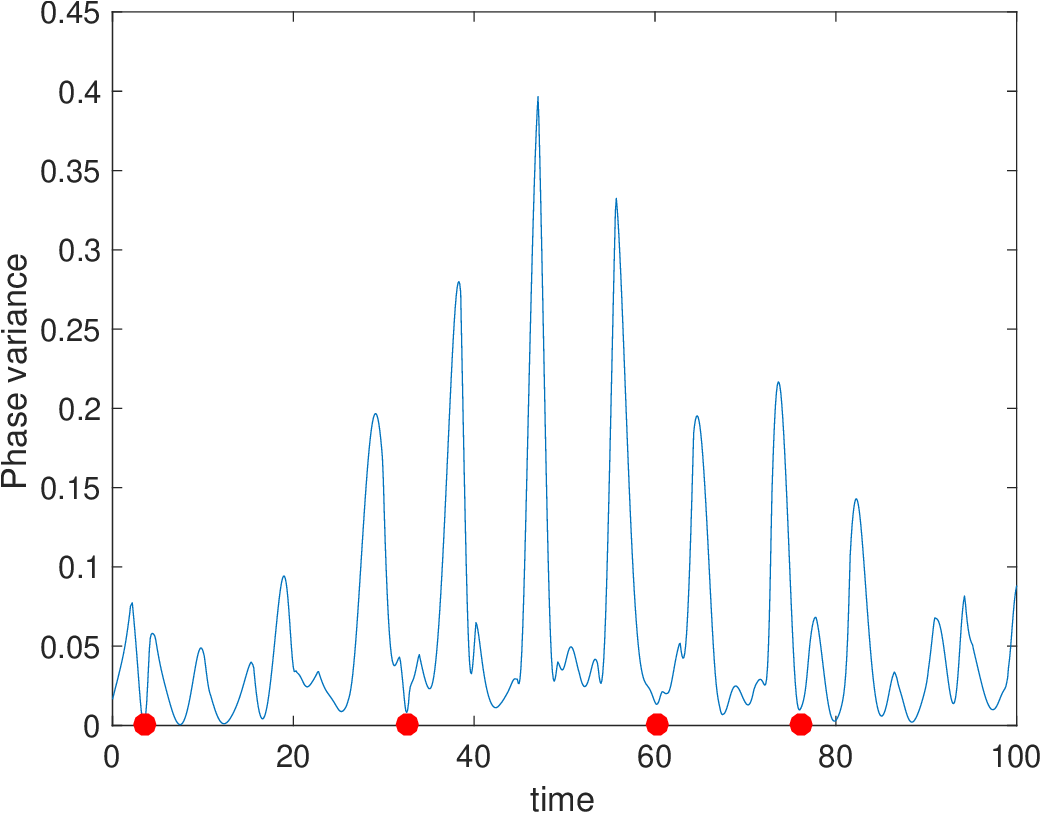}
}
\caption{HONLS evolution with SPB initial data \rf{SPB_ic}: (a) $|u(x,t)|$ for $0\leq t\leq 200$. (b) the strength $S(t)$ with the dashed red line indicating the threshold for a rogue wave; 
  (c) the phase variance PVD$(t)$ with red dots marking rogue wave events.
}
\label{HONLS_SPB_str}
\end{figure}

Figures~\ref{HONLS_SPB_str}(a) and ~\ref{HONLS_SPB_str}(b)  depict  the time evolution of the surface $|u(x,t)|$ and the corresponding 
 strength function $S(t)$ for the SPB initial data \rf{SPB_ic} over the interval $0 < t < 200$.  The dashed red line in the strength plot indicates the threshold for a rogue wave. Frequent rogue wave events occur throughout this time span, with the first occurring at $t = 3.5$, and due to the conservative nature of the HONLS equation, will  continue to emerge for $t > 200$.

\vspace{6pt}
\noindent {\bf The Floquet spectral evolution up to the first rogue wave event:}

\begin{figure}[hpt!]
  \centerline{
    \includegraphics[width=0.225\textwidth]{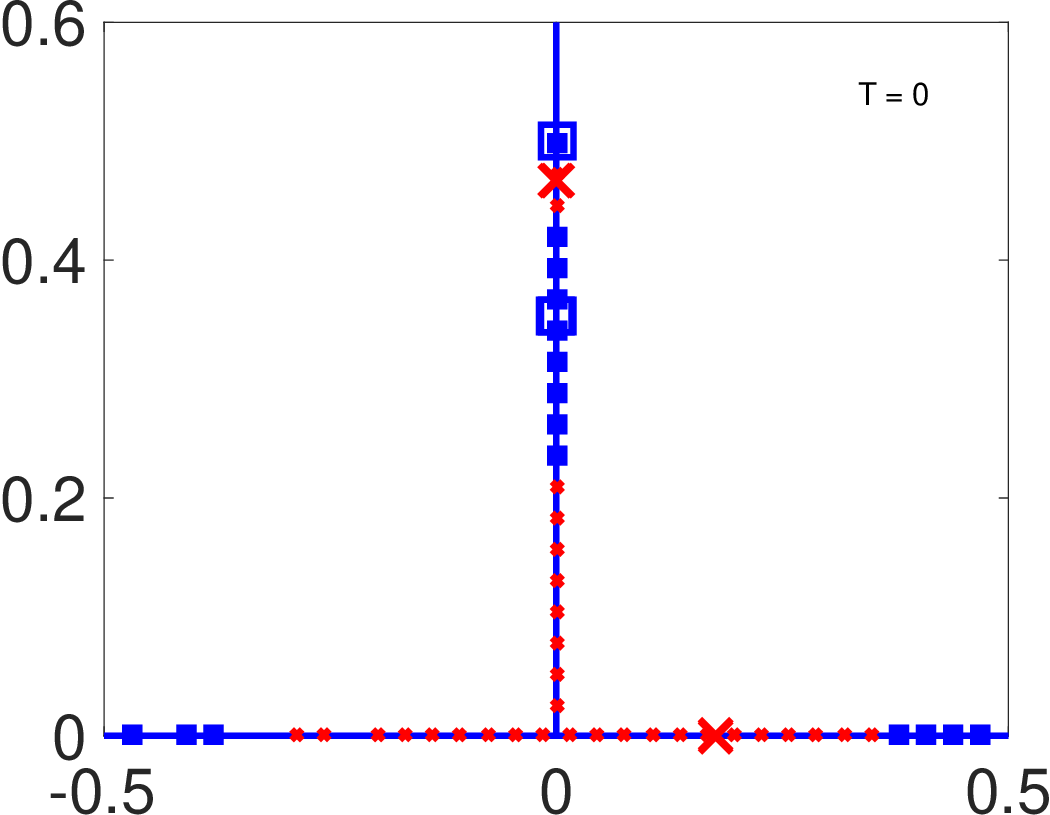}
\hspace{6pt}\includegraphics[width=0.225\textwidth]{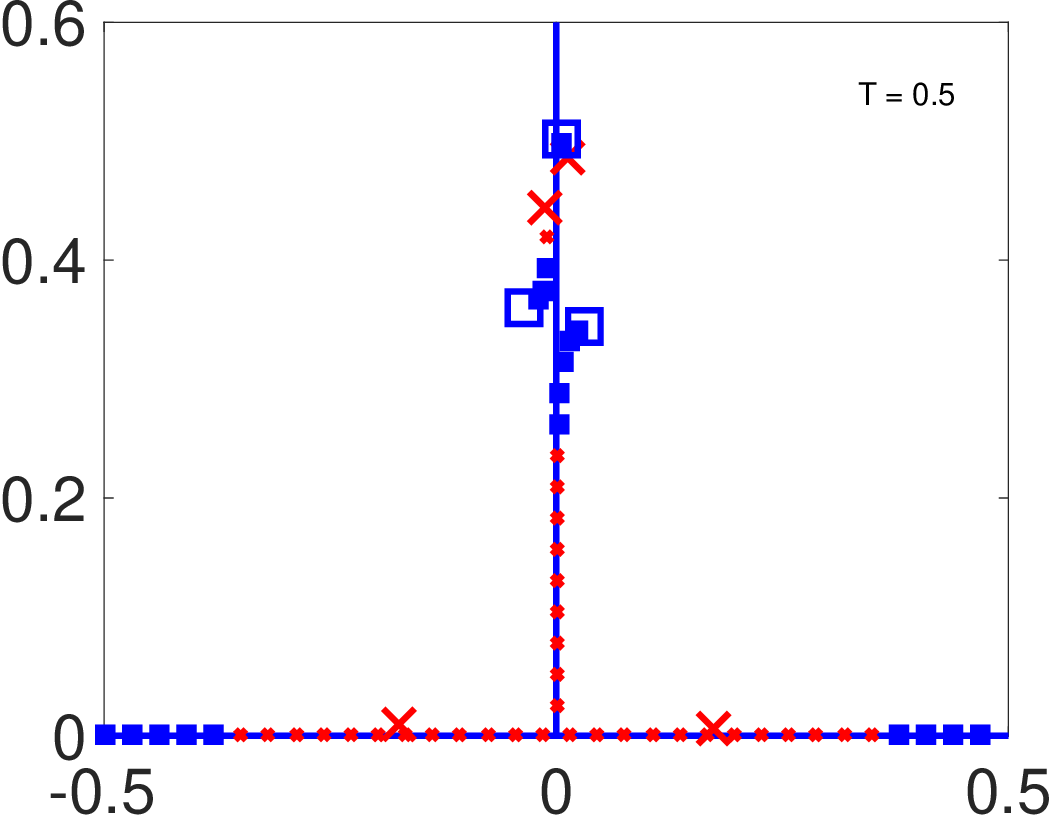}
\hspace{6pt}\includegraphics[width=0.225\textwidth]{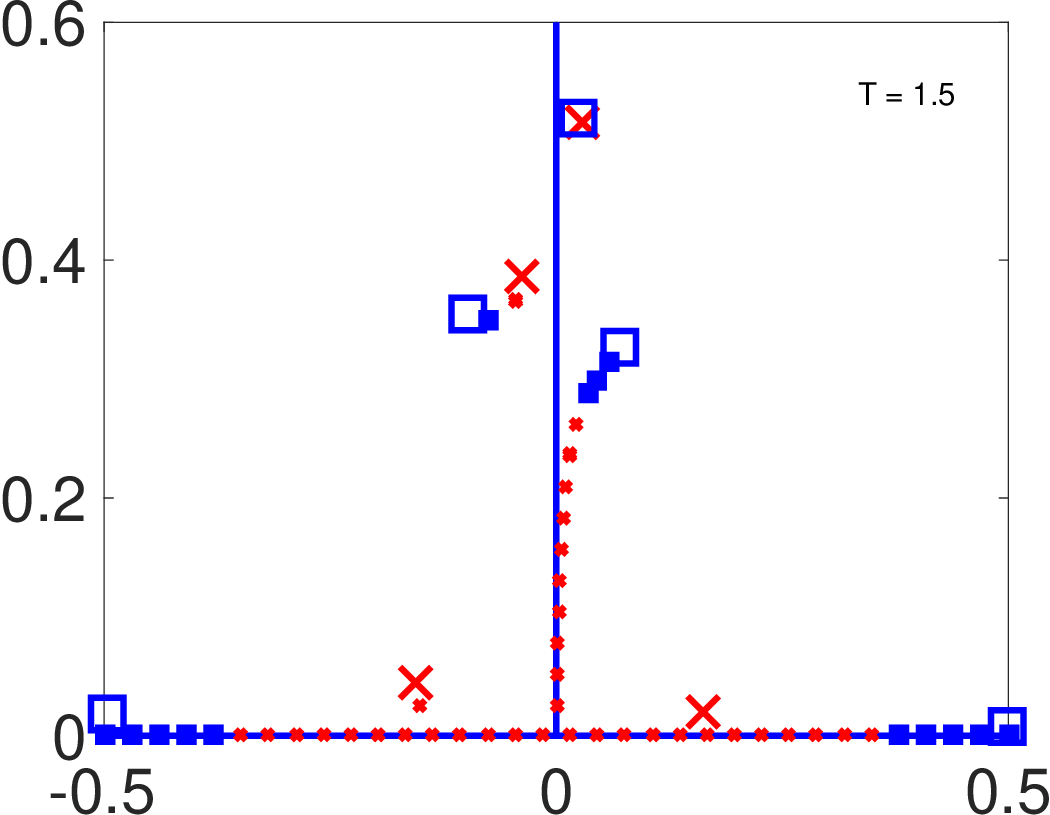}
\hspace{6pt}\includegraphics[width=0.225\textwidth]{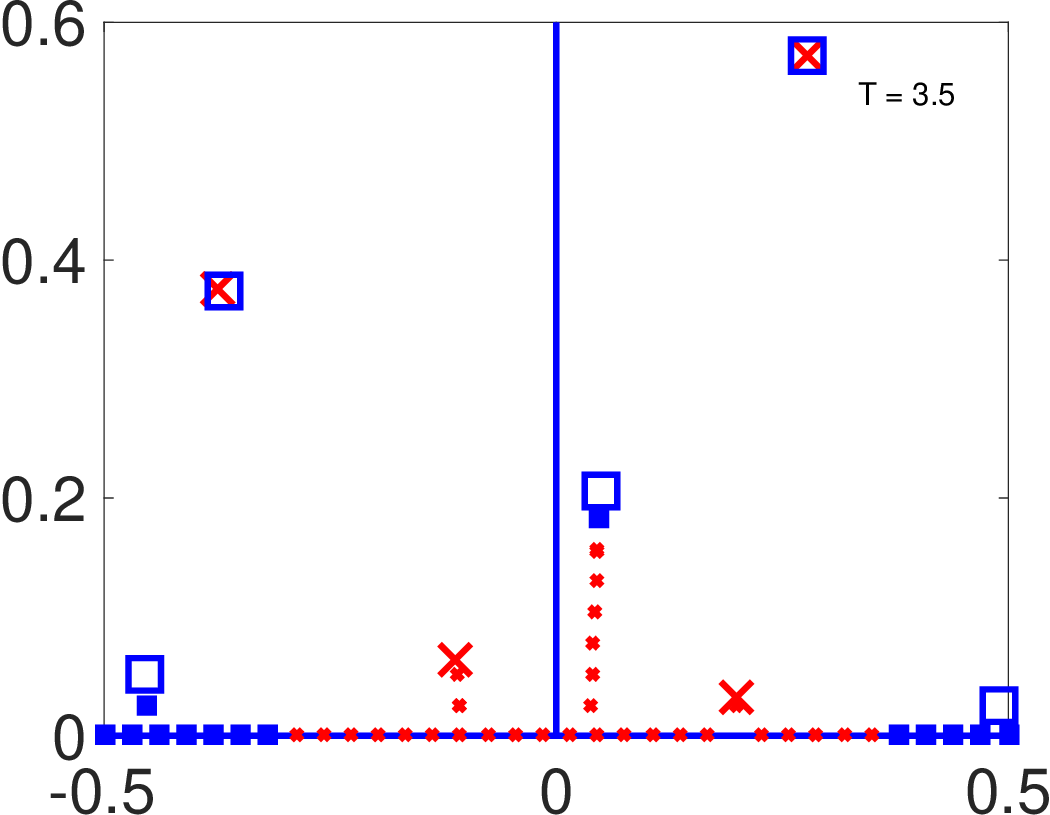}
  }
  \caption{HONLS evolution with SPB initial data \rf{SPB_ic}: Floquet spectra at (a) $t = 0$, (b ) $t =  0.5$ (where the characteristic asymmetric spectrum is clearly visible), (c) $t = 1.5$ (one-mode soliton-like state), and
    (d) $ t = 3.5$ (two-mode SRW event).}
  \label{ic_SPB_spec}
  \end{figure}

The Floquet spectrum of the SPB initial condition
is the same as for the underlying Stokes wave, 
consisting of a single band of spectrum with end point at $\lambda_0^s$ ($\in$ $\sigma^s$) indicated by a box, and two imaginary double points $\lambda_1^d$ and $\lambda_2^d$, indicated by an `$\bigtimes$' and a ``box'' respectively, given in
Figure~\ref{ic_SPB_spec}(a).

The spectral evolutions for all three models
-- HONLS,  V-HONLS, and NLD-HONLS -- are qualitatively similar and closely aligned
until $t \lessapprox 3.5$.
The spectrum rapidly evolves to
the characteristic
 asymmetric configuration shown in Figure~\ref{ic_SPB_spec}(b)  at approximately
 $t \approx 0.5$.
 This short time spectral behavior is supported by
 perturbation analysis for
 the HONLS and NLD-HONLS models \cite{SI2021}.
 
 The two upper spectral bands separate asymmetrically from the
 imaginary axis: the uppermost  band, denoted as $\gamma_1$, initially migrates
 into the right quadrant while the other, $\gamma_2$, shifts into the left quadrant. The  band $\gamma_1$ 
 rapidly contracts with its length $|\gamma_1|$ meeting the soliton-like criterion \rf{s_state}, indicating entry into a one soliton-like state at $t = 1.5$ (Figure~\ref{ic_SPB_spec}(c)).

\begin{figure}[htp!]
\centerline{
\includegraphics[width=0.25\textwidth]{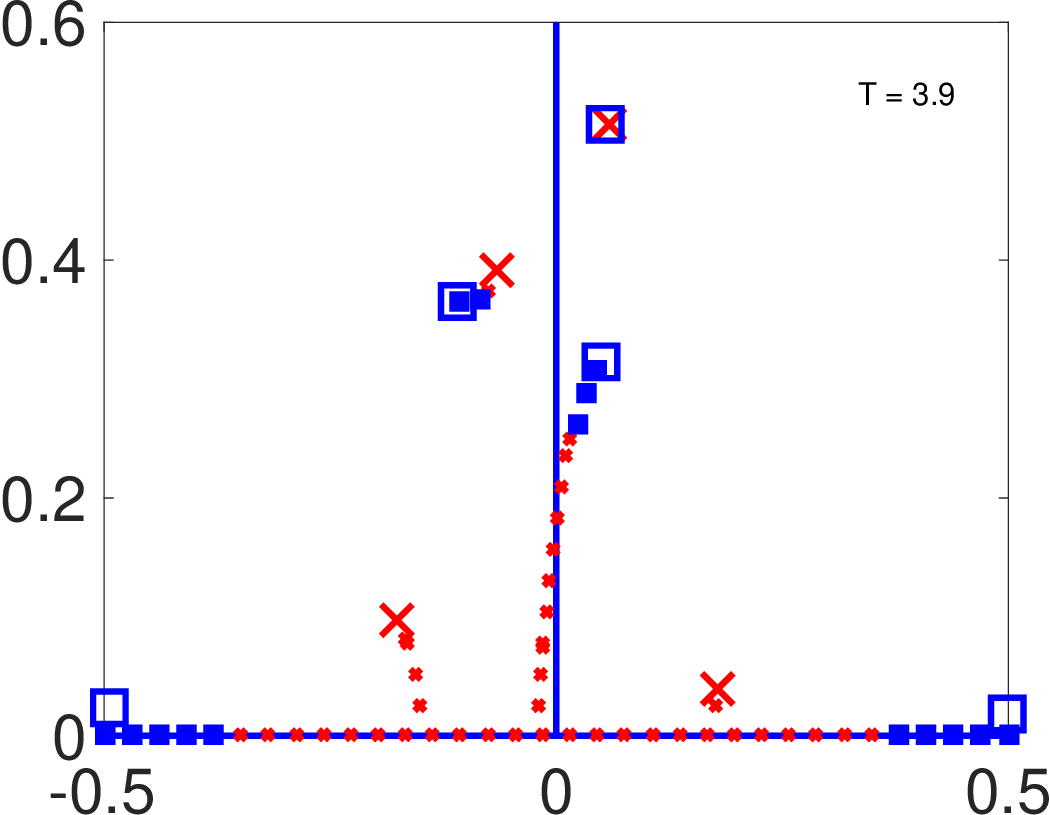}
\hspace{6pt}\includegraphics[width=0.25\textwidth]{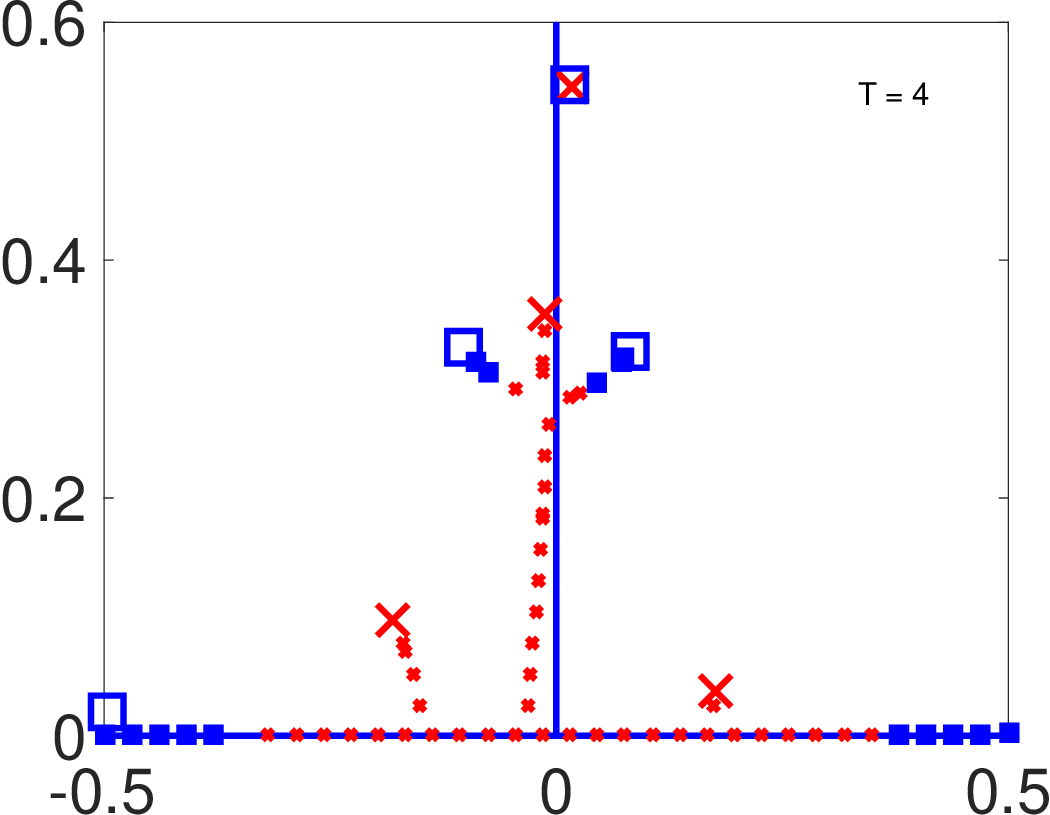}
\hspace{6pt}\includegraphics[width=0.25\textwidth]{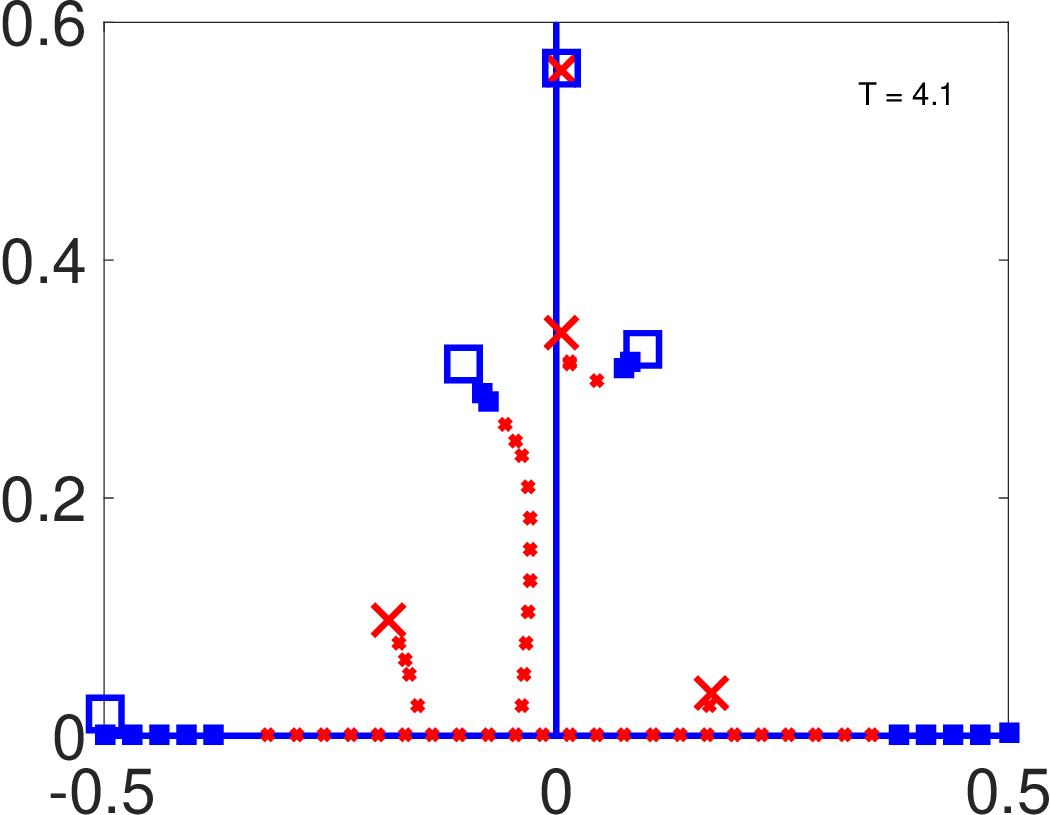}
}
\vspace{6pt}
\centerline{
\includegraphics[width=0.25\textwidth]{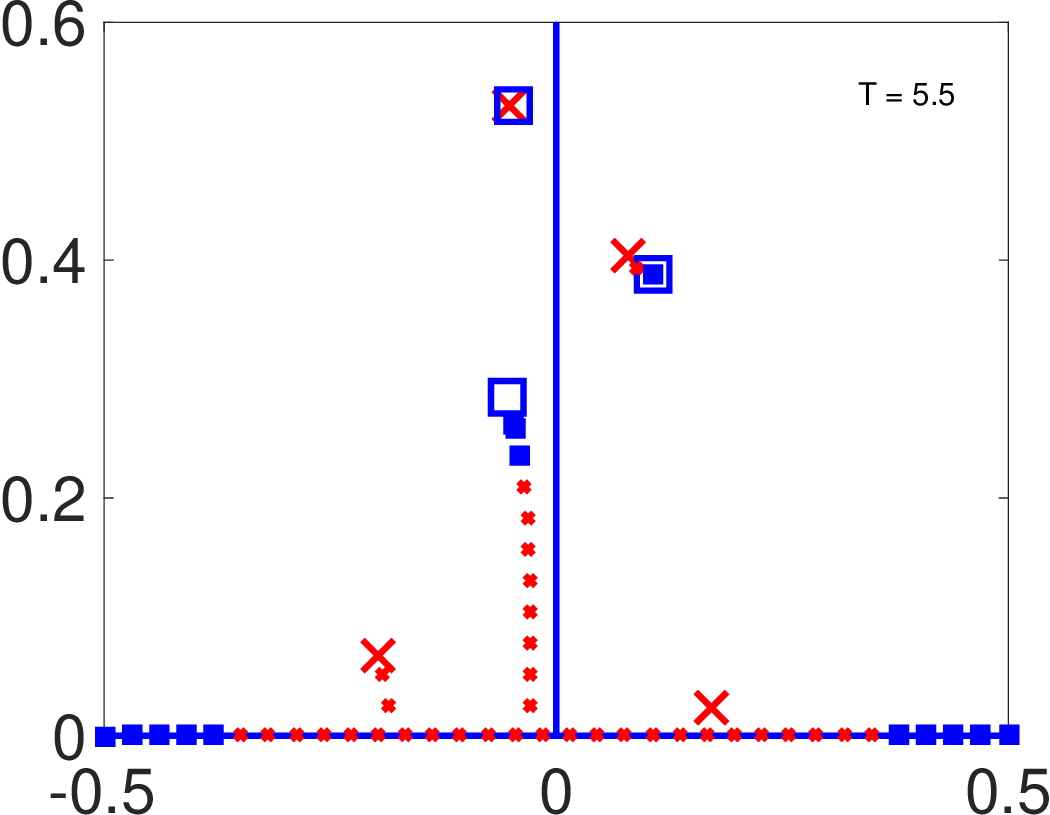}
\hspace{6pt}\includegraphics[width=0.25\textwidth]{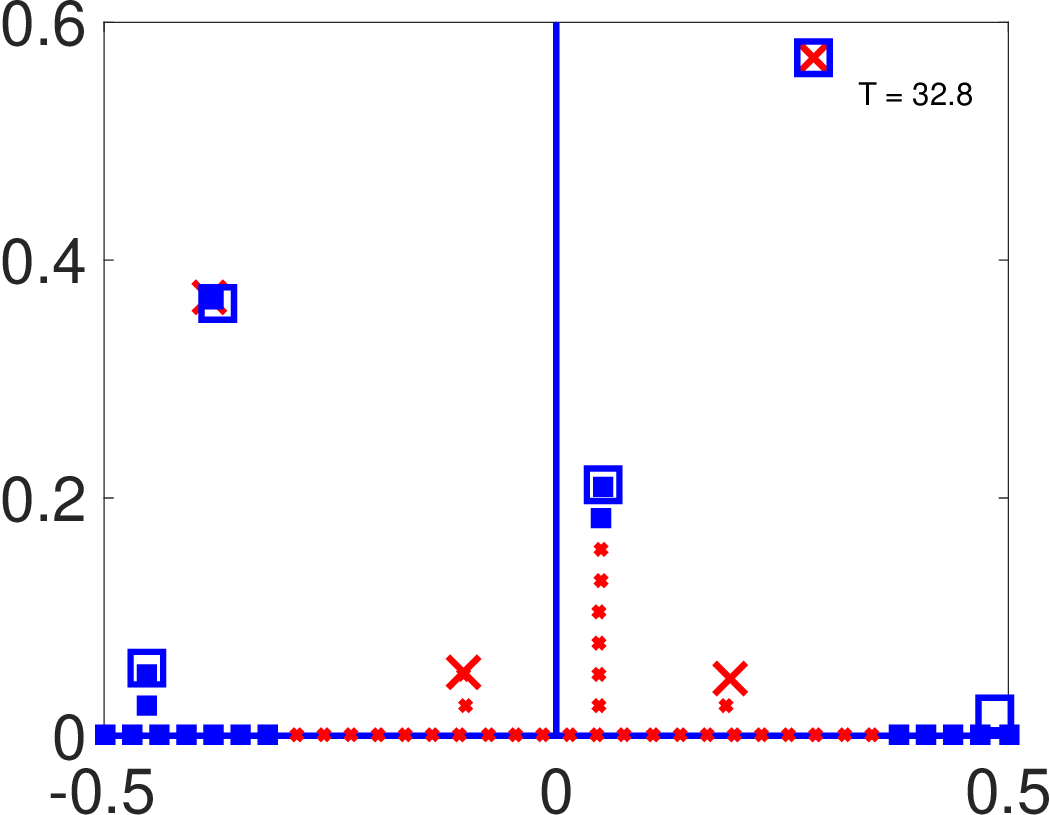}
  \hspace{6pt}\includegraphics[width=0.25\textwidth]{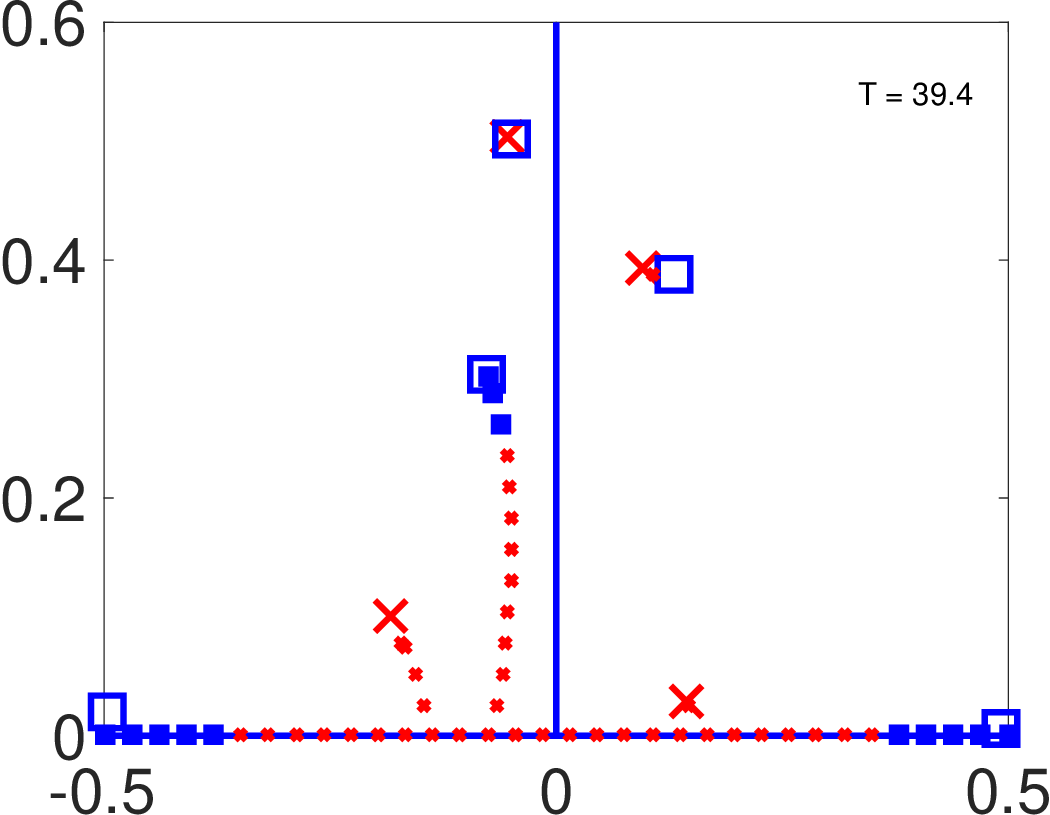}
}
\vspace{6pt}
\centerline{\includegraphics[width=0.25\textwidth]{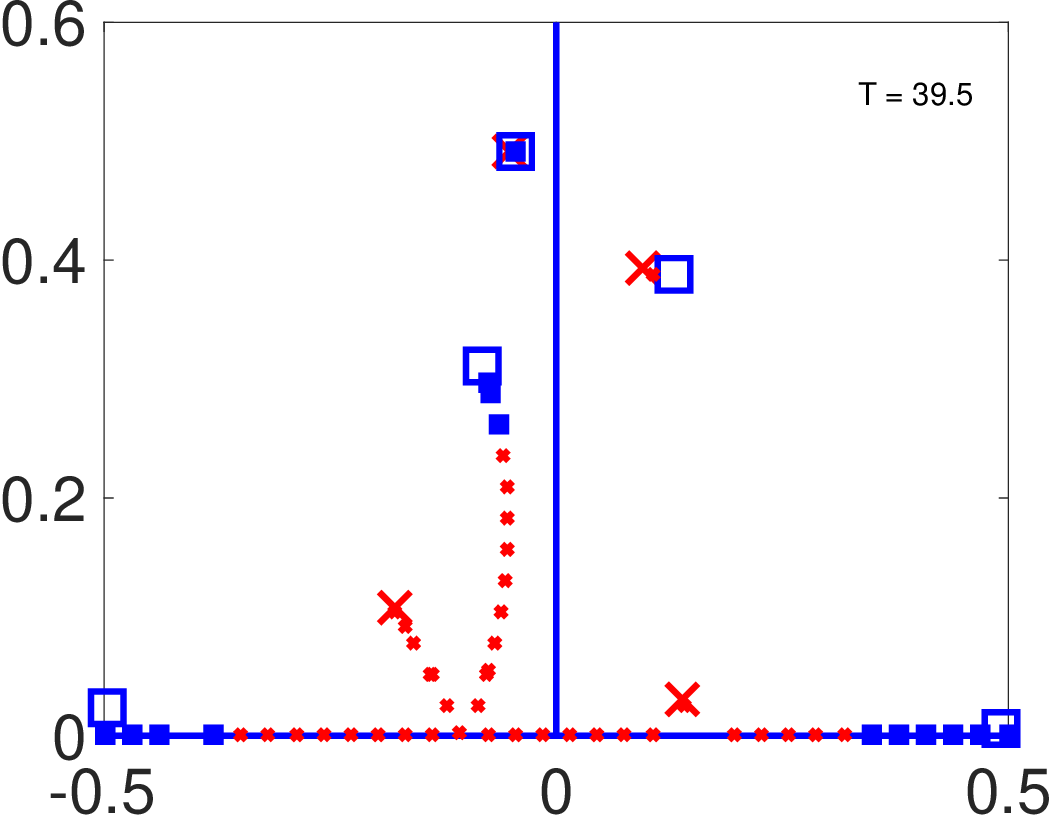}
\hspace{6pt} \includegraphics[width=0.25\textwidth]{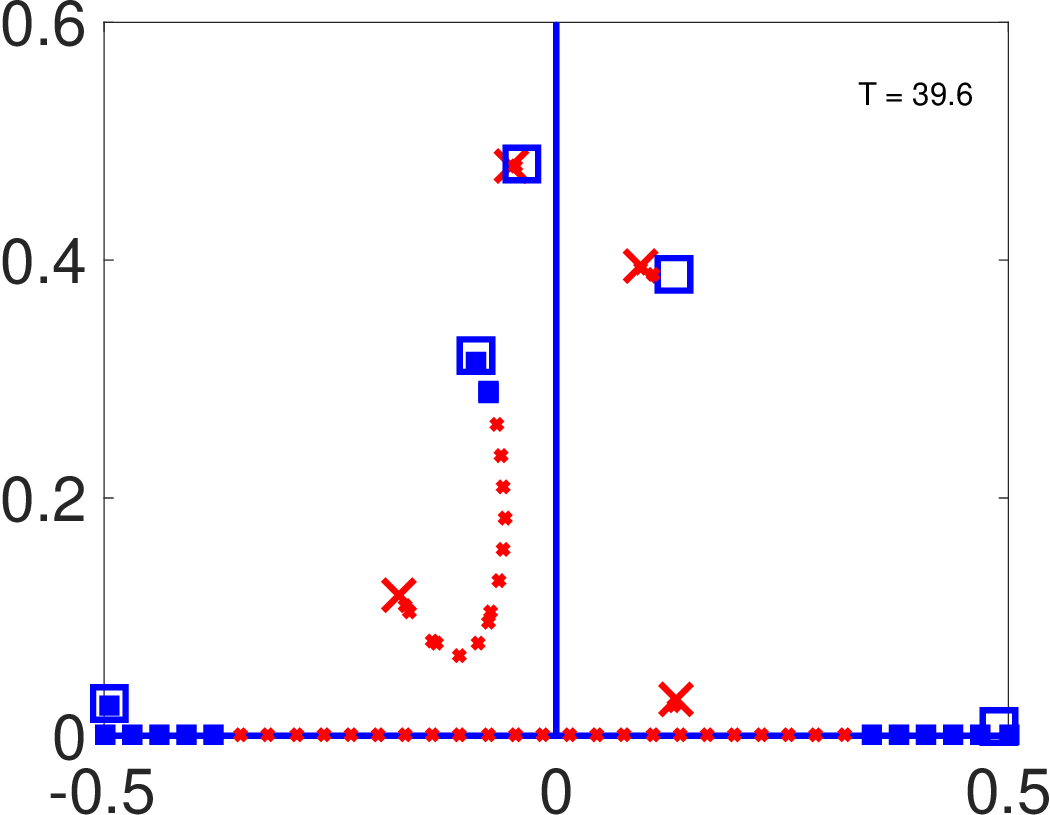}
\hspace{6pt}\includegraphics[width=0.25\textwidth,height=0.14\textheight]{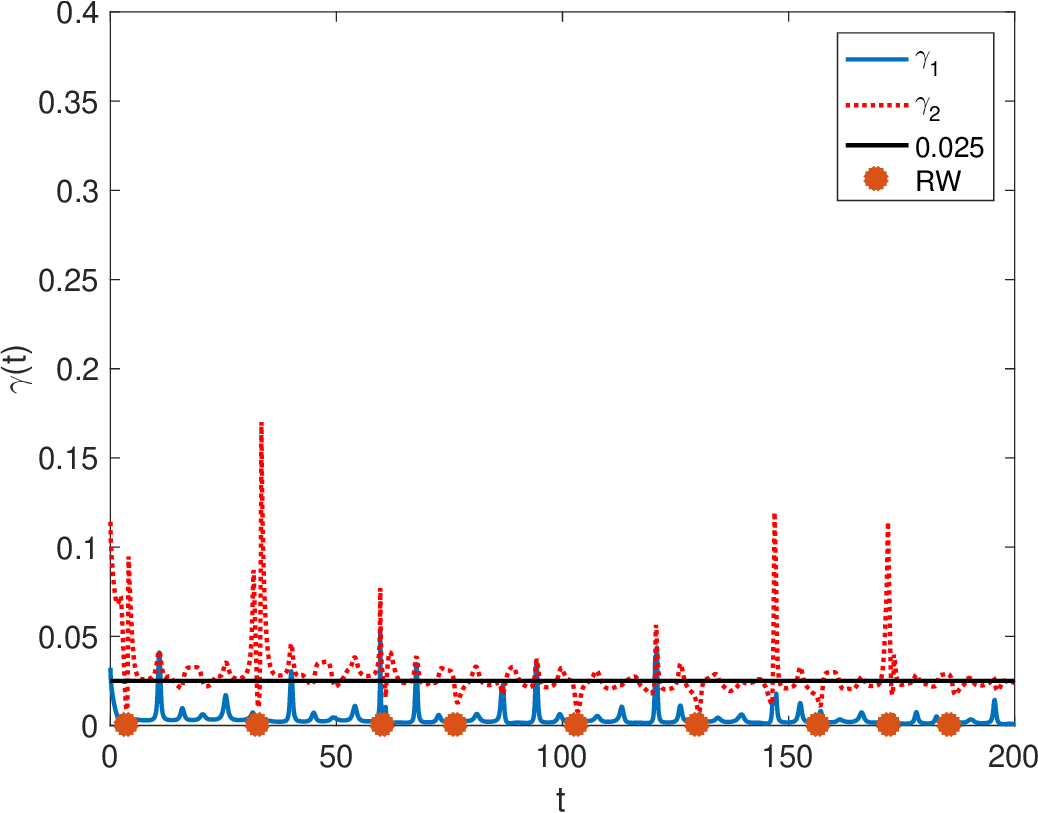}
}
\caption{HONLS evolution with SPB initial data \rf{SPB_ic}:  Floquet spectra at  (a) $t= 3.9$, (b) $t=4.0$,
  (c)$t = 4.1$, (d) = $t = 5.5$, (e) $t = 32.8$,
  (f) $t=39.4$, (g) $t = 39.5$, (h) $t = 39.6$, and 
(i) the band lengths $|\gamma_1(t)|$ and $|\gamma_2(t)|$ with the horizontal line indicating the soliton-like threshold
  and  red dots marking rogue wave events.}
\label{HONLS_SPB_spec}
\end{figure}

 The length $|\gamma_2|$ typically exceeds
 the  soliton-like threshold. However, at the time of the first rogue wave event,
 $t = 3.5$, $|\gamma_2|$ undergoes a pronounced contraction (see Figure \ref{ic_SPB_spec}(d)),
 indicating the solution enters a transient two-mode soliton-like state
 (see also Figure~\ref{HONLS_SPB_spec}(i)). As a result, this rogue wave can be
 interpreted as a two-mode SRW.

 Following this initial rogue wave event, the spectral evolutions of the three models  diverge.

\vspace{6pt}
\noindent {\bf Later stage HONLS dynamics:} 
 The two-mode SRW configuration at $t = 3.5$ is short lived:  $\gamma_2$ quickly re-expands and goes through a complex critical point crossing at $t = 4$
 as shown in Figures~\ref{HONLS_SPB_spec}(a)-(c).  
 Throughout the simulation, further complex critical point transitions 
 occur, followed by the band $\gamma_1$  switching quadrants
 (Figures~\ref{HONLS_SPB_spec}(d)), which reflects  a transition in the 
 associated  mode's propagation direction, e.g.  from left-
 to right-traveling. 
 Notably, $|\gamma_1|$ remains below the soliton-like threshold for
 much of the evolution (see Figure~\ref{HONLS_SPB_spec}(i)) and tends to stay in the left quadrant.

Rogue waves continue to form  throughout the simulation.
As each rogue wave forms, $\gamma_2$ briefly contracts in length,
resulting in   a persistent one-mode soliton-like regime which intermittently becomes a two-mode regime as the 2-mode SRW forms.
A representative example appears at $t = 32.8$ in Figure~\ref{HONLS_SPB_spec}(e).

Real critical point transitions also play a role.
Figures~\ref{HONLS_SPB_spec}(f)-(h) show how  two bands
(one linked to the carrier wave and the other linked to a third nonlinear mode) merge at a degenerate real critical point before a new band separates into
the upper complex plane. This new band,
Figure~\ref{HONLS_SPB_spec}(h),  corresponds to a multi-mode background state with a more  diffuse spatial structure. The spectrum  eventually reverses
back to the earlier  configuration.

 The  evolution of the band lengths $|\gamma_1(t)|$
 and $|\gamma_2(t)|$ and their  connection to SRW formation
 is summarized in   Figure \ref{HONLS_SPB_spec}(i).
The horizontal line denotes  the soliton-like threshold,
 based on criterion \rf{s_state} and
red dots indicate rogue wave events.
Most of the time, the spectrum is in a one-mode
  soliton-like configuration,
  representing a  hybrid structure,
  partially  localized,  with only one spectral component in
 a soliton-like regime and critical point transitions occurring in the other nonlinear modes.
  
   Each rogue wave coincides with both bands being below the soliton-like threshold, although the second band only briefly  dips below this level at the moment the rogue wave forms.
   The third band in Figure~\ref{HONLS_SPB_spec}(g) is excluded from this plot, as it appears only intermittently, remains large, and is
   absent during   rogue waves formation.
These results support interpreting
the rogue waves as  two-mode SRWs within intermittent, transient two-mode soliton-like regimes.

\begin{figure}[htp!]
  \centerline{
\includegraphics[width=0.3\textwidth]{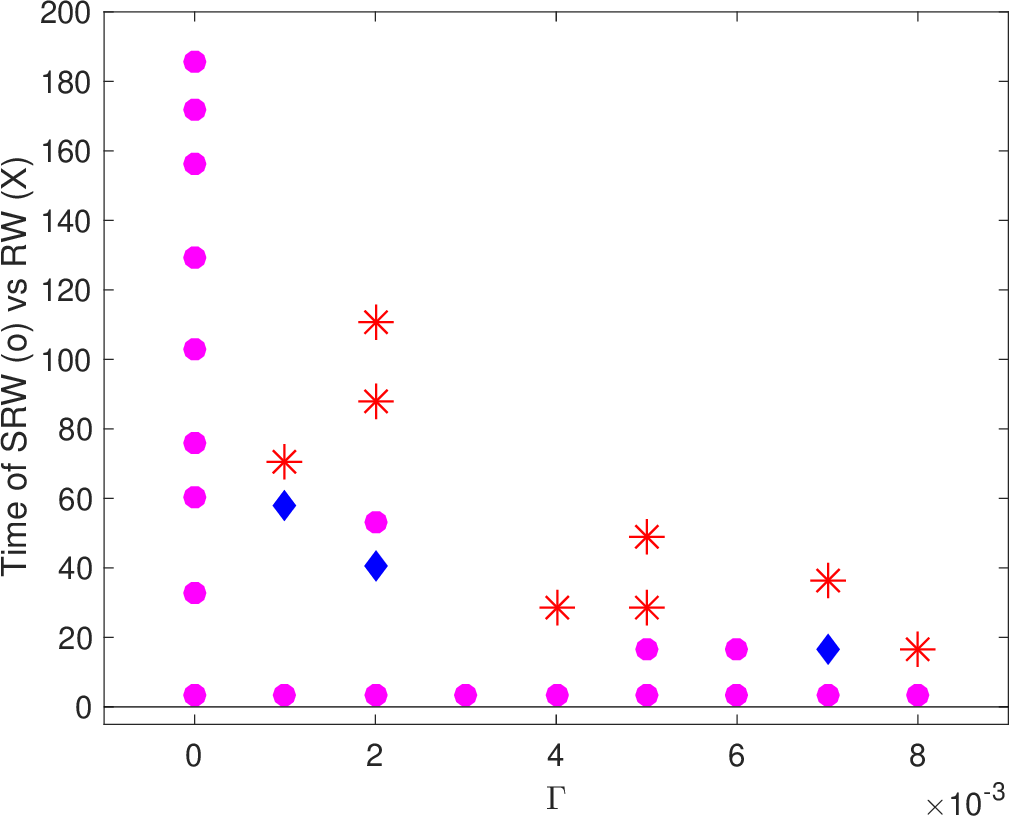}
\hspace{6pt}\includegraphics[width=0.3\textwidth]{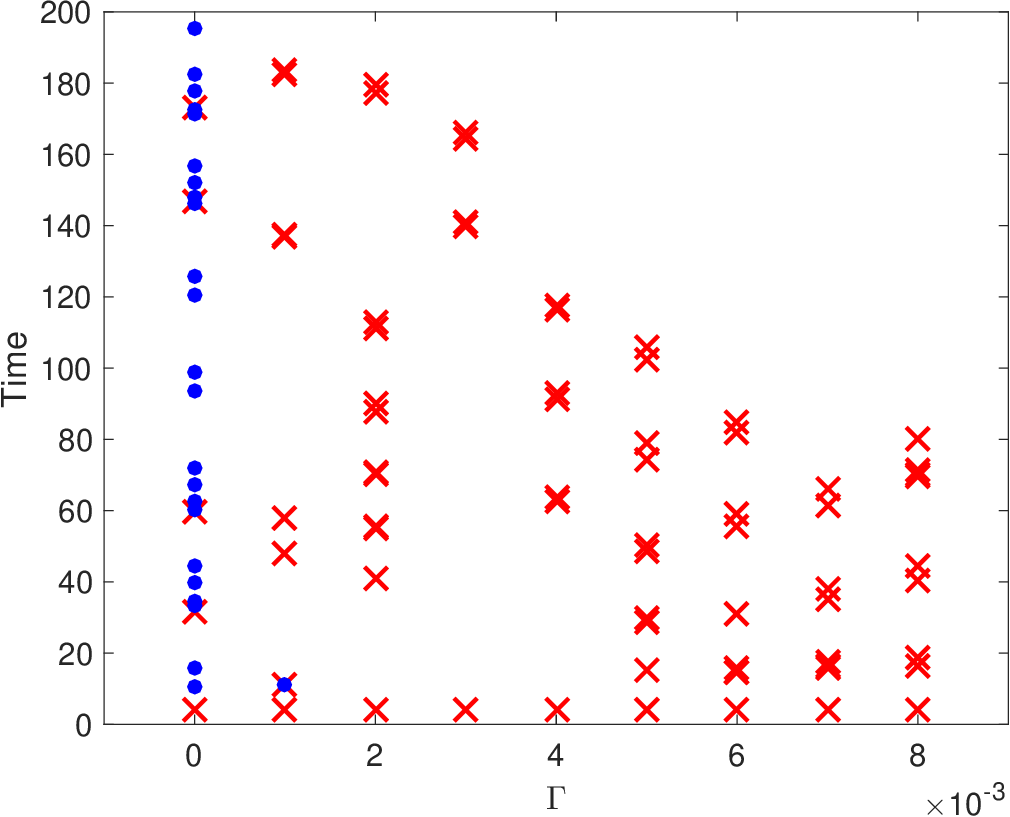}
\hspace{6pt}\includegraphics[width=0.3\textwidth]{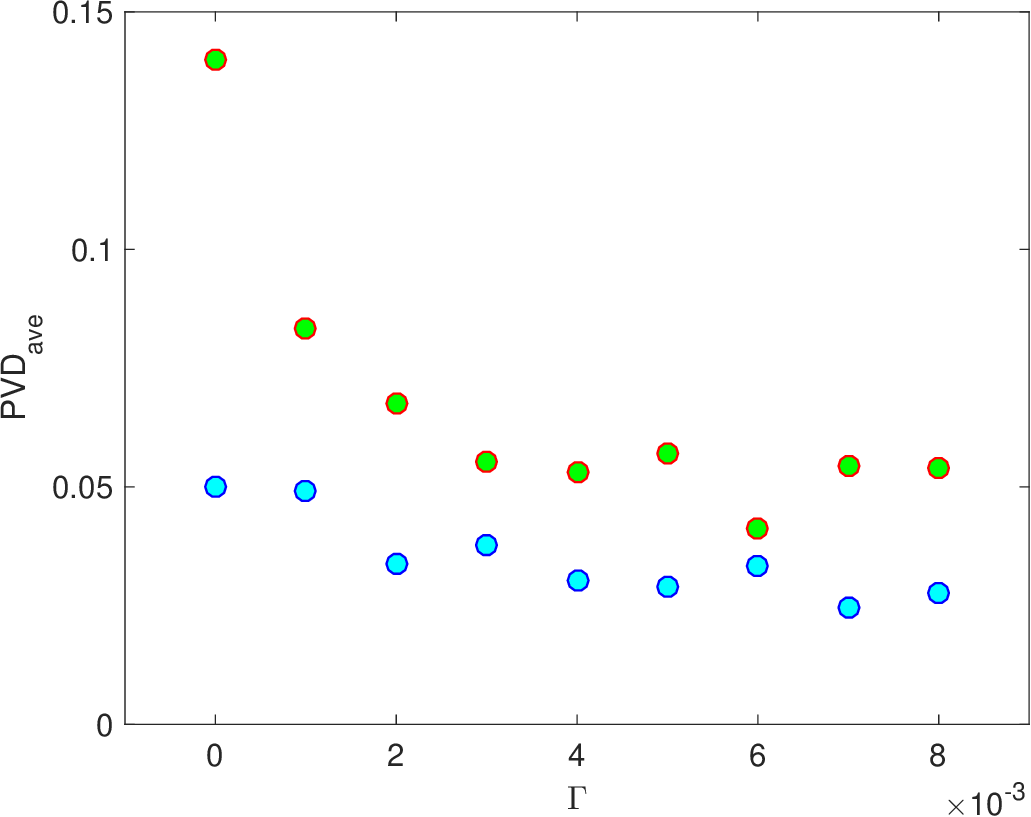}
}
  \caption{HONLS evolution ($\Gamma = 0$) and V-HONLS evolutions  for varying damping parameter $\Gamma$: Timelines using SPB initial data 
    showing (a) SRW and generic rogue wave events (Magenta dots: 2-mode SRWs, blue diamonds: 1-mode SRWs,
 and stars:  generic rogue waves), and 
  (b) critical point transitions: blue dots for real,  red '$\bigtimes$' for complex; 
 (c) average PVD for SPB (light blue) and Stokes (green) initial data.
  }
  \label{VHONLS_vary_Gamma}
\end{figure}

 Timelines of  rogue wave events and critical point transitions  for SPB initial data are shown 
 in  Figure~\ref{VHONLS_vary_Gamma}(a)-(b), which
 include the HONLS case when $\Gamma = 0$.  
In Figure~\ref{VHONLS_vary_Gamma}(a),
 magenta dots correspond to  two-mode SRWs, blue diamonds to
one-mode SRWs, and stars represent generic,
  non soliton-like rogue waves. 
 In Figure~\ref{VHONLS_vary_Gamma}(b), blue dots indicate real critical point transitions while red `$\bigtimes$' marks indicate  complex critical point crossings.

  The evolution of the Floquet spectrum 
  suggests the system  typically remains close to  one-mode soliton-like states with frequent brief excursions close to a two-mode
  soliton-like state. This interpretation is
further supported by the phase variance diagnostic (PVD).
Figure~\ref{HONLS_SPB_str}(c) shows the evolution of the PVD for steep SPB initial data. In particular, small values of the PVD  are observed
at the times of SRW events, indicated by red dots in the PVD evolution plot.
Figure~\ref{VHONLS_vary_Gamma}(c) shows the average PVD for the HONLS ($\Gamma =0$) remains low.
This reflects
the enhanced phase alignment and localization of SRWs  in the 
dynamics driven by steep waves.
The spikes in the PVD arise from the intermittent appearance of the
third, dephased band which emerges  when the system is not in a SRW state.

\vspace{6pt}
\noindent{\bf Summary of Results (HONLS, Steep wave initial data):}
For SPB initial data, the HONLS evolution spends most of the time in
a partially localized one-mode soliton-like state.
Two-mode SRWs arise intermittently.
The two-mode soliton-like regime is destabilized by  critical point crossings
(both real and complex), leading to
repeated transitions between localized and more diffuse multi-mode
configurations.

Unique to the HONLS evolution, the Floquet spectrum also
 shows the emergence of a third band which arises from the interaction of the carrier wave with the third nonlinear mode and
 resulting in a multi-mode background state with a
 more  diffuse structure.
 Despite these spectral transitions, the average phase variance (PVD) remains low, reflecting the
 dominant influence  of the persistent one-mode soliton-like state.

\subsubsection{Steep Wave Initial Data: V-HONLS equation}

Having established the baseline behavior in the undamped system, we now study
how uniform viscous damping, as introduced in the V-HONLS model, alters both the generation of SRWs and the underlying background dynamics, as reflected in the evolution of the Floquet spectrum.

 The HONLS simulations were conducted over the interval $0 < t <200$ 
 to capture the sustained chaotic dynamics characterized by persistent critical
 point crossings and recurring rogue wave events.
 In contrast, damping in the  V-HONLS model eventually
 suppresses rogue wave formation entirely, with such events disappearing
 before $t = 100$
for all damping strengths -- except in the case of 
$\Gamma = 0.002$ where an additional  rogue wave occurs at $t = 115$.

Interestingly, the number of rogue wave events does not decrease monotonically with increasing damping. This non-monotonic behavior is due to the underlying chaotic background, which introduces sensitivity to initial conditions and parameter variations, leading to irregularities in rogue wave emergence despite an increase in the
dissipation strength.

\begin{figure}[htp!]
\centerline{
\includegraphics[width=0.3\textwidth]{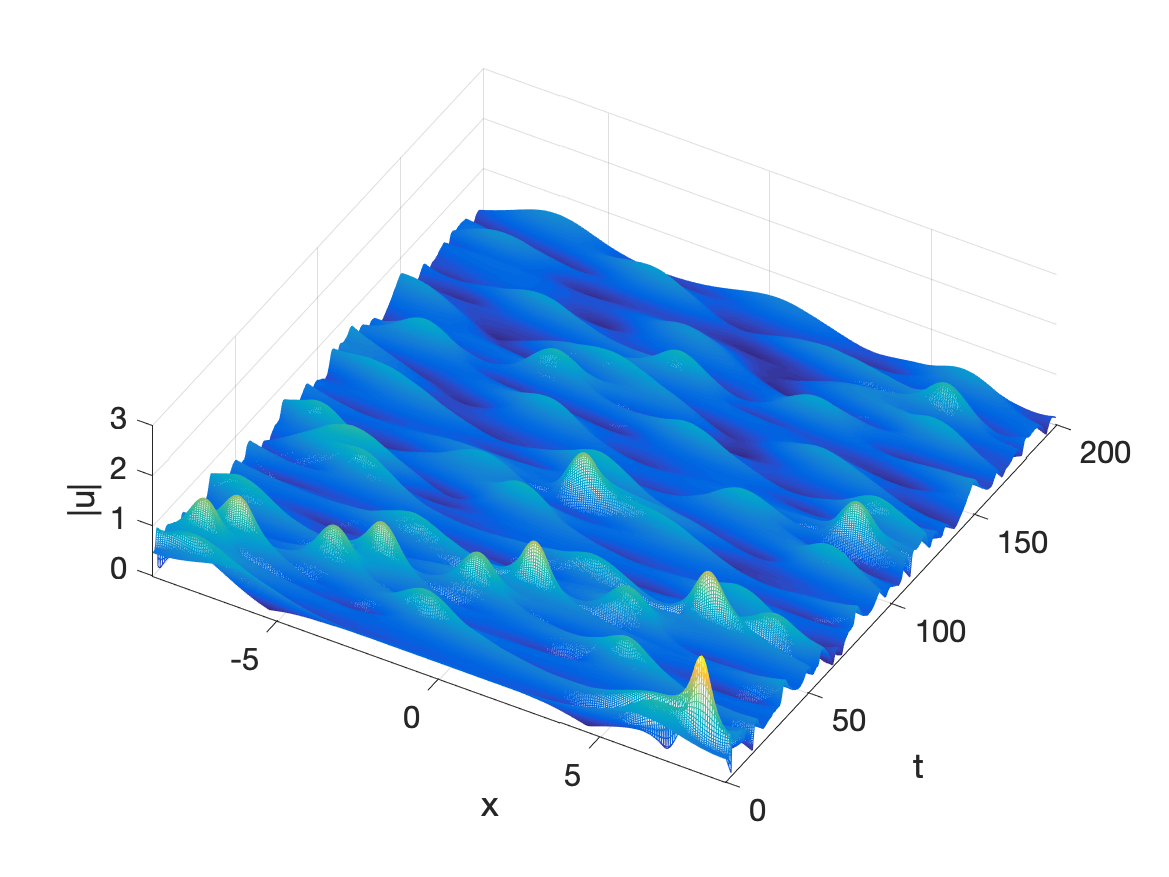}
\hspace{6pt}\includegraphics[width=0.25\textwidth]{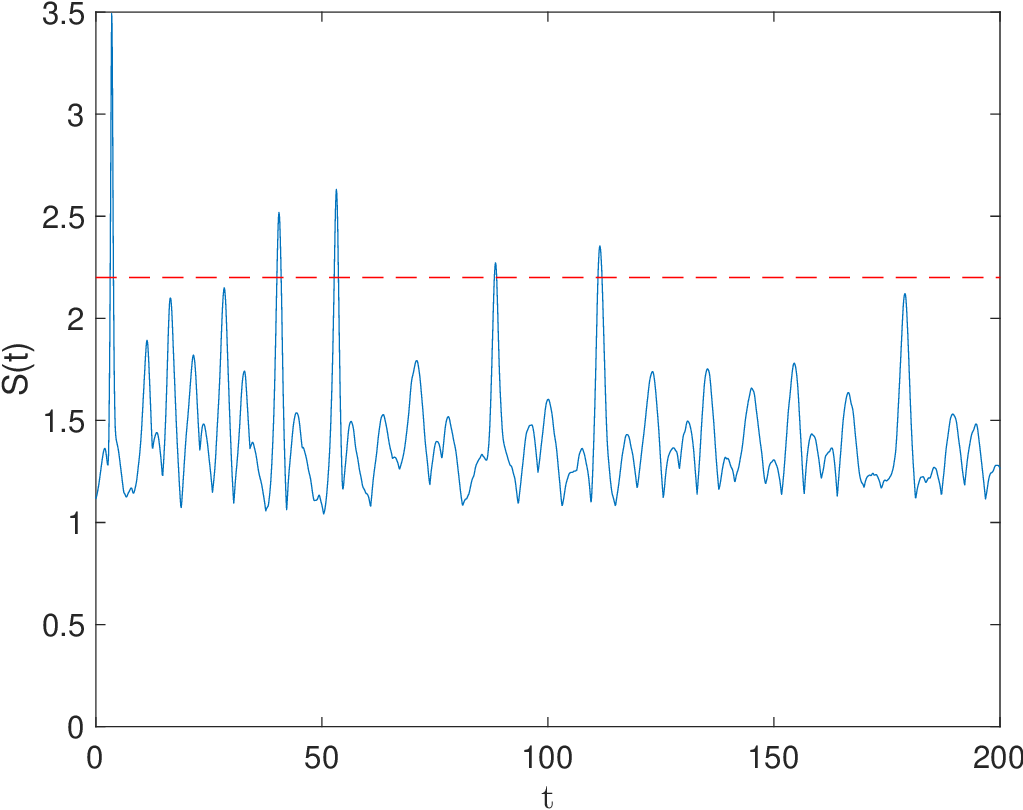}
\hspace{6pt}\includegraphics[width=0.25\textwidth]{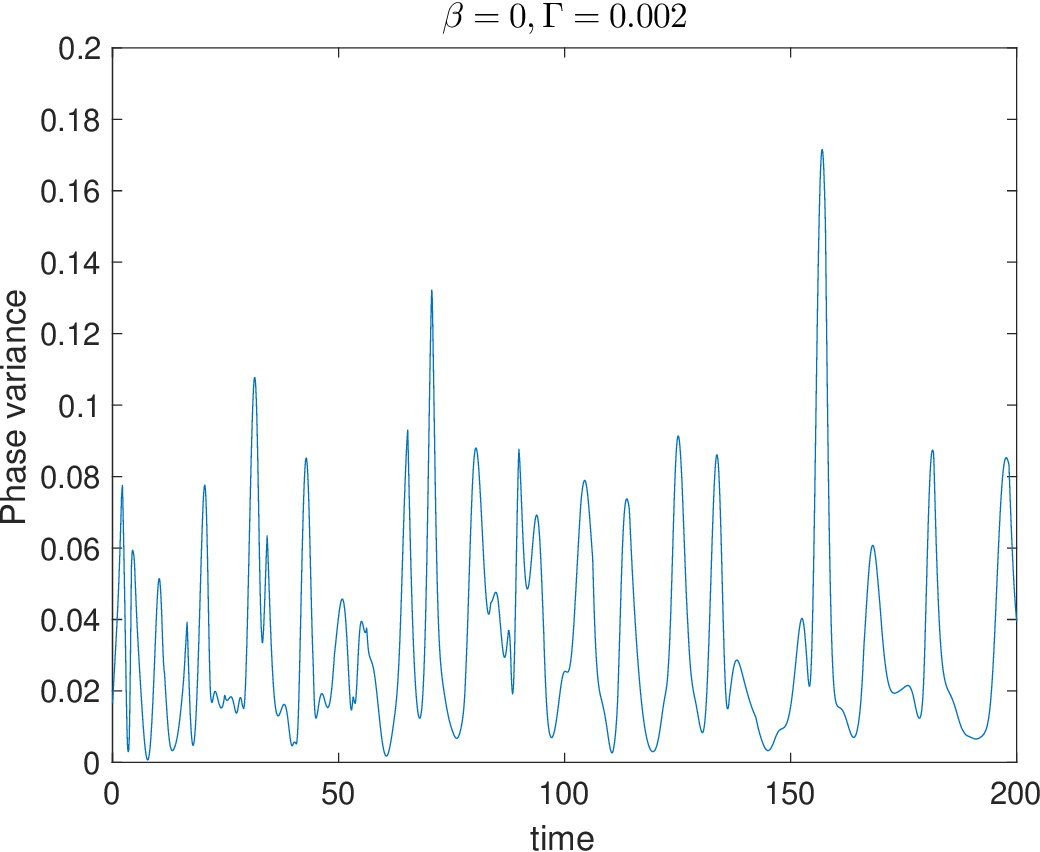}
}
\caption{V-HONLS with $\Gamma = 0.002$ and SPB initial data \rf{SPB_ic}: (a) $|u(x,t)|$ for $0\leq t\leq 200$, (b) the strength $S(t)$,
  (c) the phase variance PVD$(t)$.}
\label{VHONLS_SPB_str}
\end{figure}   

\begin{figure}[htp!]
\centerline{
\includegraphics[width=0.225\textwidth]{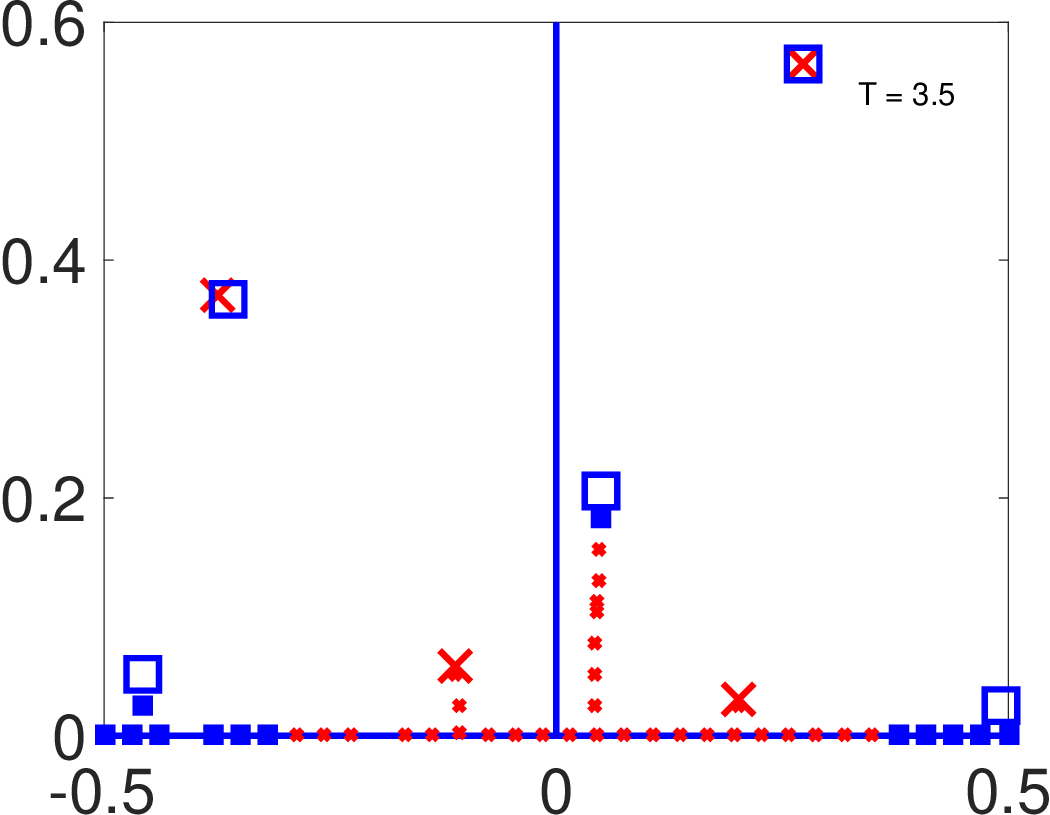}
\hspace{6pt}\includegraphics[width=0.225\textwidth]{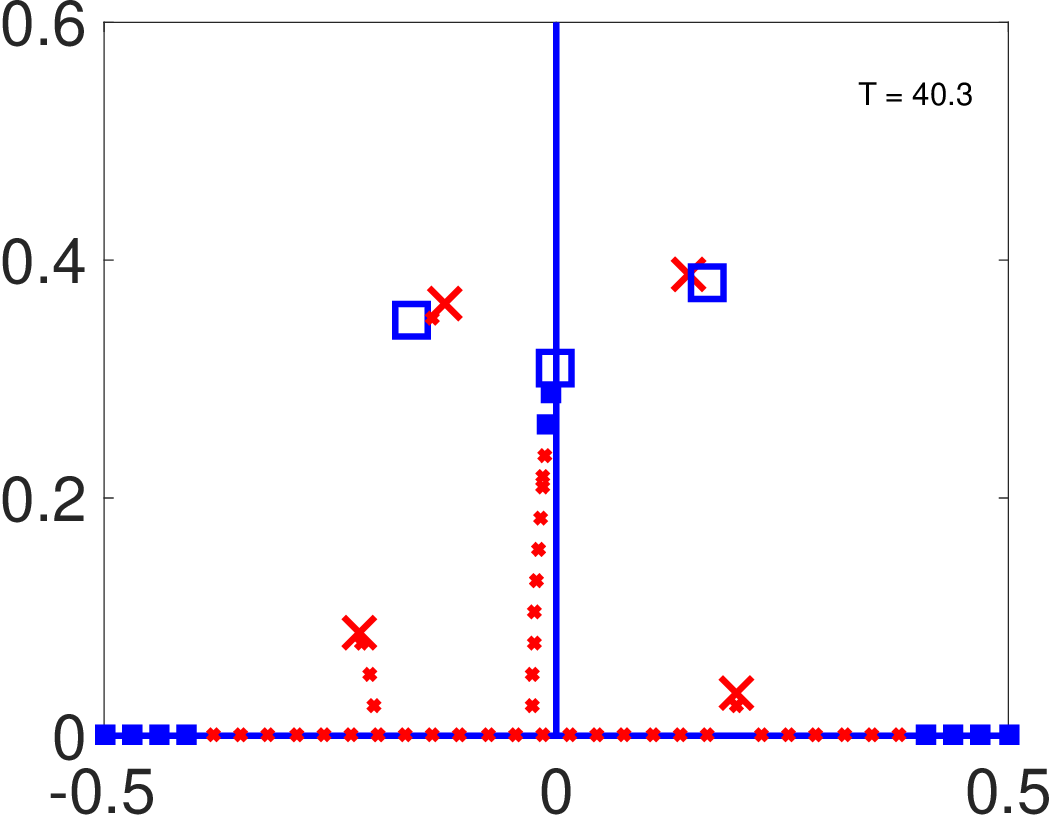}
\hspace{6pt}\includegraphics[width=0.225\textwidth]{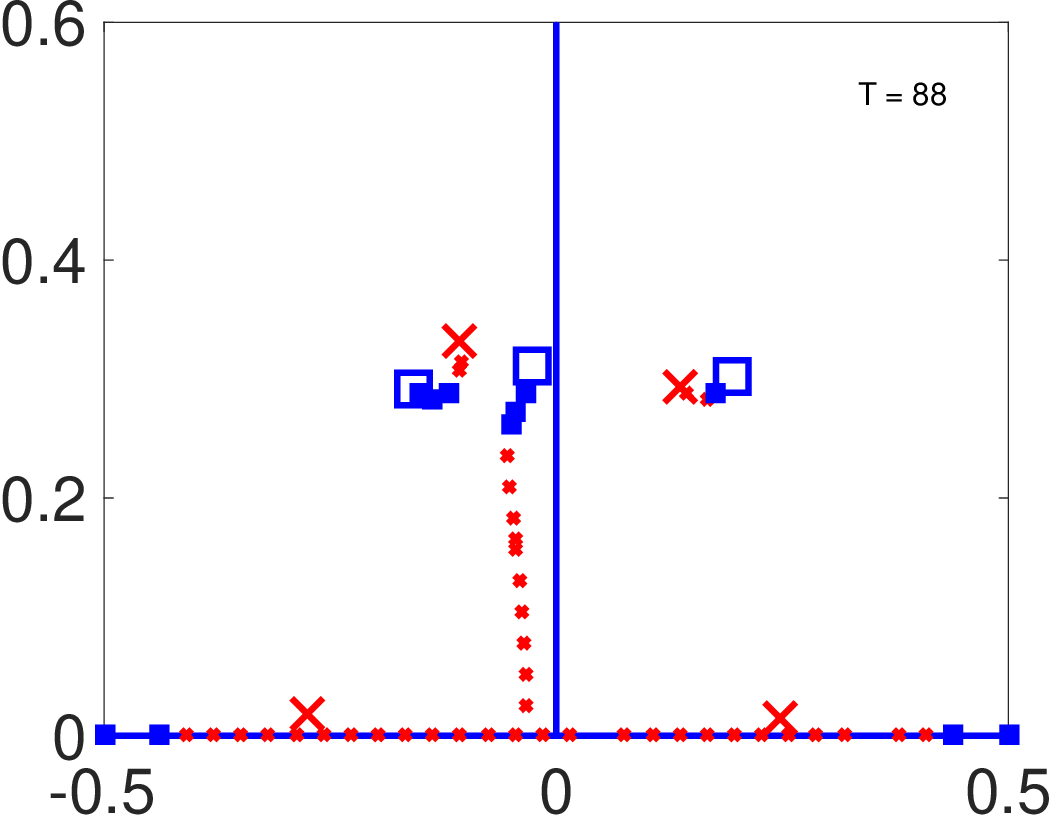}
\hspace{6pt}\includegraphics[width=0.225\textwidth,height=0.125\textheight]{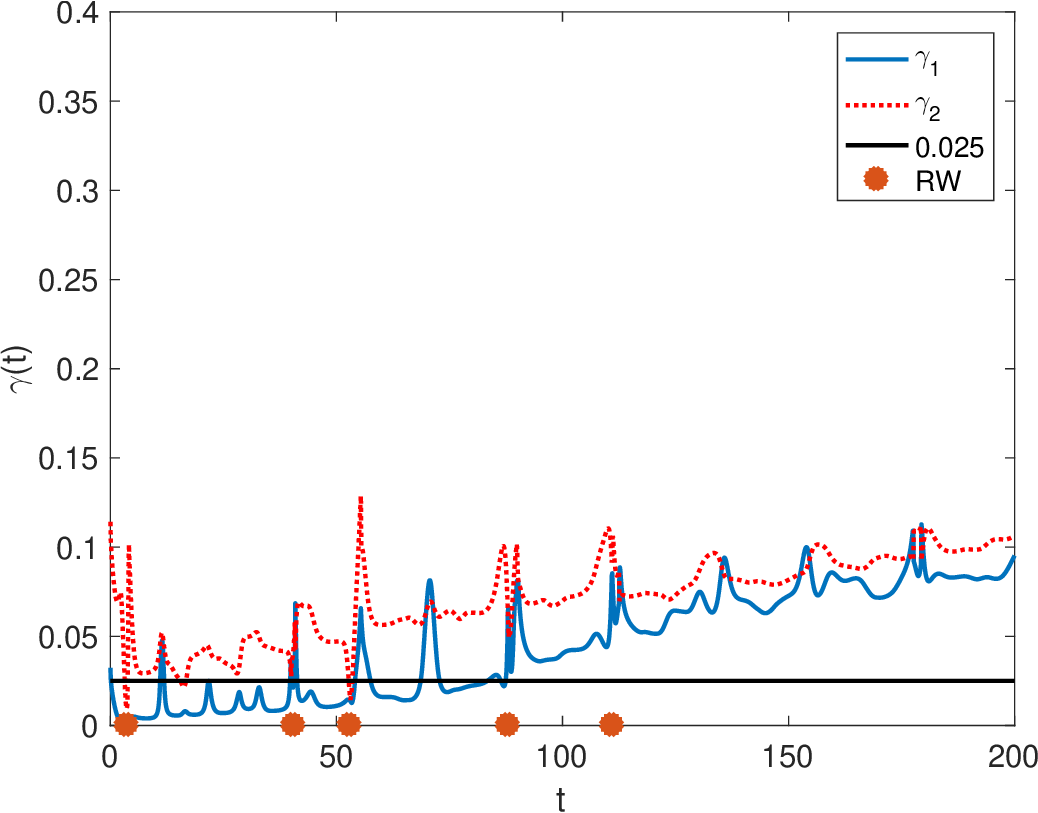}
}
\caption{V-HONLS evolution with $\Gamma = 0.002$ and SPB initial data \rf{SPB_ic}: Floquet spectra at (a) $t=3.5$, (b) $t = 40.3$ (1mode SRW) and  (c) $t = 88$ (generic RW); (d) the band lengths $|\gamma_1(t)|$ and $|\gamma_2(t)|$ with the horizontal line indicating the  soliton-like threshold
  and  red dots marking rogue wave events. 
}
\label{VHONLS_SPB_spec}
  \end{figure}

\vspace{6pt}
\noindent {\bf The Floquet spectral evolution up to the first rogue wave event:}
Figure~\ref{VHONLS_SPB_str}(a)  shows  the time evolution of the surface $|u(x,t)|$  for the SPB initial data \rf{SPB_ic} over the interval $0 < t < 200$.  For short times, $t < 3.5$,  the spectral evolution under V-HONLS  closely mirrors that of the undamped HONLS simulation. This similarity is particularly evident
at  $t = 3.5$, when  the V-HONLS enters a two-mode SRW configuration, as seen
by comparing
Figure~\ref{ic_SPB_spec}(d), and \ref{VHONLS_SPB_spec}(a).

\vspace{6pt}
\noindent {\bf Later stage V-HONLS dynamics:} 
 The strength plot in  Figure  ~\ref{VHONLS_SPB_str}(b)  indicates that, compared to the HONLS evolution for SPB data, the V-HONLS produces fewer rogue waves.
Figure~\ref{VHONLS_SPB_spec}(d) which tracks the band length evolution and also
 marks the rogue wave events with red dots,   shows that the first
 three SRWs are within a fluctuating one- or two-mode soliton-like regime(a representative one-mode SRW is given in
 Figure~\ref{VHONLS_SPB_spec}(b)). In contrast, the last two  are 
 broader, generic rogue waves without a coherent localized structure (Figure~\ref{VHONLS_SPB_spec}(c)).

 This classification is summarized in
 Figure~\ref{VHONLS_vary_Gamma}(a) which provides a timeline of the rogue wave
 events as a function of $\Gamma $ for SPB initial data. Magenta dots denote 2-mode SRWs, blue diamonds
indicate 1-mode SRWs, and stars represent generic, non soliton-like rogue waves.
In contrast to the HONLS, where every rogue wave was a two-mode SRW, 
these results suggest that  viscosity enhances mode interaction and promotes the formation of less structured, diffuse  multimodal rogue waves.

As the simulation progresses, the Floquet spectrum in V-HONLS  undergoes
frequent
reconfigurations, characterized  by
repeated formation and annihilation of complex critical points.
The critical
point crossings in V-HONLS occur almost exclusively at  complex points
and the transitions are  qualitatively
similar to those observed at complex points in the HONLS case (e.g.
Figures~\ref{HONLS_SPB_spec}(a)-(c)).
While the total
number of crossings may be fewer, the increase in  complex critical point crossings reflects increased interaction between the first two nonlinear modes.

A timeline of these spectral transitions
as a function of the damping parameter $\Gamma$ is shown in
Figure~\ref{VHONLS_vary_Gamma}(b). Transitions at real critical points are marked with  blue dots, while those at complex critical points are indicated with a  red `$\bigtimes$'
In some cases, multiple transitions occur in rapid succession, causing
overlapping `$\bigtimes$' markers to appear darker or elongated.
Notably, as $\Gamma$ varies,
 all transitions occur at complex critical points, except for
a single real transition  when $\Gamma = 0.001$.
These frequent crossings
highlight the role of damping in shaping  the Floquet spectral complexity and nonlinear mode interactions.

Despite the increase in complex critical point crossings as compared with the HONLS case, the phase variance PVD does
not show a corresponding increase.
In fact, Figure~\ref{VHONLS_SPB_str}(c) shows that the PVD is  lower for $\Gamma = 0.002$ and Figure~\ref{VHONLS_vary_Gamma}(c) confirms that the average PVD remains
below that of the HONLS case, generally decreasing with increasing $\Gamma$
(although not monotonically).
The lower PVD  values can be  attributed to the damping induced suppression of the 
third, dephased nonlinear mode,  which was clearly present in the HONLS
spectrum, and which contributed significantly to loss of coherence and elevated phase variance.

\vspace{6pt}
\noindent{\bf Summary of Results (V-HONLS, Steep wave initial data):}
For SPB initial data, the V-HONLS evolution initially mirrors the HONLS, briefly forming a two-mode SRWs at early times ($t \approx 3.5$). However, viscosity-driven damping quickly disrupts this coherence: rogue waves become fewer and increasingly lose soliton-like structure, evolving instead into broader, diffuse multimodal events.

The Floquet spectrum undergoes frequent reconfigurations dominated by complex critical point crossings, reflecting enhanced interactions among the first two nonlinear modes. Unlike HONLS, real critical point transitions are very rare. Despite this spectral complexity, the average phase variance (PVD) remains lower than in HONLS, due to damping suppressing the third, dephased nonlinear mode that had previously contributed to phase disorder in the HONLS.

\subsubsection{Steep Wave Initial Data: NLD-HONLS equation}
 Finally, we analyze the effects of mean-flow damping on SRW development,
 comparing results from the  NLD-HONLS with those of the V-HONLS.
 This comparison highlights both shared and  model-specific distinct features,
 providing insight into the mechanisms that support or inhibit SRW formation across the  two models.

\begin{figure}[htp!]
\centerline{
\includegraphics[width=0.3\textwidth]{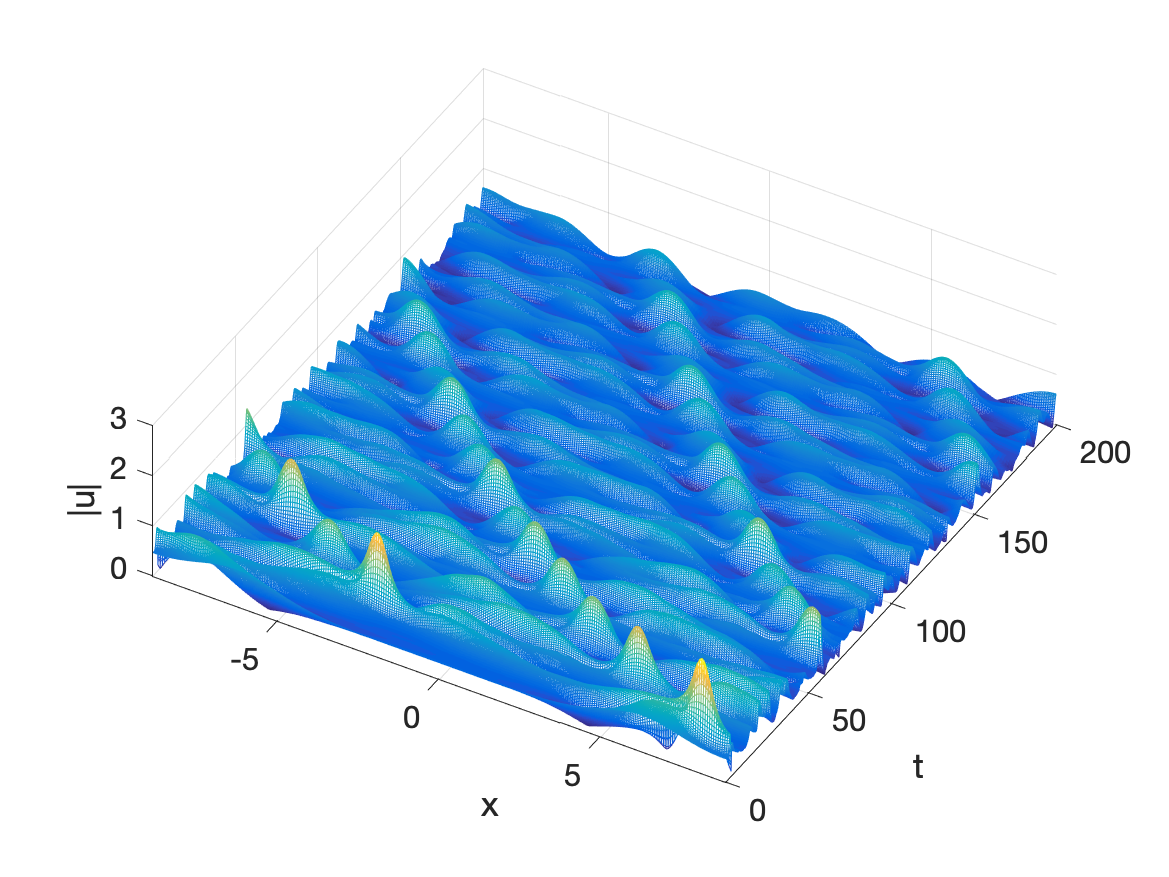}
\hspace{6pt}\includegraphics[width=0.25\textwidth]{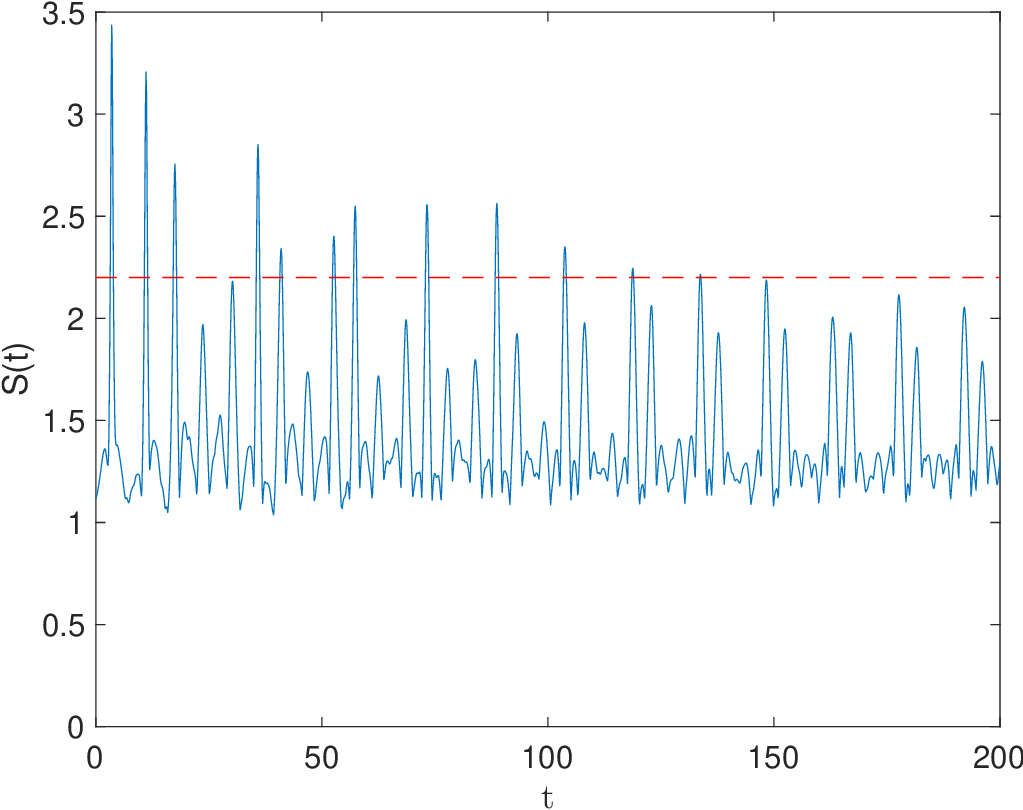}
\hspace{6pt}\includegraphics[width=0.25\textwidth]{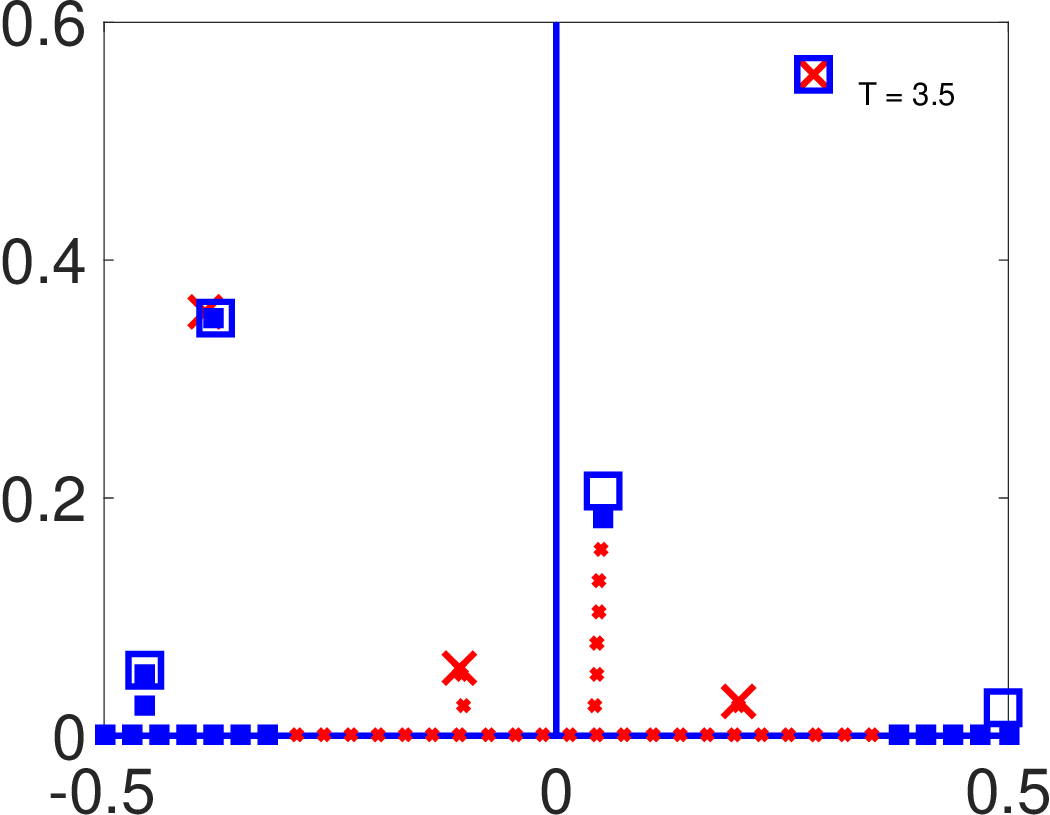}
}
\vspace{6pt}
\centerline{
    \includegraphics[width=0.25\textwidth]{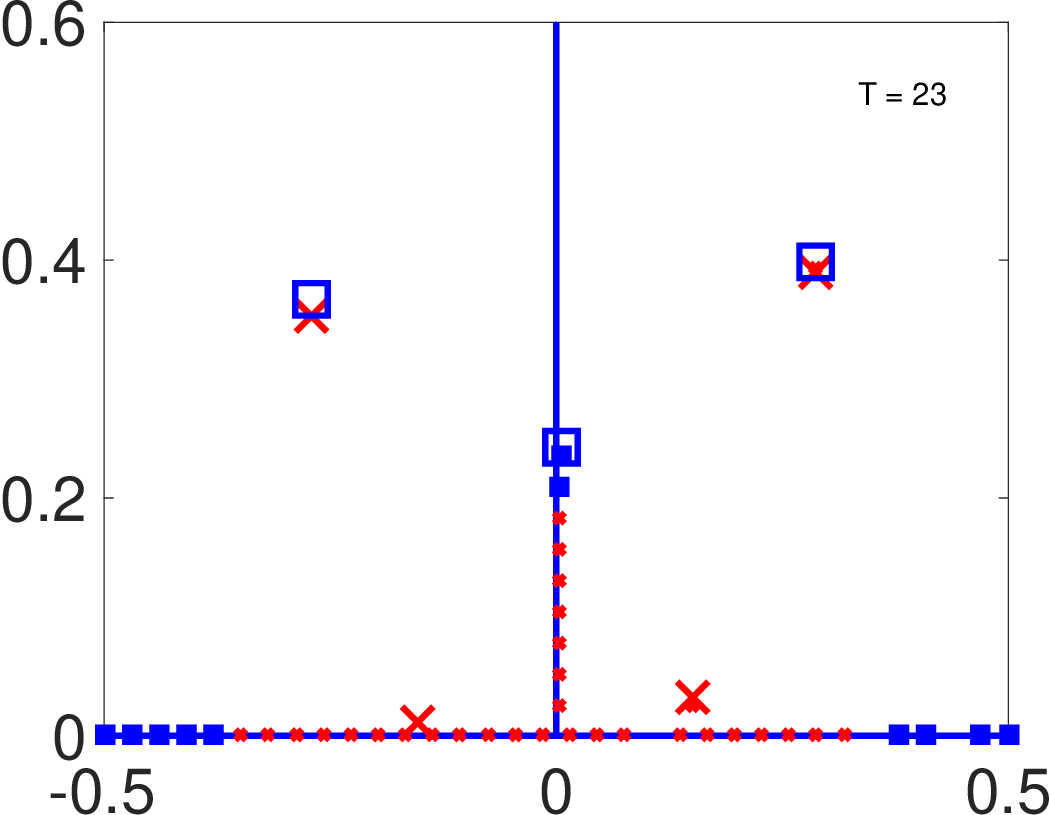}
\hspace{6pt}\includegraphics[width=0.25\textwidth]{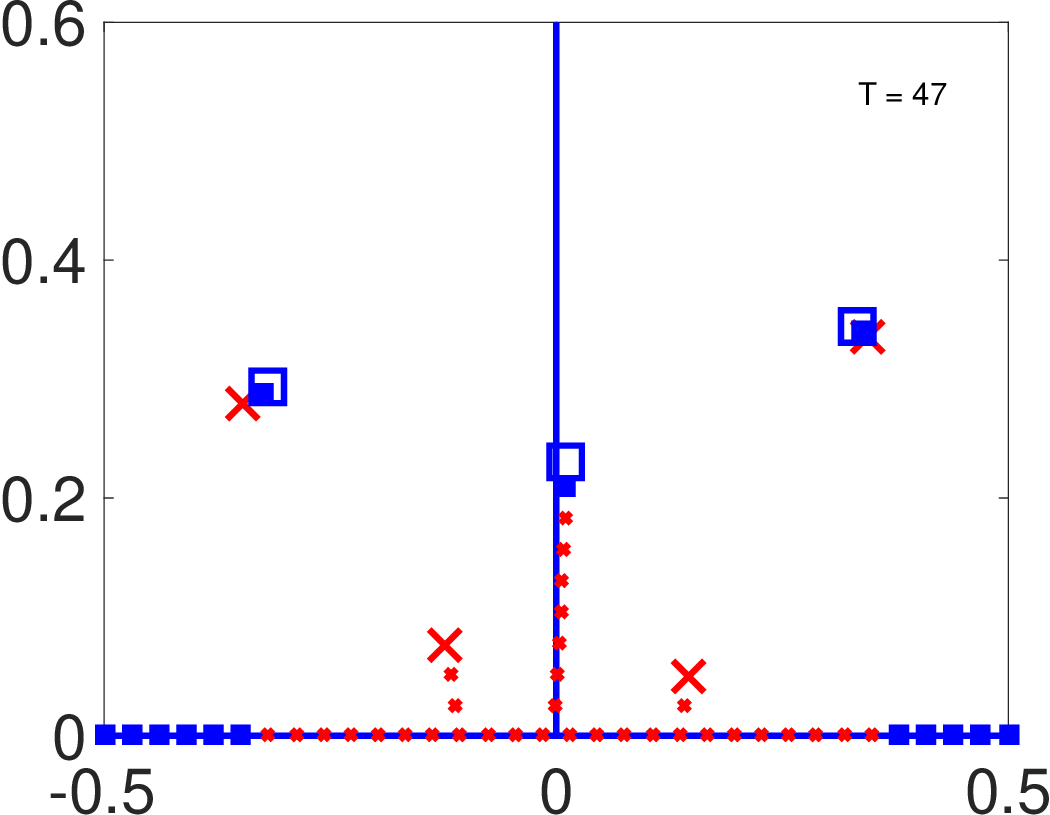}
\hspace{6pt}\includegraphics[width=0.25\textwidth]{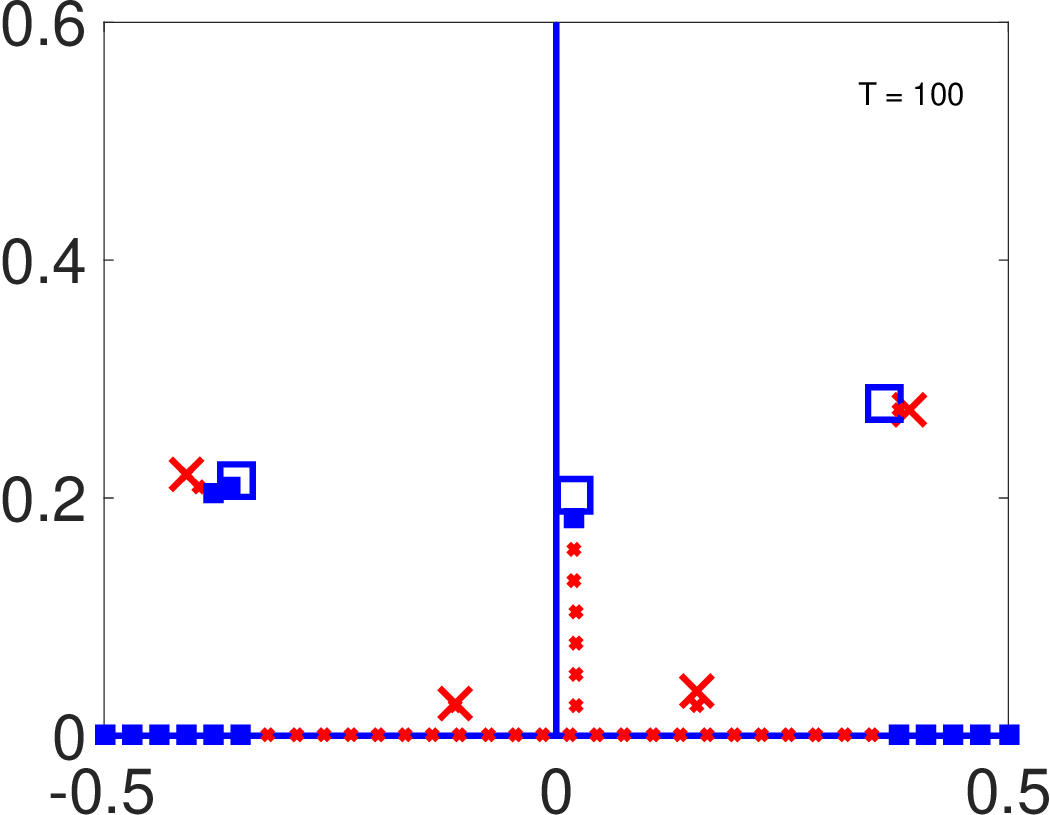}
}  
\caption{NLD-HONLS evolution with $\beta = 0.1$ and SPB initial data \rf{SPB_ic}: (a) $|u(x,t)|$ for $0\leq t\leq 200$, (b) the strength
  $S(t)$, and Floquet spectra at  c) $t=3.5$ (d) $t=23$,
  (e) $t=47$, (f) $t=100$.}
\label{NLD_SPB_spec}
\end{figure}

\begin{figure}[htp!]
\centerline{
\includegraphics[width=0.3\textwidth]{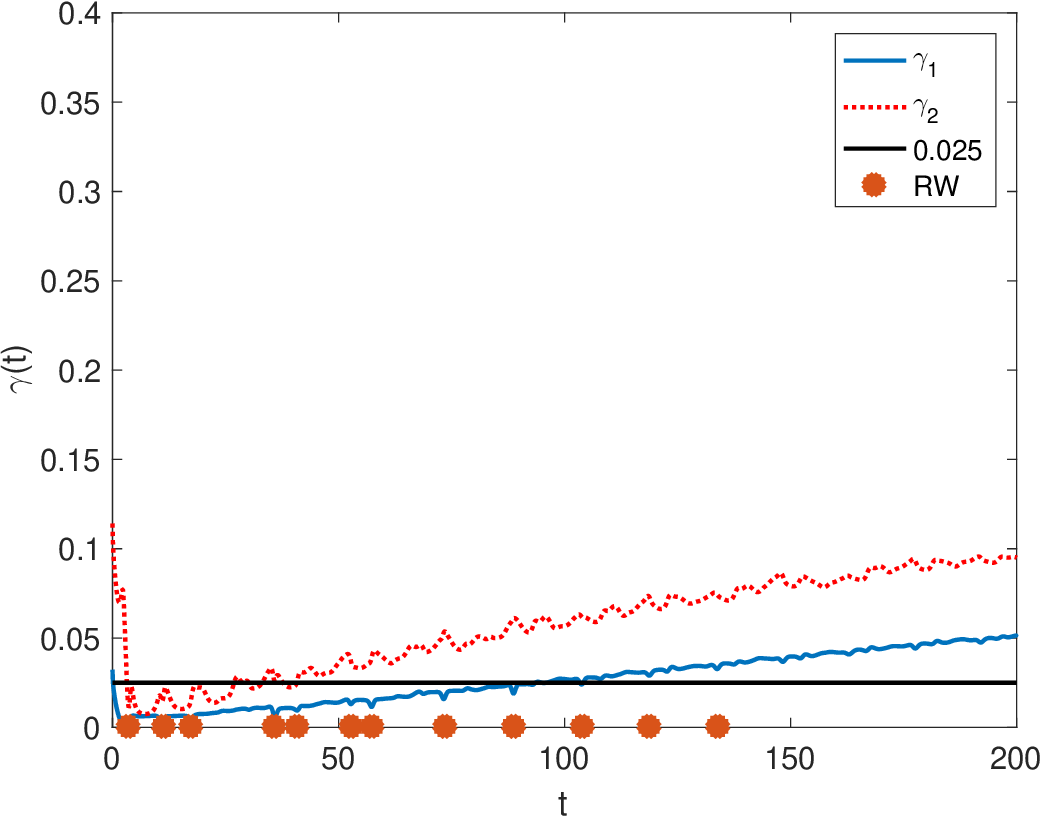}
\hspace{6pt}  \includegraphics[width=0.3\textwidth]{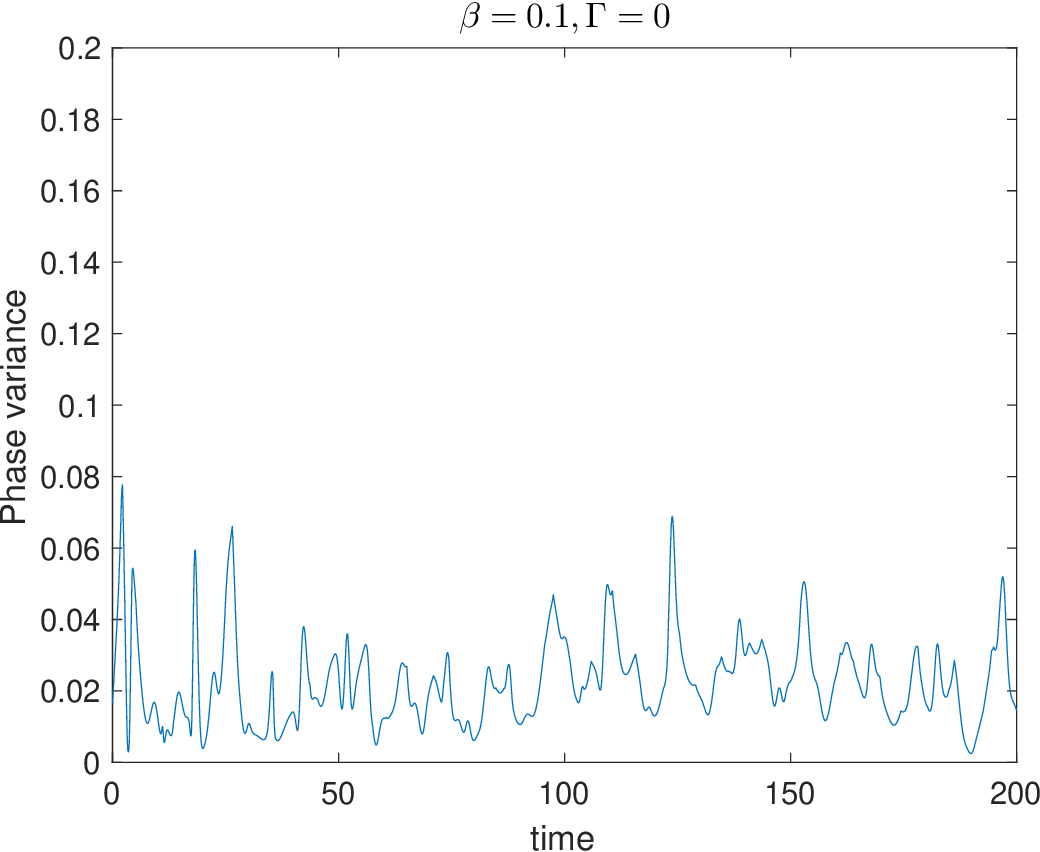}
\hspace{6pt}\includegraphics[width=0.3\textwidth]{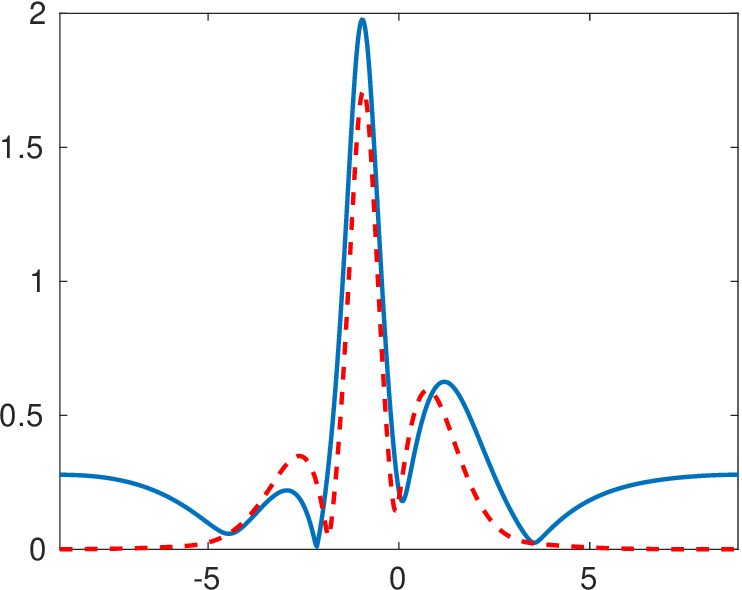} 
}
\caption{NLD-HONLS evolution with $\beta = 0.1$ and SPB initial data \rf{SPB_ic}: (a) Evolution of the band lengths $|\gamma_1(t)|$ and $|\gamma_2(t)|$. The horizontal line marks the  soliton-like threshold
  and the  red dots denote  rogue wave events.
 (b) Phase variance PVD$(t)$.
(c)  Comparison of $|u(x,t)|$ (solid line) with a  two-soliton analytical solution (dashed line)
    obtained using the midpoints
    of the spectral bands $\gamma_1$ and $\gamma_2$.}
\label{NLD_SPB_PVD}
\end{figure}

\vspace{6pt}
\noindent {\bf The Floquet spectral evolution up to the first rogue wave event:}
Figures~\ref{NLD_SPB_spec}(a) and (b) provide the evolution of the surface $|u(x,t)|$ and the
strength $S(t)$, respectively, for the two mode SPB initial data \rf{SPB_ic} over the interval $0 < t < 200$.  The strength plot reveals frequent rogue wave events up to approximately  $t = 137$.

As in the case of the V-HONLS model, the spectral evolution under NLD-HONLS  follows closely that of the undamped HONLS simulation for $0 \leq t \leq 3.5$ 
(see Figures~\ref{ic_SPB_spec}((a)-(d)).
In particular, compare Figure~\ref{NLD_SPB_spec}(c) with Figures~\ref{ic_SPB_spec}(d).

\vspace{6pt}
\noindent {\bf Later stage NLD-HONLS dynamics:} 
However, beyond the first rogue wave event, the spectral evolution  in
the NLD-HONLS model differs strikingly from that observed in the previous
models. 
Notably, the  two-mode
soliton-like state is not transient:  it persists until $t \approx 40$ accompanied by repeated formation of two-mode SRWs (see Figure~\ref{NLD_SPB_spec}(d)) .
Further, the Floquet spectral decomposition remains highly organized and critical point transitions are not observed.
 After $ t = 40$ 
one of the bands expands sufficiently, indicating a
transition back to a
a one soliton-like state. A representative spectrum  of
a one-mode SRW at
$t = 47$ is shown in Figure~\ref{NLD_SPB_spec}(e).  As time evolves, both  spectral
bands continue to grow, leading to
a spectral configuration consistent with a generic five-phase
solution, see Figure~\ref{NLD_SPB_spec}(f).

Importantly, no complex or real critical point transitions
are observed  for $t >0$ in this spectral evolution -- 
a notable contrast to the behavior seen in the HONLS and V-HONLS.

This spectral analysis reveals not only that the observed rogue waves are indeed SRWs, but also that the waveform, as characterized by the Floquet spectral decomposition, remains in soliton-like configurations for prolonged periods.
  Figure \ref{NLD_SPB_PVD}(a) summarizes this behavior by showing the time evolution of the band lengths, $|\gamma_1(t)|$
 and $|\gamma_2(t)|$.
  The red dots along the $t$-axis indicated rogue wave events,
  which strongly correlate with the soliton-like spectral regime.

  Unlike the HONLS and V-HONLS models, where soliton-like states are
  brief and transient,  the NLD-HONLS system exhibits a remarkably
  sustained structure: a two-mode soliton-like state  persists up to
  approximately $t \approx 40$, followed by a one-mode soliton-like state
  that continues up to about $t \approx 90$. Only after this time
  do the rogue waves, such as the one at $t = 113$, begin to deviate
  from the SRW profile, exhibiting  broader and less localized
  features.

\vspace{6pt}
\noindent {\sl Observation.} For SPB initial data, the spectral evolution  under NLD-HONLS exhibits a remarkably structured
progression for all $\beta\neq 0$,
illustrating the  organizing influence  of nonlocal mean-flow damping.
The system transitions through a sequence of coherent,
well-structured states with minimal Floquet spectral disorder:
{\sl Both complex double points split} $\rightarrow$ one-mode
 {\sl soliton-like}$\rightarrow$ two-mode
 {\sl soliton-like} $\rightarrow$ one-mode {\sl soliton-like}
 $\rightarrow$ {\sl generic N-phase spectrum.}
 
   Figure~\ref{NLD_SPB_PVD}(c) compares the numerical solution $|u(x,t)|$ with the analytical two-soliton solution \rf{two-soliton}, constructed using the spectral data obtained from the Floquet decomposition of the numerical solution.  
   In formula \rf{two-soliton}  $\lambda_1$ and $\lambda_2$
   correspond to the midpoints of the two small bands in the upper half pane at $t = 3.5$.
   Although the NLD-HONLS soliton-like state  is over a non uniform, finite-amplitude background,  it's structure near the peak
   closely matches the two-soliton analytical profile.

\begin{figure}[htp!]
\centerline{
\includegraphics[width=0.3\textwidth]{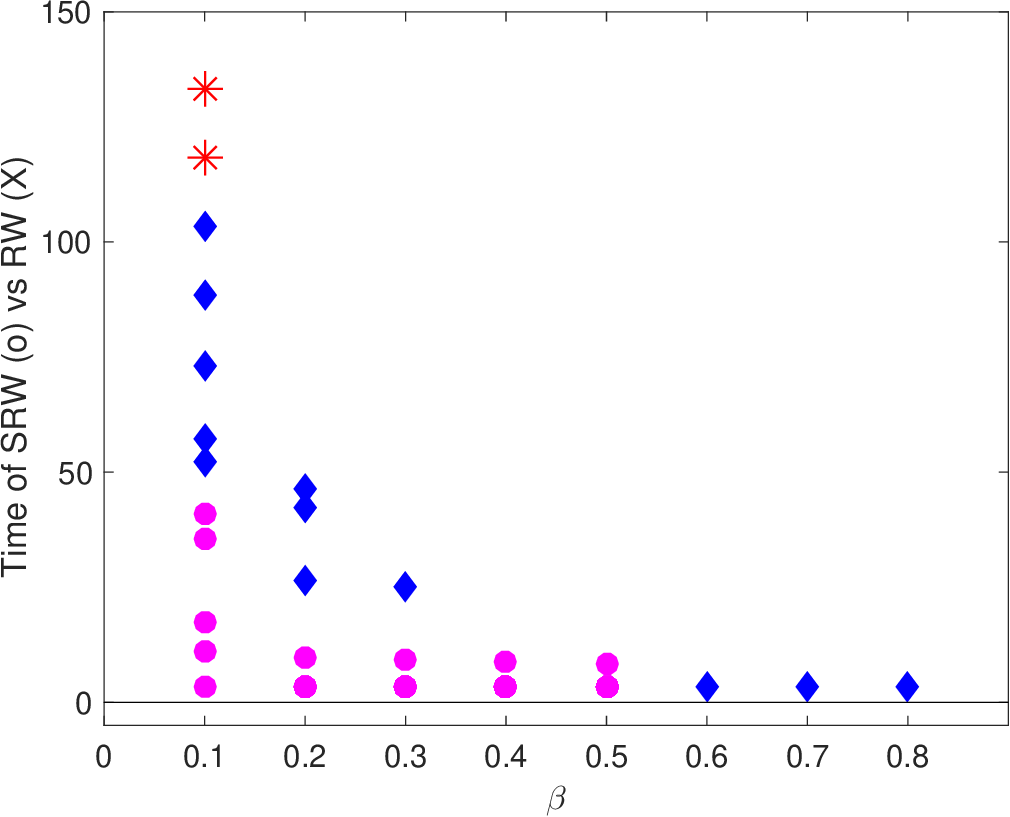}
\hspace{6pt}\includegraphics[width=0.3\textwidth]{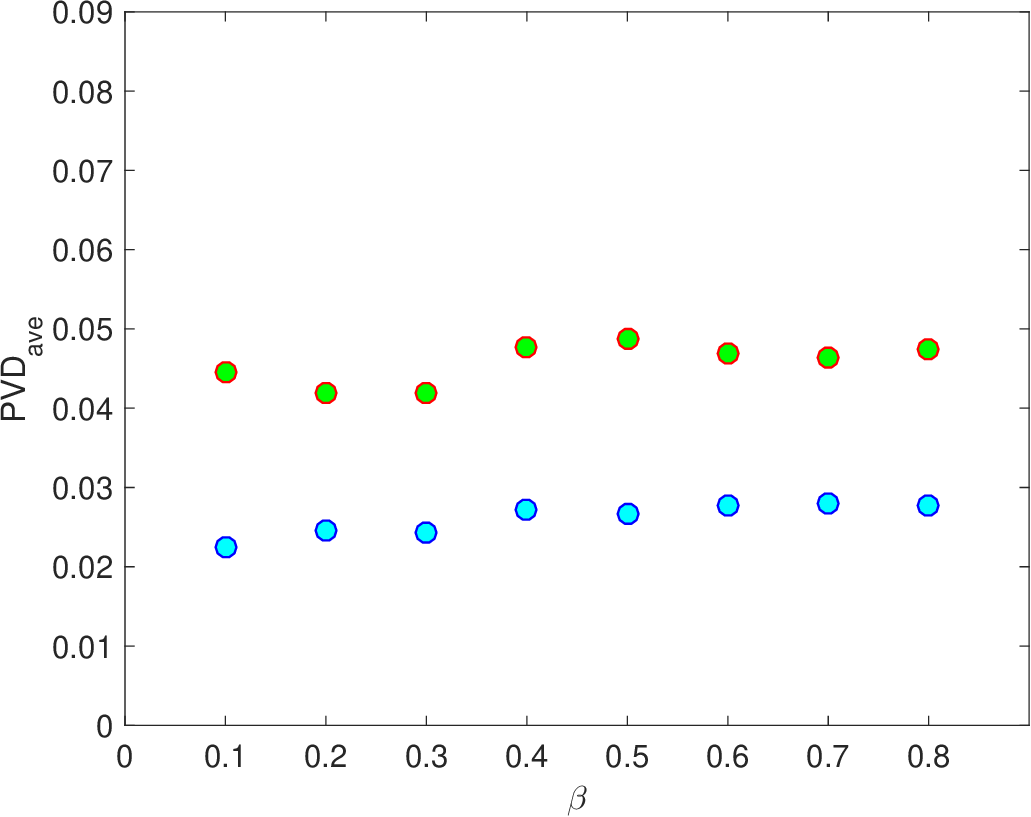} 
}
\caption{NLD-HONLS evolutions for varying damping parameter $\beta$;
  (a) Timeline for SRW and generic rogue wave
 events for SPB initial data \rf{SPB_ic}.
 Magenta dots indicate the occurrence of 2-mode SRWs, blue diamonds 1-mode SRWs,
    and stars mark generic rogue waves. (b) Average PVD for SPB (light blue) and Stokes (green) initial data.
}
\label{NLD_vary_Beta}
\end{figure}

  The organization and robustness of the Floquet spectral configuration  reflects an underlying dynamical coherence, further supported by low values of the phase variance  PVD diagnostic. 
  Figure \ref{NLD_SPB_PVD}(b) shows the evolution of the  PVD, with a maximum value of approximately  $\mbox{PVD}_{max}\approx 0.08$, well below that observed in  the HONLS and V-HONLS models.
  Figure~\ref{NLD_vary_Beta}(b) further confirms that the average PVD
  remains below that of the HONLS case (Figure~\ref{VHONLS_vary_Gamma}(c))
  and is typically lower than
  that of the V-HONLS model for small values of $\Gamma$ 
and  comparable for larger values of $\Gamma$.

  A timeline for SRW and generic rogue wave
  events as $\beta$ varies, for both SPB
  is shown in Figure~\rf{NLD_vary_Beta}(a).
  As $\beta$ increases, the total number of rogue wave events
  decreases with  generic rogue waves observed  after
  $t = 100$   only for $\beta = 0.1$. Notably,  not only are all 
  rogue waves of the SRW type,  but the waveform and its associated
  Floquet spectrum remain consistently in a soliton like state for $0 < t <90$

  Figure~\rf{NLD_SPB_PVD}(b) shows the phase variance remains consistently low, PVD$(t) < 0.08$,  across all values of $\beta$.
  This low PVD is a hallmark of highly localized SRWs, confirming that mean-flow damping promotes coherent structures and suppresses mode interaction.

An interesting question is why the NLD-HONLS model produces
more SRWs, despite its
damping targeting steep, highly nonlinear regions
of the wave field. The key lies in the selective nature of
the damping which  acts strongly only where the field becomes
locally steep. Since the damping is nearly negligible when the
field is close to the background Stokes wave, the seeds of
localized structures embedded in the background are less
affected, much like in the undamped HONLS. When broad,
high-amplitude modulations start to steepen over extended
regions, the nonlinear damping suppresses them,
preventing large-scale
disruption of the background.  In this way, the NLD-HONLS
 “protects” the background coherence,
which supports modulational growth and sustains localized
extreme events,
while selectively damping broader, potentially destabilizing structures.

{\bf Summary of Results (NLD-HONLS, Steep wave initial data):} 
For SPB initial data, the spectral evolution under NLD-HONLS
shows that nonlocal mean-flow damping strongly organizes and
stabilizes the dynamics: the system remains in coherent,
soliton-like spectral states for prolonged intervals, rather
than exhibiting brief, transient soliton-like states as seen in HONLS and
V-HONLS.

 In contrast to the HONLS and V-HONLS simulations,
 a defining feature of the  NLD-HONLS evolution is
 the persistent  organization of the 
 Floquet spectrum. For all $\beta \neq 0$, spectral   bands  remain distinct, well-separated,  and free of critical point
 crossings throughout the 
entire time evolution.
This is consistently
corroborated by low phase variance (PVD) across all $\beta$.

{\sl Pre-wave-breaking Interpretation:} The soliton-like rogue waves (SRWs) observed in the NLD-HONLS model may  be interpreted as coherent states that are consistent with the pre–wave-breaking focusing regime reported in hydrodynamic studies \cite{BF1967, BP1998, OOSB2001, SB2002, BCYS2007}. In these events, nonlinear focusing drives the wave envelope toward a steep, localized structure, while the nonlinear mean-flow damping term acts most strongly near the crest to regulate  energy, suppress spectral disorder, and promote coherence. This regulated focusing process results in a highly organized state that aligns with the final stage of nonlinear focusing prior to physical wave breaking.  We emphasize that the NLD-HONLS does not model the breaking process itself; rather, it captures a coherent focusing regime that precedes breaking, characterized by strong energy localization and spectral coherence.

The emergence and sustained persistence of tiny, well-separated  Floquet spectral bands provide a diagnostic of this pre–breaking regime. While similar band contraction and localization occur transiently in the HONLS and V-HONLS models, only the NLD-HONLS sustains this behavior over extended times,
indicating that
the associated coherent spectral state is a distinctive feature of the
nonlinear mean-flow damping mechanism.

The contraction of Floquet bands therefore serves as  a
  spectral indicator of  localized, coherent focusing dynamics.
  This interpretation is consistent with experimental observations showing that strongly modulated wave groups approach near-breaking steepness through a phase of increased coherence and spectral narrowing during the final stages of modulational focusing. In this sense, the NLD-HONLS provides a mathematically tractable model for the envelope-scale dynamics associated with the pre-wave-breaking regime, offering a conceptual link between nonlinear spectral theory and hydrodynamic experiments.

\subsection{Moderately Steep Wave Initial Data and the Formation of Rogue Waves}
    %{HONLS equation: Perturbed Stokes Initial Data}
We now test the robustness of SRW formation in a broader, more physically generic setting using 
perturbed Stokes wave initial data which is less structured.

\subsubsection{Moderately Steep Wave Initial Data: HONLS equation}

\begin{figure}[htp!]
\centerline{
  \includegraphics[width=0.3\textwidth]{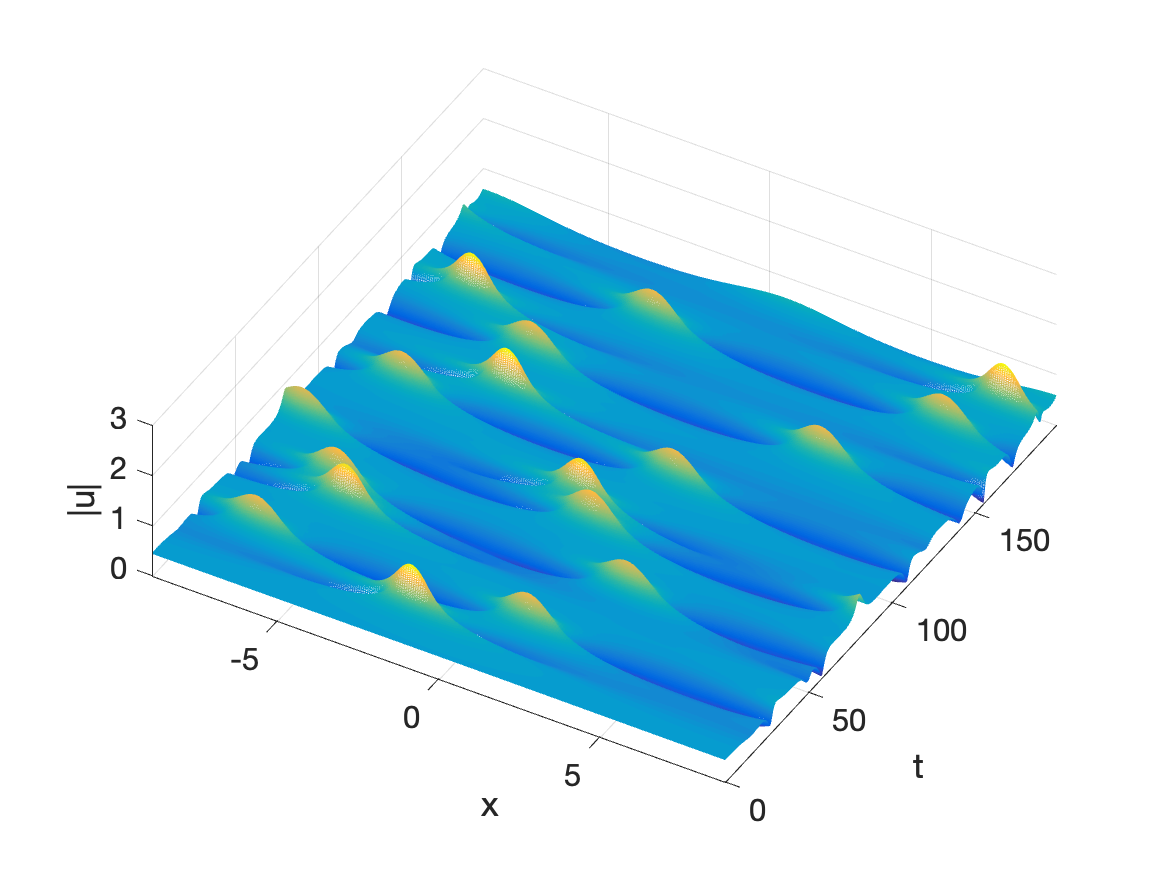}
\hspace{6pt}\includegraphics[width=0.3\textwidth]{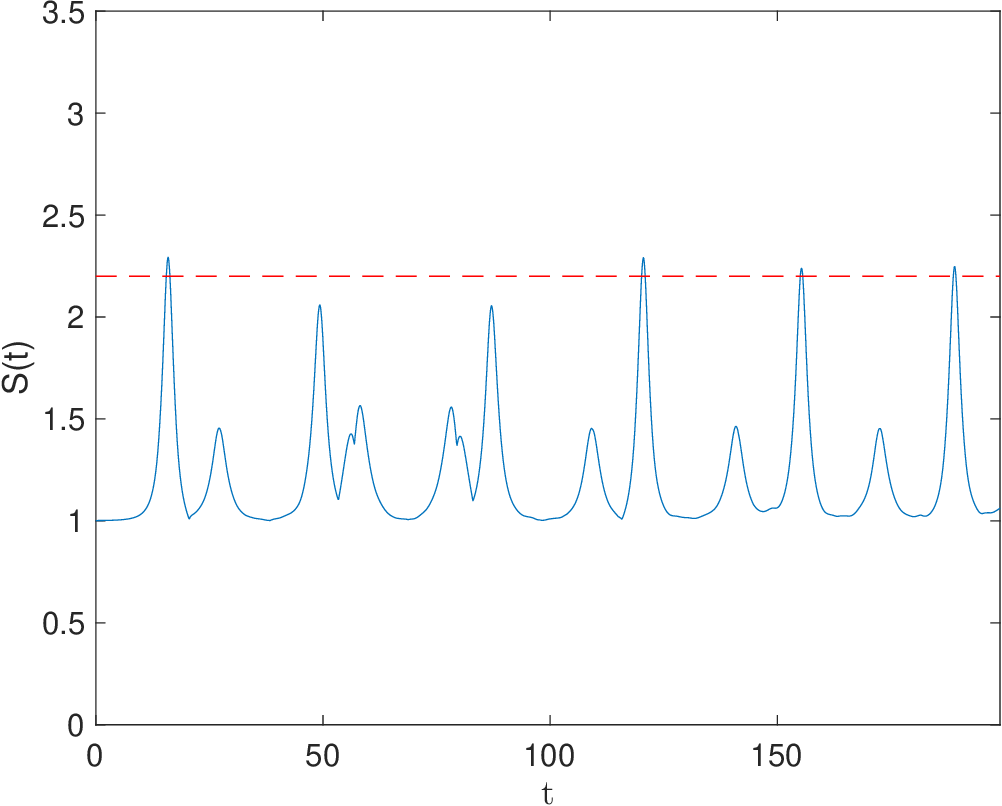}
\hspace{6pt}\includegraphics[width=0.3\textwidth]{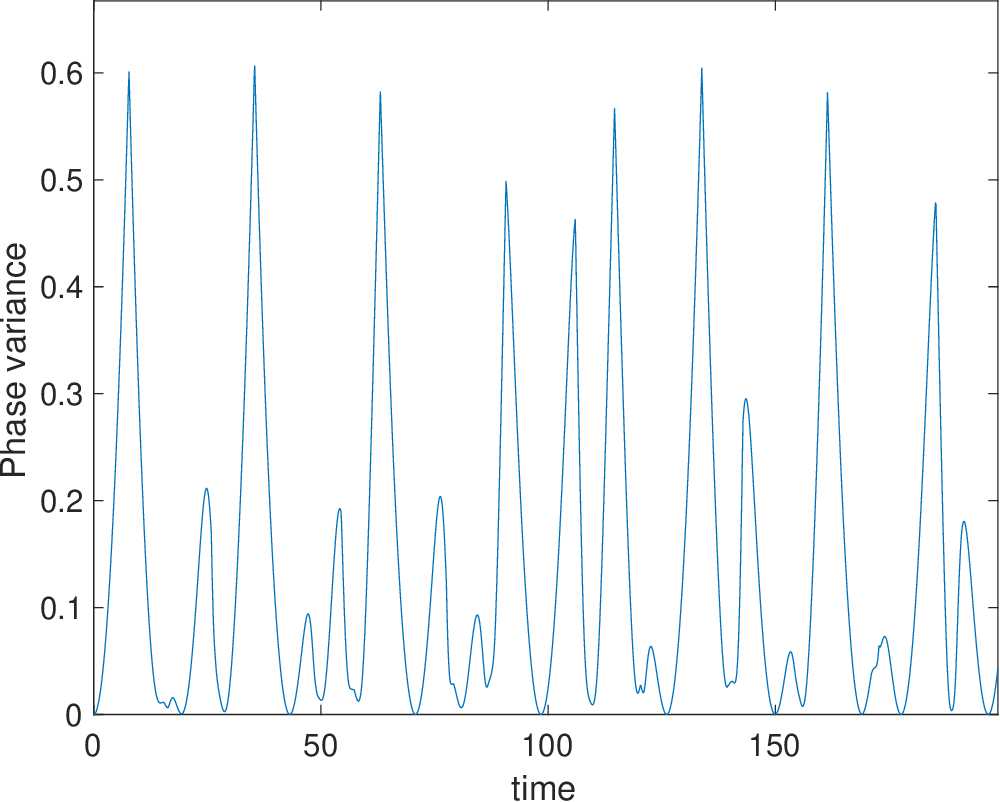}
}
\caption{HONLS evolution with Stokes initial data \rf{Stokes_ic}: (a) $|u(x,t)|$ for $0\leq t\leq 200$.
  (b) the strength $S(t)$ with the dashed red line indicating the threshold for a rogue wave;
and (c) the phase variance PVD$(t)$}
\label{HONLS_Stokes_str}
\end{figure}

Figures~\ref{HONLS_Stokes_str}(a) and ~\ref{HONLS_Stokes_str}(b)  show  the evolution of the surface $|u(x,t)|$ and the
strength $S(t)$ for perturbed Stokes initial data \rf{SPB_ic} over the interval $0 < t < 200$.  The strength plot indicates that, compared to the results for SPB initial data, 
rogue waves  are both fewer in number and significantly 
weaker in strength. This difference arises due to the nature of the initial conditions: SPB  data is for  localized and well structured waves,   while
perturbed Stokes initial data requires the system to generate structure dynamically via the modulational instability, leading to less coherence and weaker rogue waves.

 \begin{figure}[hpt!]
  \centerline{
    \includegraphics[width=0.225\textwidth]{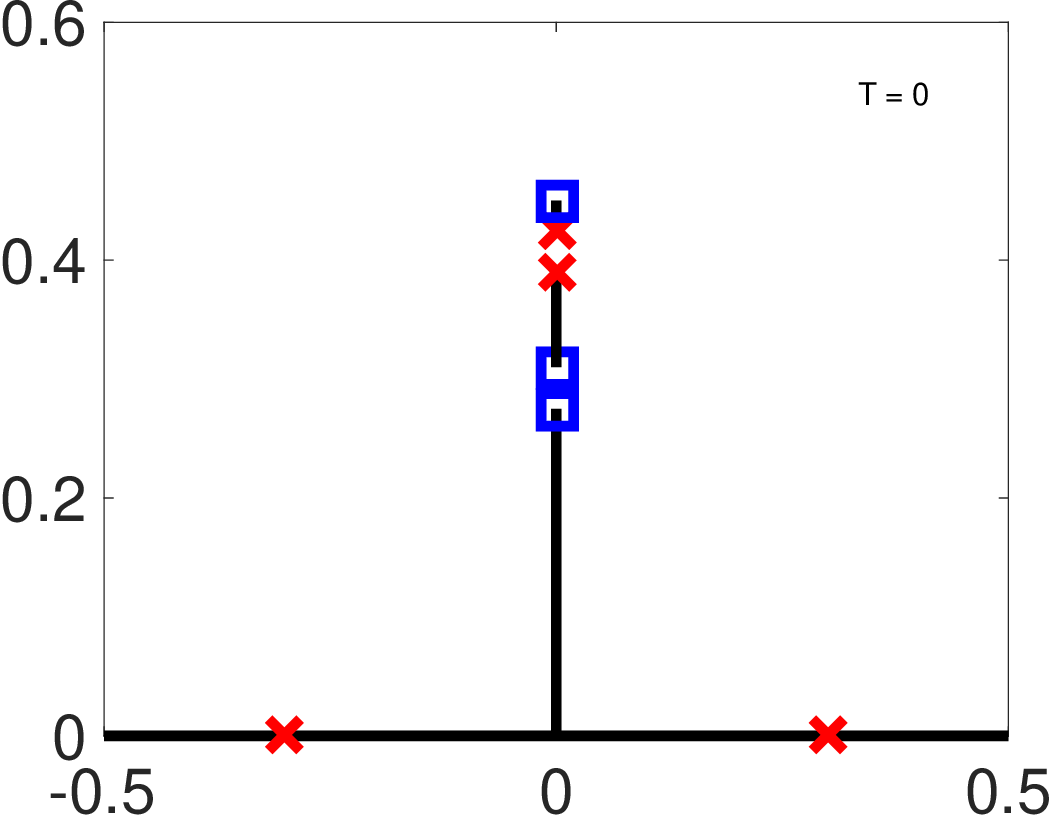}  
\hspace{6pt}\includegraphics[width=0.225\textwidth]{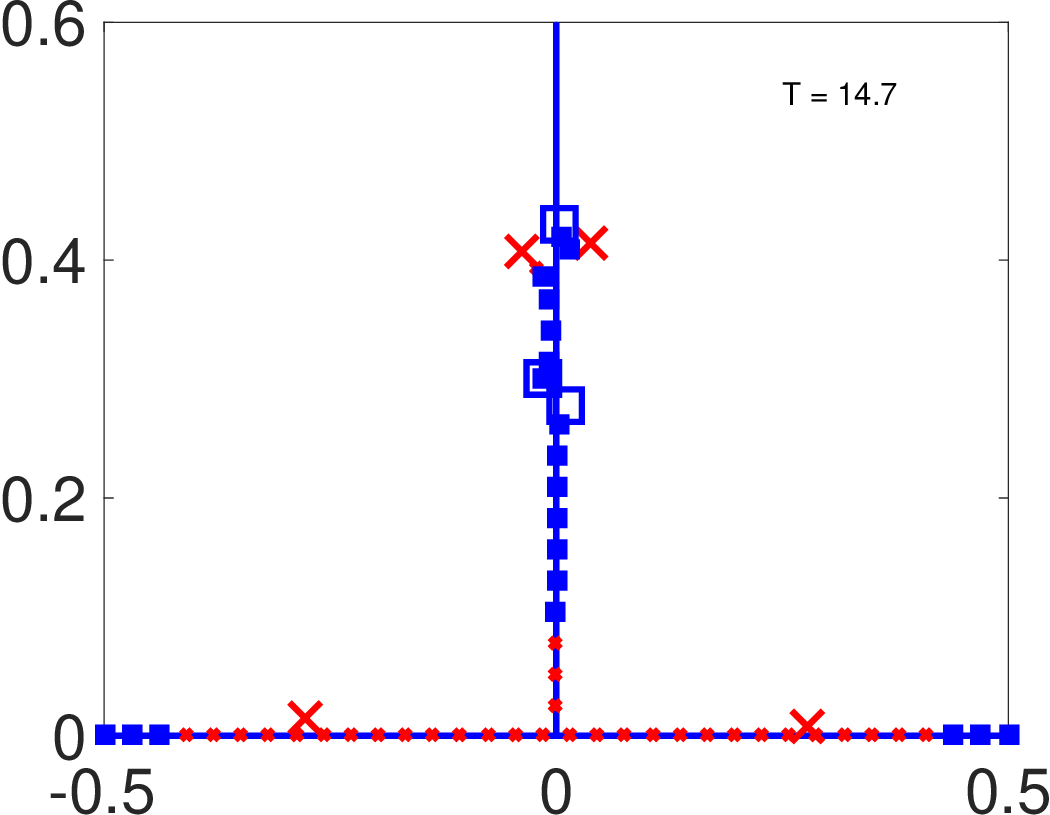}
\hspace{6pt}\includegraphics[width=0.225\textwidth]{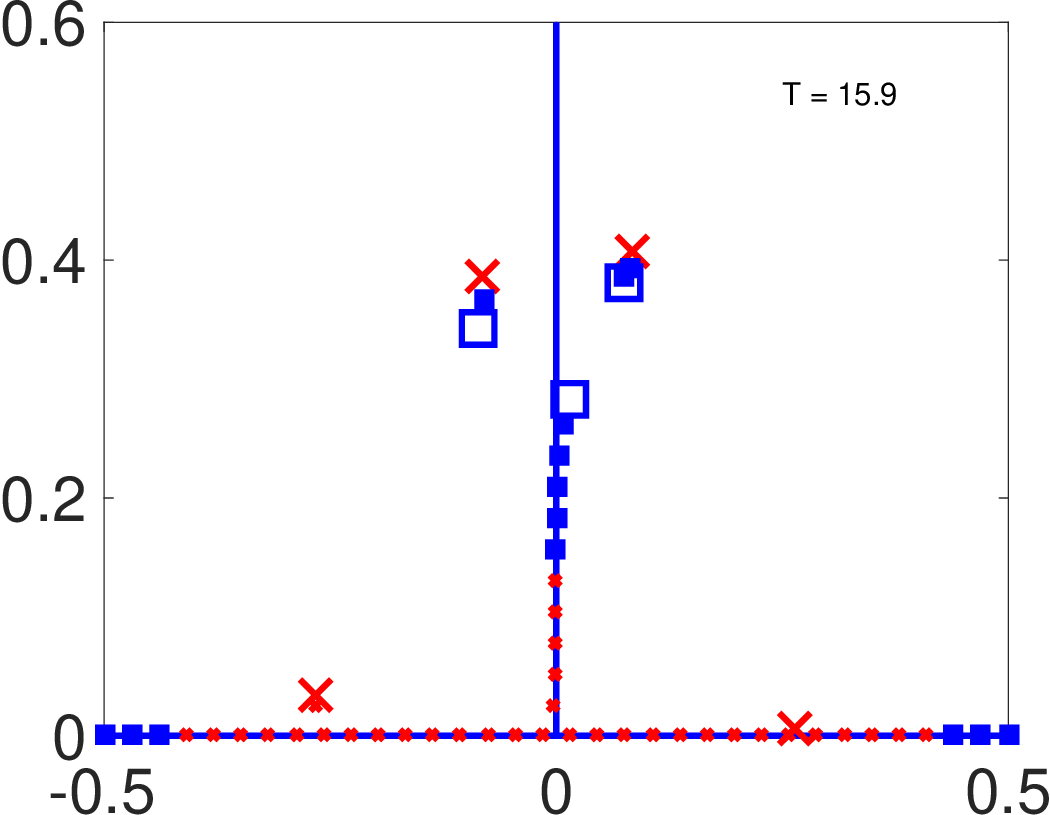}
  }
  \caption{HONLS evolution with perturbed Stokes initial data \rf{Stokes_ic}: Floquet spectra at (a) $t = 0$   and (b) $t = 14.7 $ (the characteristic asymmetric spectrum),   and
    (c) $ t = 15.9$ (generic rogue wave event).}
  \label{ic_Stokes_spec}
  \end{figure}

\vspace{6pt}
\noindent {\bf The Floquet spectral evolution up to the first rogue wave event:}
 Given initial data \rf{Stokes_ic}, a perturbation analysis shows that each double point of the unperturbed Stokes wave ( in Figure~\ref{ic_SPB_spec}(a)  given by the '$\bigtimes$' and the box) splits into two simple points along the imaginary axis, creating
two spectral gaps: the first of order  $ {\cal O}(\alpha)$, and the second of order   $ {\cal O}(\alpha^2)$ \cite{AHS1996}.
As a result, the Floquet spectrum for perturbed Stokes wave initial data in the upper half plane consists of a band along the imaginary axis with two small gaps. This is illustrated 
in Figure~\ref{ic_Stokes_spec}(a), which is specifically
drawn to make these  small gaps easier to see.

\begin{figure}[htp!]
\centerline{
\includegraphics[width=0.25\textwidth]{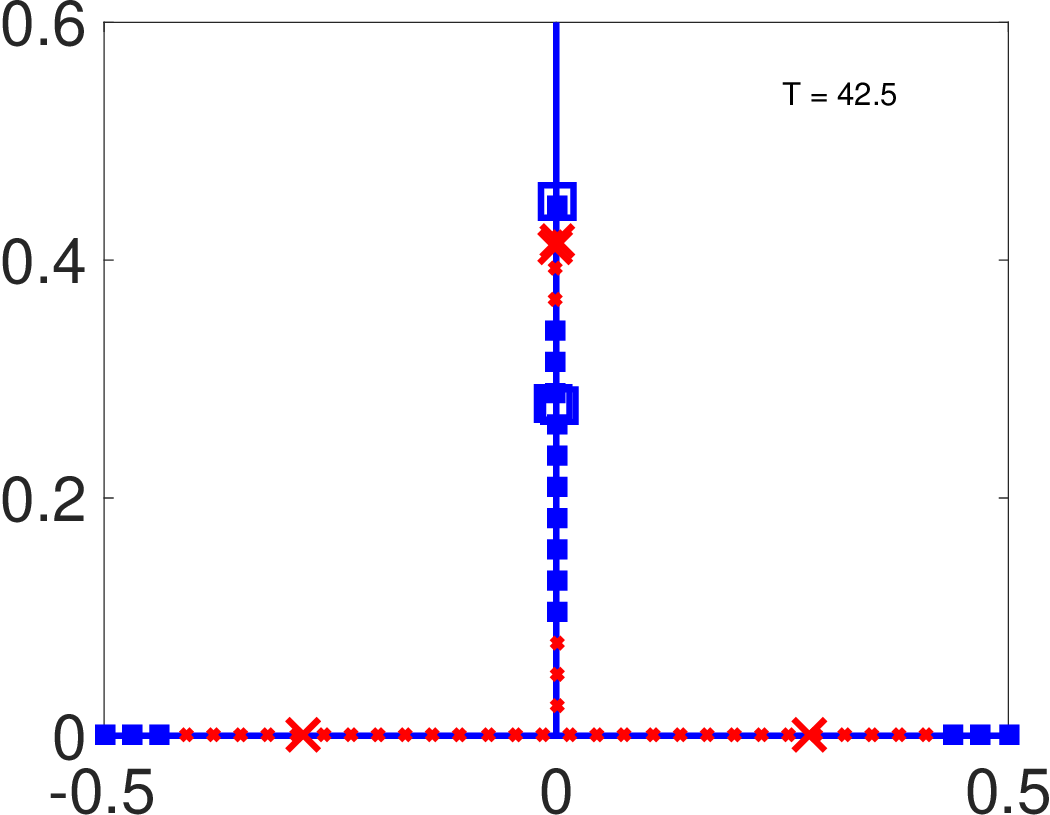}
\hspace{6pt}\includegraphics[width=0.25\textwidth]{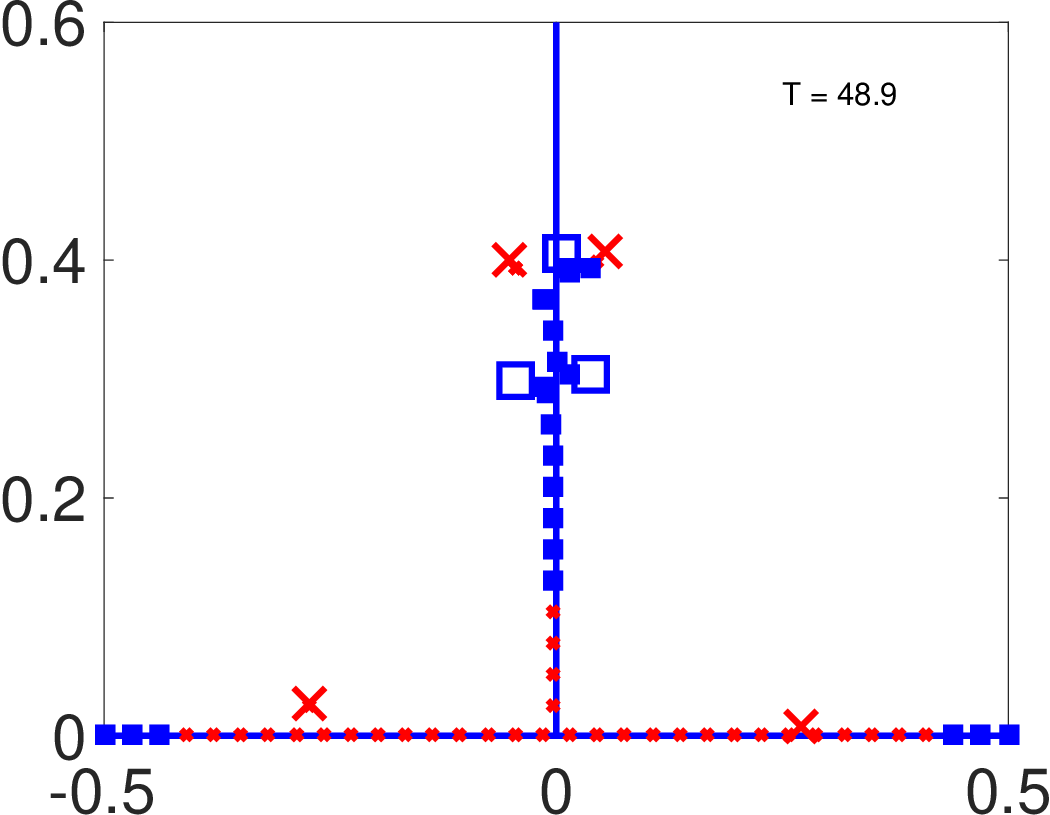}
\hspace{6pt}\includegraphics[width=0.25\textwidth]{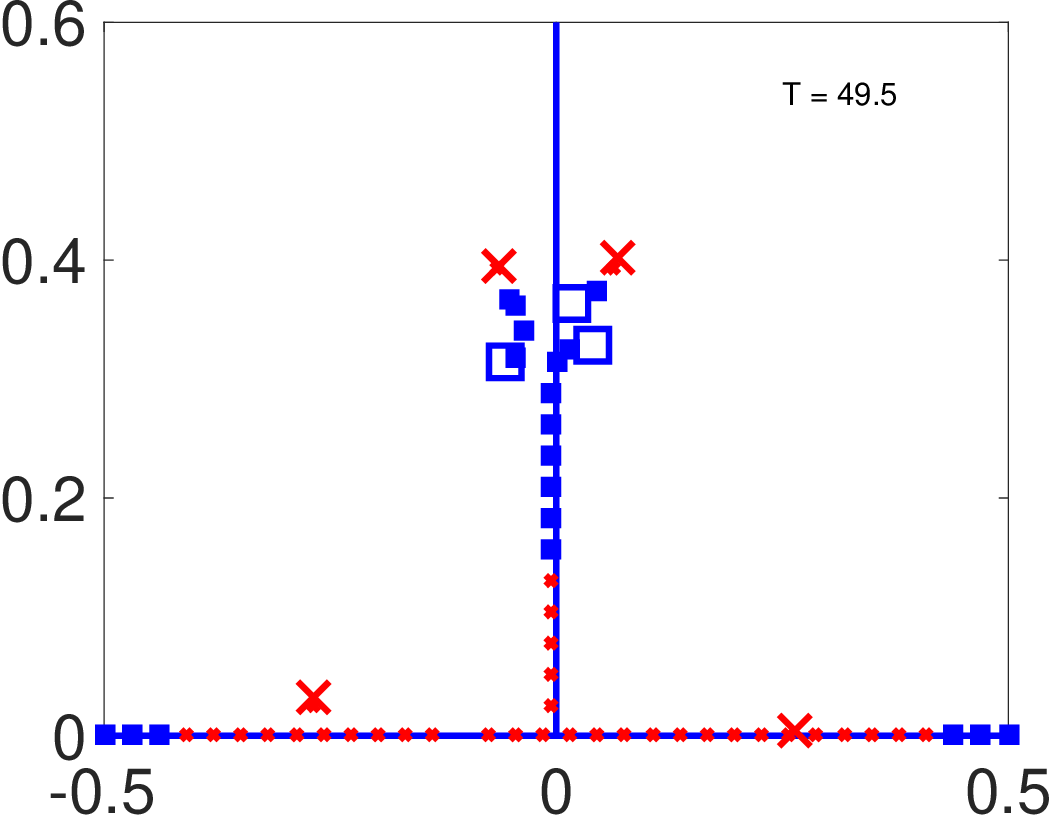}
}
\vspace{6pt}
\centerline{
\includegraphics[width=0.25\textwidth]{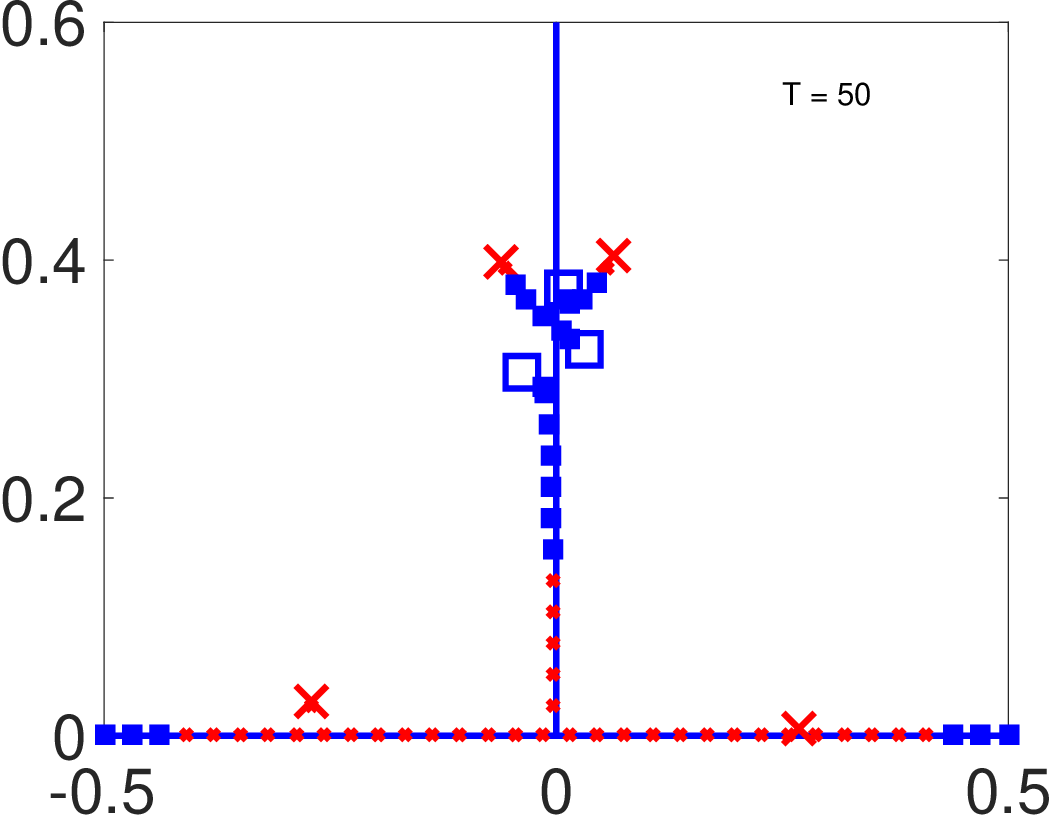}
\hspace{6pt}\includegraphics[width=0.25\textwidth]{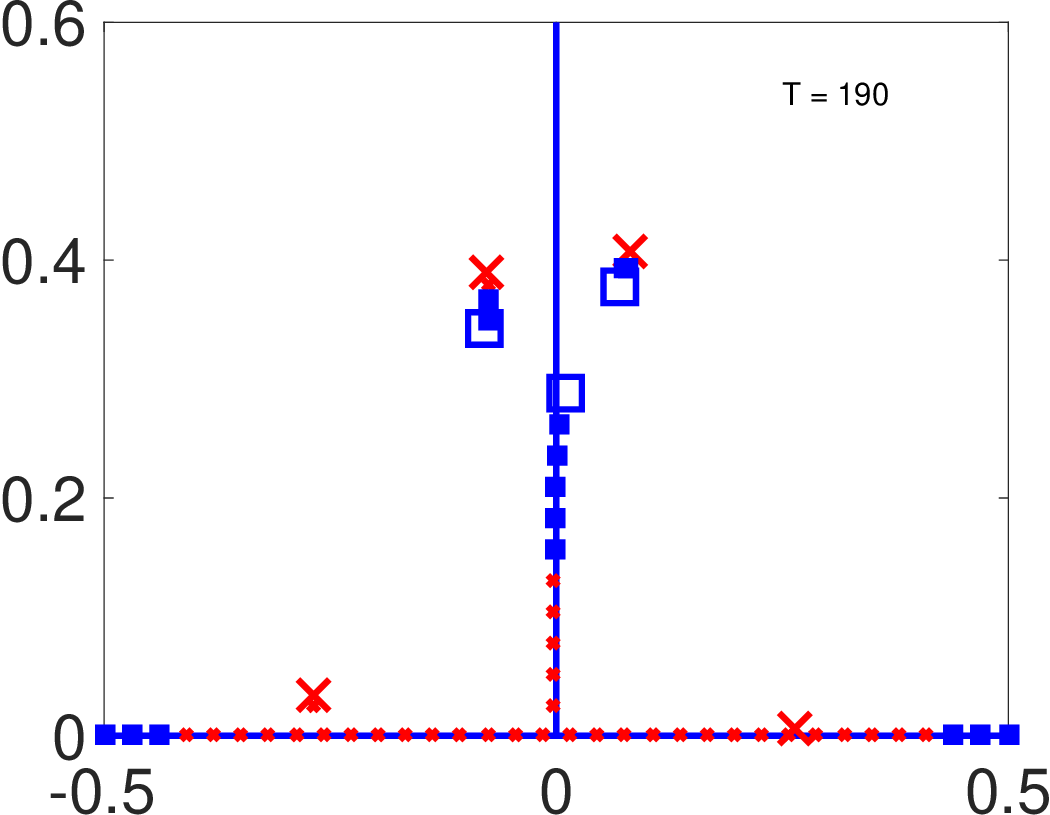}
\hspace{6pt}\includegraphics[width=0.25\textwidth,height=0.14\textheight]{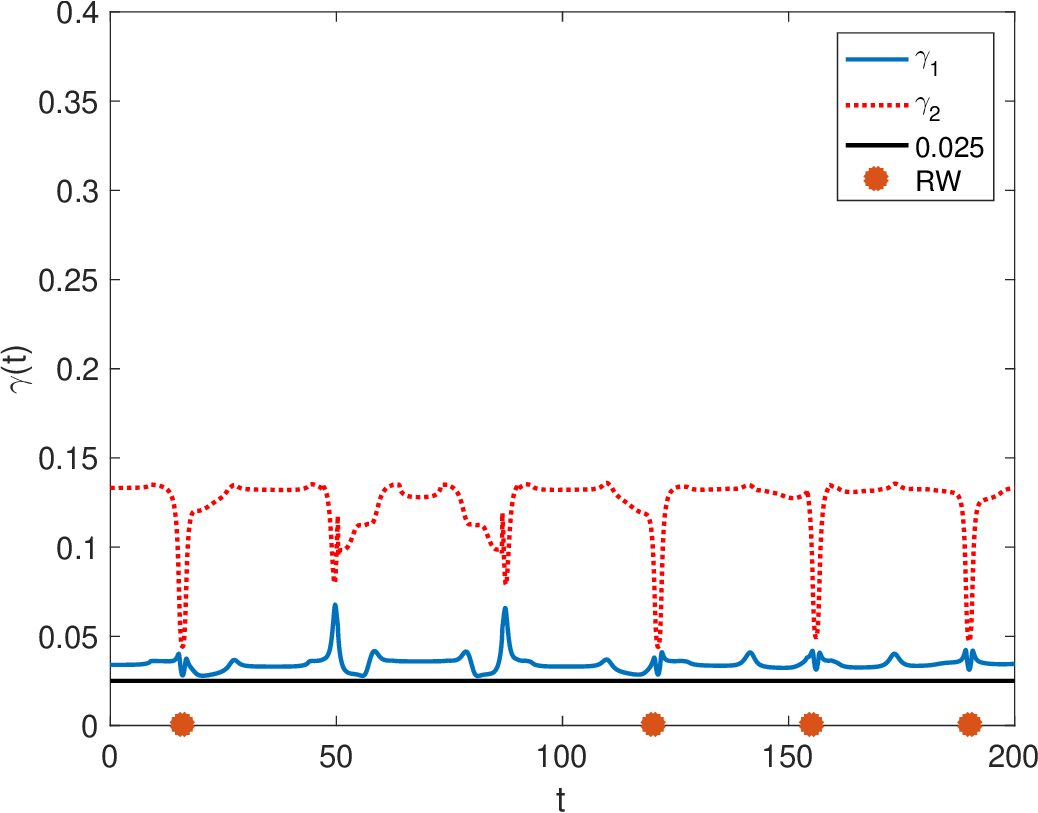}
}
\caption{HONLS evolution with Stokes initial data \rf{Stokes_ic}:
 Floquet spectra at  (a) $t = 42.5$, (b) $t= 48.9$, (c) $t=49.5$, (d) $t = 50$, and (e) $t = 190$.
(f) The band lengths $|\gamma_1(t)|$ and $|\gamma_2(t)|$ with the horizontal line indicating the threshold for a soliton-like state
  and  red dots marking rogue wave events.}
\label{HONLS_Stokes_spec}
\end{figure}

The spectrum evolves very gradually
 in the early stages of the simulation, as the solution takes time to build
 structure. Nevertheless, the spectral evolutions for all three models -- HONLS, V-HONLS, and NLD-HONLS -- are qualitatively
similar and closely aligned until $t \lessapprox 16$, with  the asymmetric configuration  becoming clearly visible around $t = 14.7$, as
shown in Figure~\ref{ic_Stokes_spec}(b).
Subsequently, the uppermost  band, $\gamma_1$, migrates
into the right quadrant, while  $\gamma_2$ moves into the left quadrant.
At the time
of the first rogue wave  event, $t = 15.9$, the spectrum is not in an SRW configuration (see Figure~\ref{ic_Stokes_spec}(c)).  
As shown in the bandlength evolution, Figure~\ref{HONLS_Stokes_spec}(f),
neither $\gamma_1$ nor $\gamma_2$ contracts sufficiently in length to satisfy the
soliton-like criterion and the rogue wave that forms is classified as a broad generic rogue wave.

Following this initial rogue wave event, the spectral evolutions of the three models  diverge.

\vspace{6pt}
\noindent {\bf Later stage HONLS dynamics:} Figures~\ref{HONLS_Stokes_spec} (a)-(e) present key features of
the spectral evolution.
Notably, 
Figure~\ref{HONLS_Stokes_spec}(a) shows that 
under HONLS evolution,   at $t = 42.5$ the spectrum
returns to a configuration close
to its initial state,  a small perturbation of the Stokes wave, and remains near this state for an extended interval
$ 38 \le t \le 46$.
This pattern of return is repeated multiple times during the interval $0 < t < 200$,  indicating that the 
solution undergoes prolonged and  repeated excursions through phase space that stay
near the unstable Stokes wave.

\begin{figure}[htp!]
  \centerline{
\includegraphics[width=0.3\textwidth]{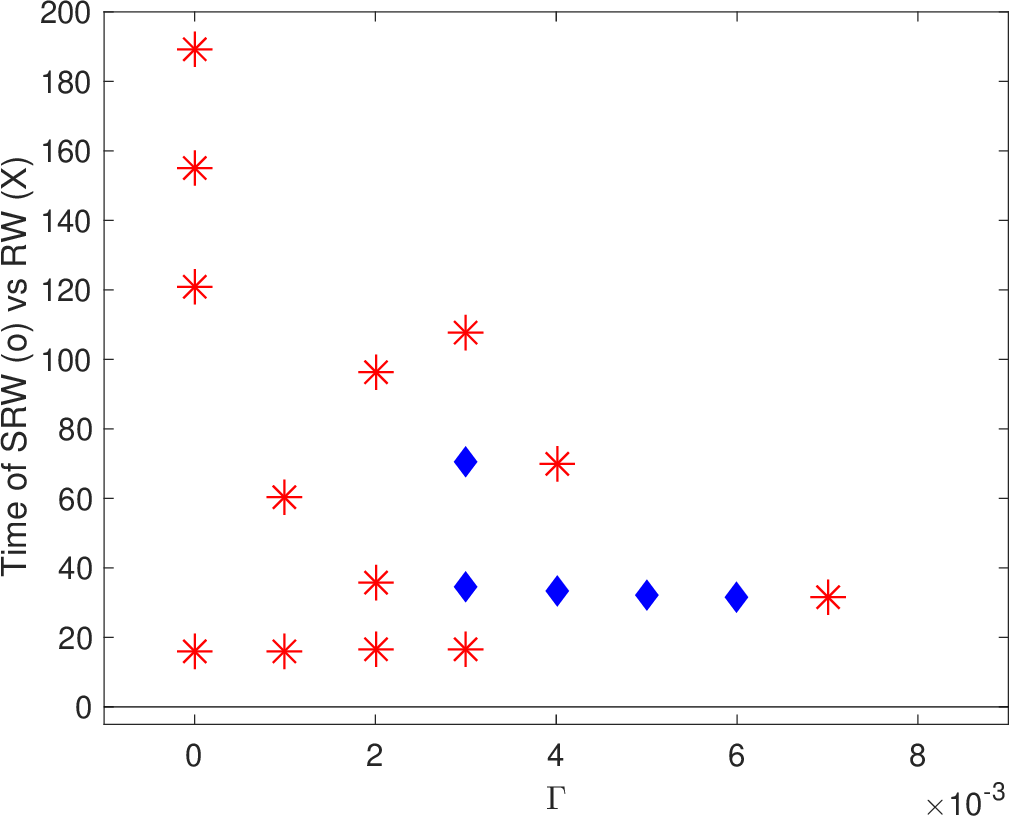}
\hspace{6pt}\includegraphics[width=0.3\textwidth]{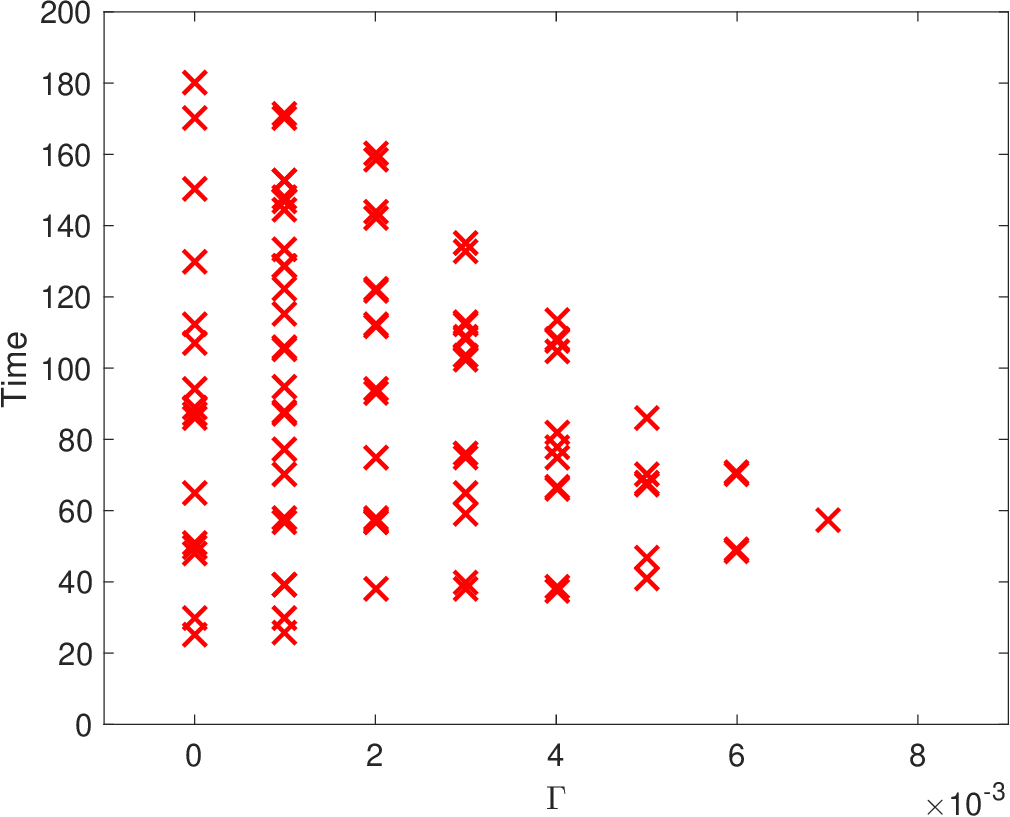}
}
  \caption{HONLS evolution ($\Gamma = 0$) and V-HONLS evolutions  for varying damping parameter $\Gamma$:  Timelines using Stokes  initial data
    \ref{Stokes_ic} 
    showing (a) SRW and generic rogue wave events (Magenta dots: 2-mode SRWs, blue diamonds: 1-mode SRWs,
 and stars:  generic rogue waves), and 
 (b) complex critical point transitions indicated by red '$\bigtimes$'.
 (there are no real transitions) }
  \label{VHONLS_Stokes_vary_G}
\end{figure}

In addition to the spectral configuration being intermittently close to that 
of the
Stokes wave,
the Floquet spectral evolution features frequent
critical point crossings 
 and
 reorganizations, Figures~\ref{HONLS_Stokes_spec}(b)-(d).
The overall spectral behavior is disordered, characterized by 
persistent bifurcations and continuous interaction between the first two nonlinear modes.
Critical point crossings in the HONLS evolution are summarized in the timeline 
shown in Figure~\ref{VHONLS_Stokes_vary_G}(b).
Significant excitation of a third nonlinear mode and real critical point transitions  -- observed in HONLS evolutions with SPB
initial data -- do not occur for Stokes wave initial data.

Figure~\ref{VHONLS_Stokes_vary_G}(a) provides a timeline for rogue wave activity for perturbed Stokes initial data, classifying events as
one- or two-mode SRWs or 
broader, non soliton-like waves. The HONLS equation corresponds
to $\Gamma = 0$. Unlike for SPB initial data,  neither
the first  rogue wave, nor any of the subsequent ones
meet the
criteria to be classified as SRWs. A representative
spectrum  of a generic rogue wave at 
$t = 190$ is shown in Figure~\ref{HONLS_Stokes_spec}(e).
These rogue waves all arise from
nonlinear mode interactions, rather than from
coherent and localized focusing which characterize SRWs.

The proximity of the solution to the Stokes wave enables
repeated reactivation of the modulational instability, which in turn regenerates new unstable modes with unaligned phases. The recurring buildup of incoherent modal content results in a higher phase variance, distinguishing the dynamics from that of the damped models or when initially structured (SPB) data was considered.

This disordered behavior is reflected in the phase variance diagnostic. As shown in Figure~\ref{HONLS_Stokes_str}(c), the  phase variance,
PVD$(t)$, is both  higher and  more erratic for Stokes initial data compared to SPB  data,
indicating reduced phase coherence and more  pronounced multi-modal wave behavior. This observation is 
further confirmed  in Figure~\ref{VHONLS_vary_Gamma}(c) for the HONLS
($\Gamma =0$), where
the averaged PVD is notably higher for Stokes initial data,
emphasizing the less
coherent dynamical evolution relative to the SPB case.

\vspace{6pt}
\noindent {\bf Summary of Results (HONLS, Moderately steep wave initial data):}
For perturbed Stokes initial data, HONLS dynamics are characterized by prolonged and repeated excursions near the unstable Stokes wave, leading to intermittent recurrences of the initial spectral configuration. Throughout the evolution, the Floquet spectrum undergoes persistent, disordered reconfigurations driven by frequent critical point crossings, although real critical point transitions and excitation of a third nonlinear mode—seen with SPB data—do not occur.

Rogue waves do arise, but none meet the criteria to be classified as soliton-like SRWs; instead, they emerge from  incoherent nonlinear mode interactions. The sustained proximity to the Stokes wave repeatedly reactivates modulational instability, continually seeding new unstable modes with unaligned phases.

As a result, the phase variance remains both elevated and highly fluctuating, reflecting lower overall coherence and stronger multimodal background activity compared to the SPB case. Overall, the HONLS evolution from Stokes initial data shows a  disordered,
less coherent dynamics dominated by persistent mode interactions.

\subsubsection{Moderately Steep Wave Initial Data: V-HONLS  equation}

\begin{figure}[htp!]
\centerline{
\includegraphics[width=0.3\textwidth]{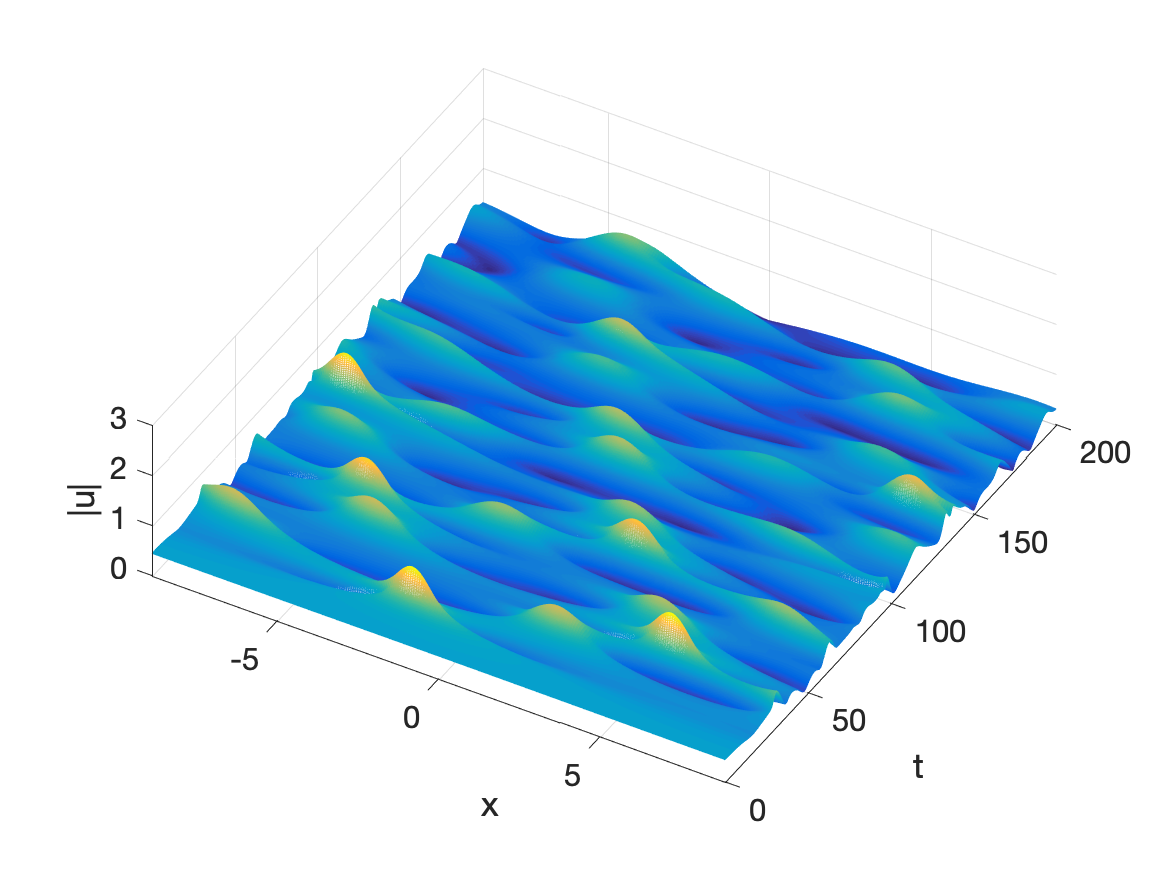}
\hspace{6pt}\includegraphics[width=0.25\textwidth]{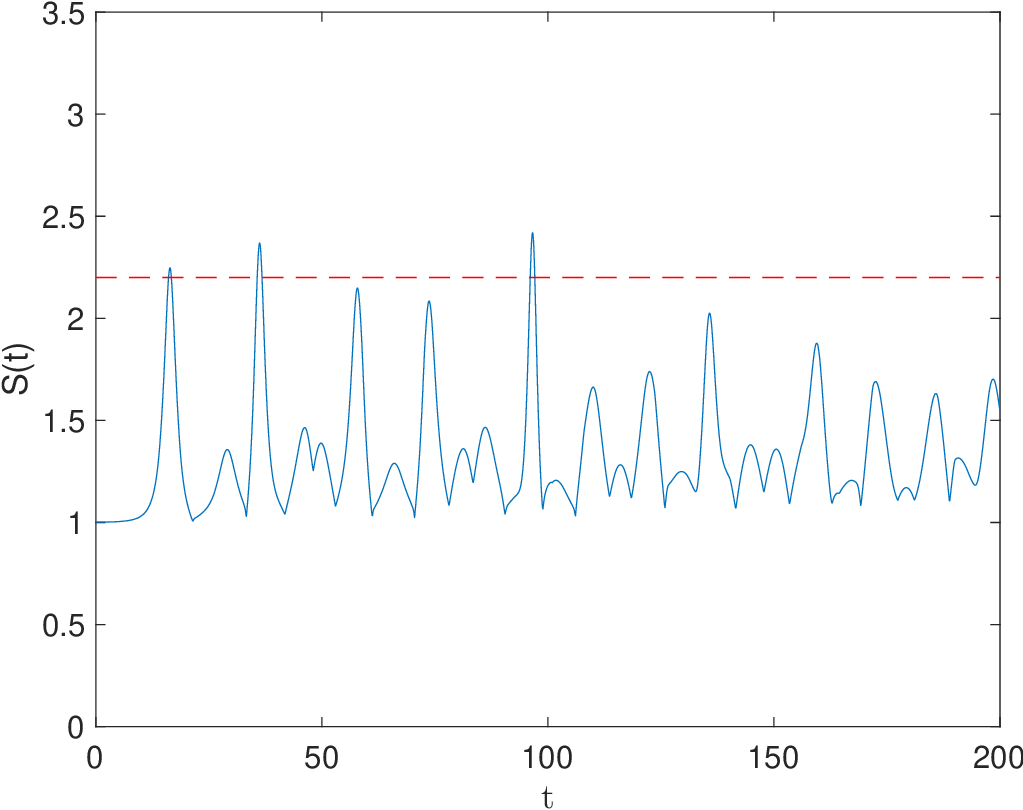}
\hspace{6pt}\includegraphics[width=0.25\textwidth]{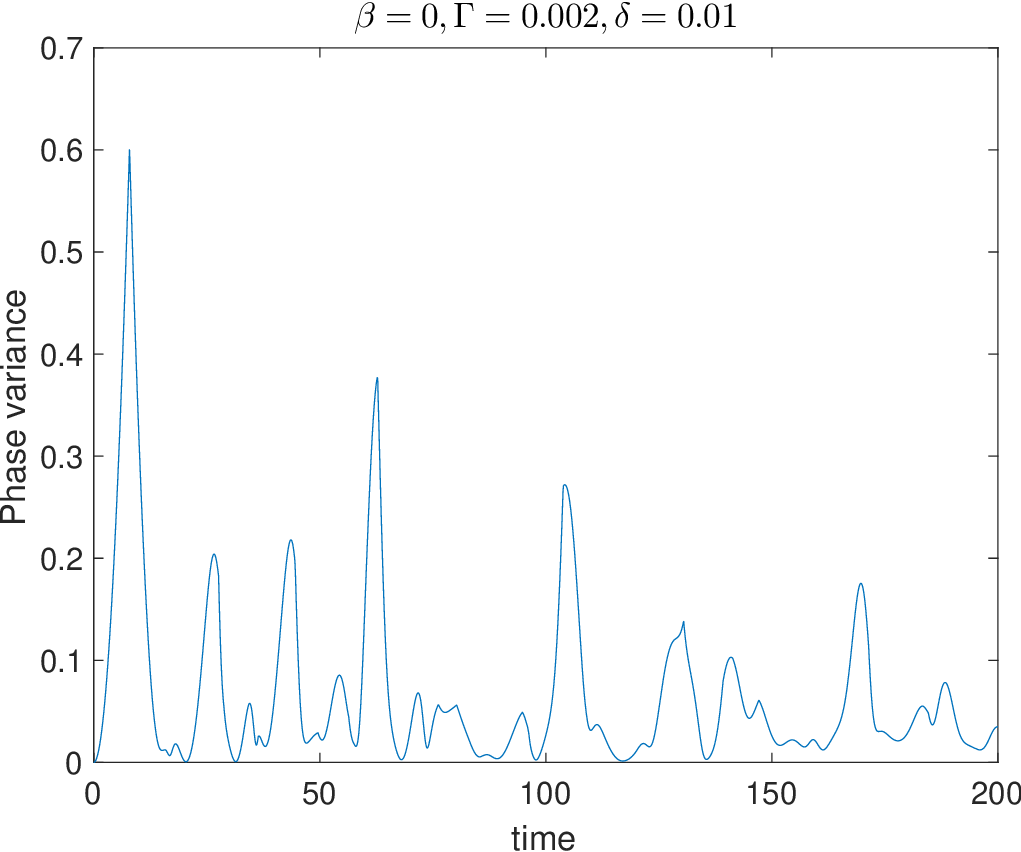}
}
\caption{V-HONLS evolution with $\Gamma = 0.002$ and Stokes  initial data
  \rf{Stokes_ic}: (a) $|u(x,t)|$ for $0\leq t\leq 200$,
  (b)  the strength $S(t)$, and (c) the phase variance PVD$(t)$. }
\label{VHONLS_Stokes_str}
\end{figure}

Figures~\ref{VHONLS_Stokes_str}(a)-(b) 
show the evolution of the surface $|u(x,t)|$ and the
strength $S(t)$ for the perturbed Stokes  initial data \rf{Stokes_ic} over the interval $0 < t < 200$ under V-HONLS with $\Gamma = 0.002$. The strength plot indicates that
rogue waves are more frequent and have greater  strength
compared to the HONLS simulations with the same initial data.
In contrast, compared to the V-HONLS
simulation with SPB initial data,  there are fewer, weaker  rogue waves.

\begin{figure}[htp!]
\centerline{
\includegraphics[width=0.225\textwidth]{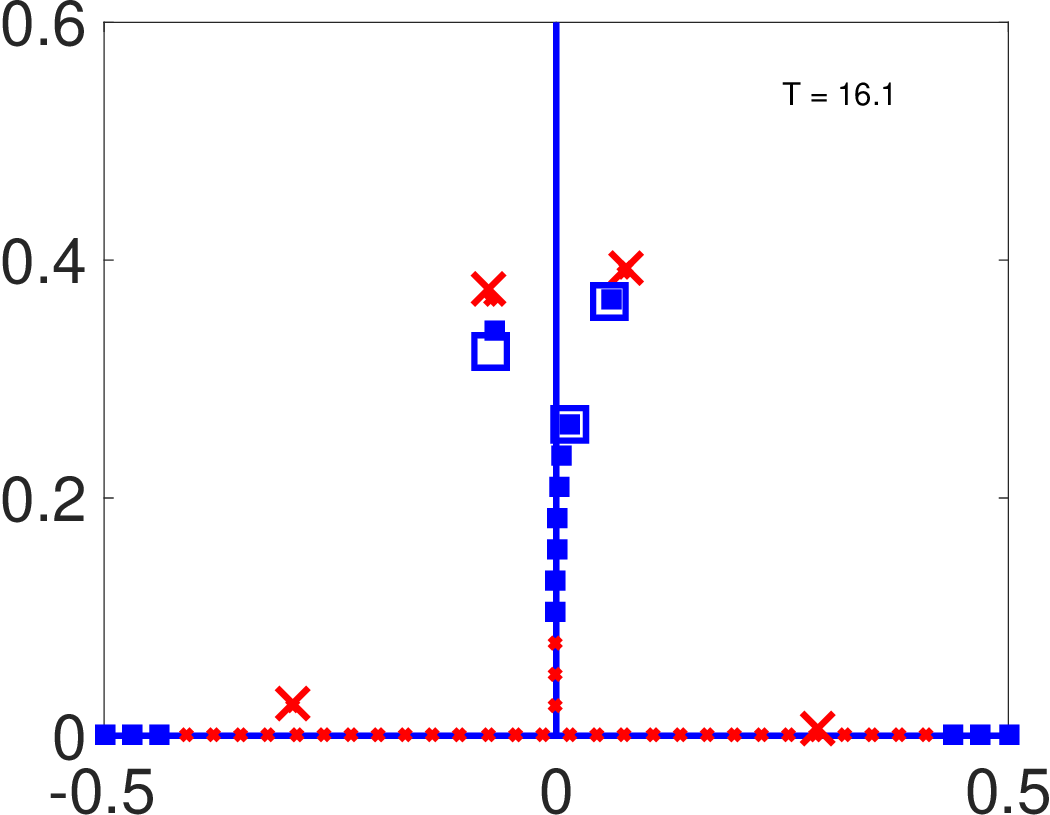}
\hspace{6pt}\includegraphics[width=0.225\textwidth]{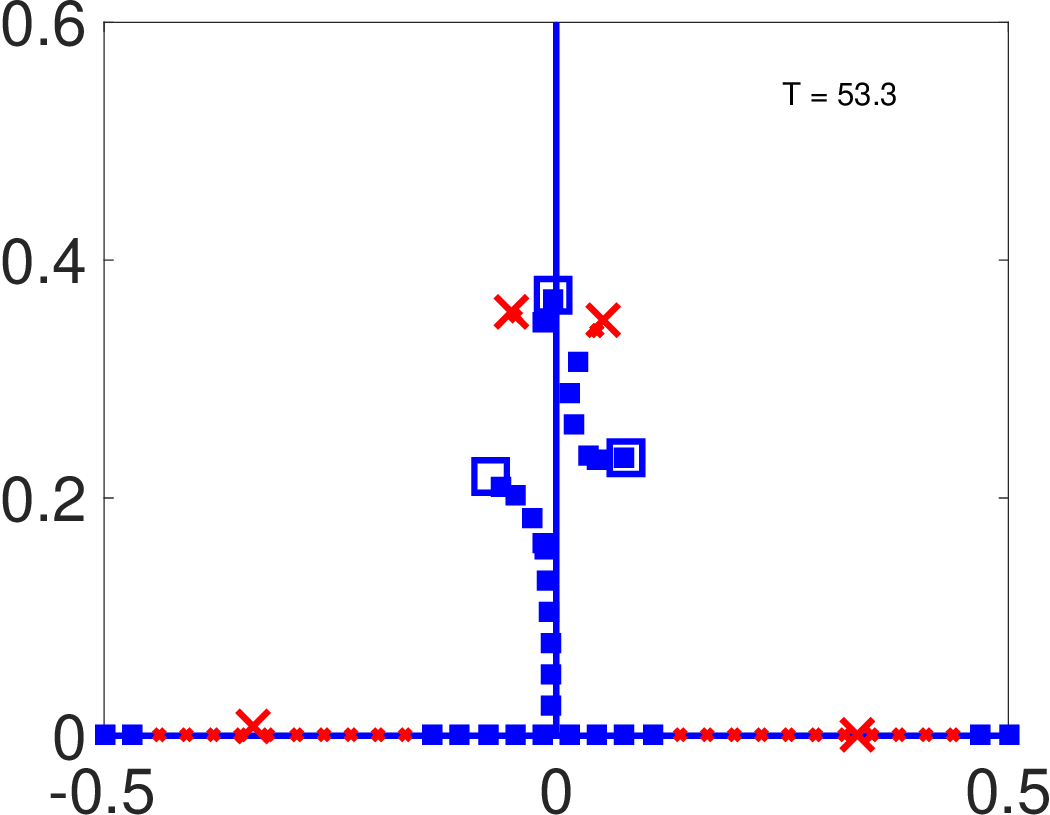}
\hspace{6pt}\includegraphics[width=0.225\textwidth]{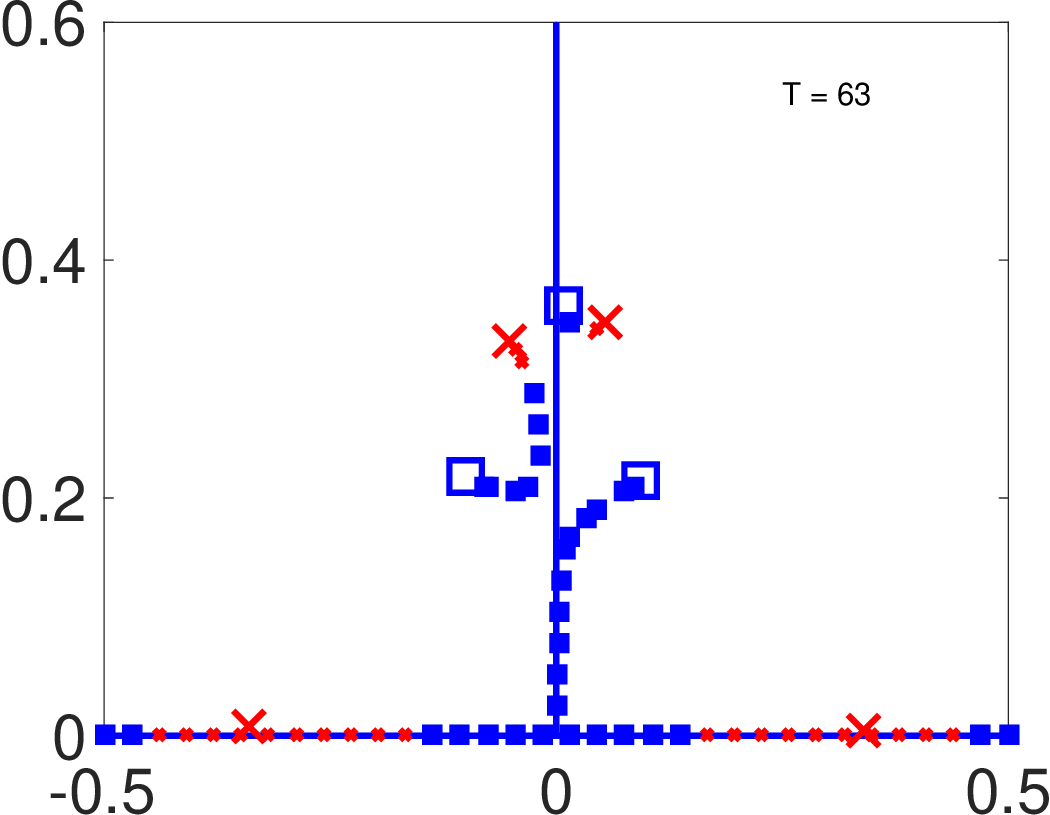}
\hspace{6pt}\includegraphics[width=0.225\textwidth,height=0.125\textheight]{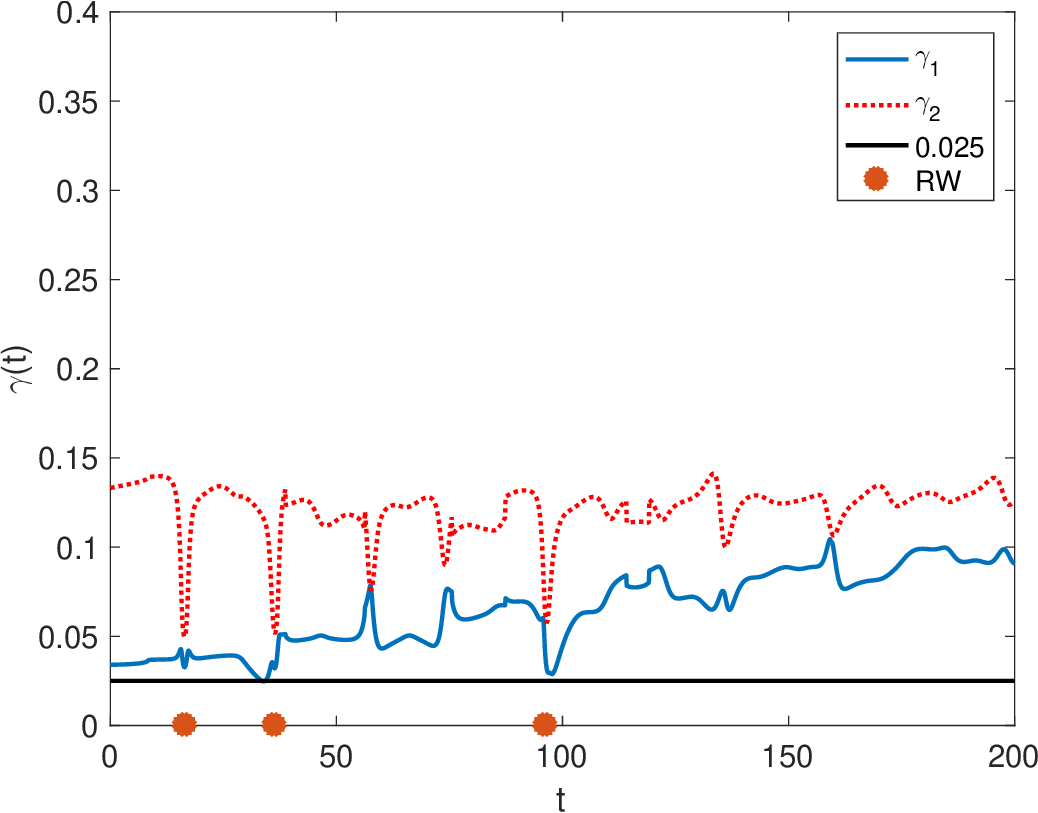}
}
\caption{V-HONLS evolution with $\Gamma = 0.002$ and Stokes initial data
  \rf{Stokes_ic}: The spectrum at
  (a) $t= 16.1$, (b) $t = 53.3$, and (c) $t = 63$, and (d) band length evolution.}
\label{VHONLS_Stokes_spec}
\end{figure}

 \vspace{6pt}
\noindent {\bf The Floquet spectral evolution up to the first rogue wave event:}
For Stokes-type initial data, the spectral evolution and dynamics  under V-HONLS qualitatively resembles those  of the  HONLS model up to approximately  $t \approx 16$.
A  generic rogue wave occurs at
$ t = 16.1$ 
in the V-HONLS simulation. The Floquet spectrum at that time, Figure~\ref{VHONLS_Stokes_spec}(a),  closely matches the spectrum observed during the corresponding event at $t = 15.9$ in the HONLS case, Figure~\ref{ic_Stokes_spec}(c).

\vspace{6pt}
\noindent {\bf Later stage V-HONLS dynamics:} 

The evolution of the band lengths in Figure~\ref{VHONLS_Stokes_spec}(d)
shows that
neither $\gamma_1$ nor $\gamma_2$ contracts sufficiently during rogue
wave events (given by the red bullets) to meet the soliton-like threshold.
Figure~\ref{VHONLS_Stokes_vary_G}(a) provides a timeline of rogue wave events as the  damping strength is varied, distinguishing between two-mode SRWs (red dots), one-mode SRWs (blue diamonds), and broader, non-soliton-like rogue waves (stars). For $\Gamma = 0.001$ and $\Gamma = 0.002$, all
observed events are generic rogue waves. Interestingly, at larger values of $\Gamma$, 
single-mode SRWs do emerge.
 Further, the number of rogue events does not decrease monotonically with increasing $\Gamma$, reflecting the influence of the unstable
background and its sensitivity to initial conditions and parameter variations.

Complex critical point crossings persist throughout the simulation and involve transitions in both unstable modes, as seen in Figures~\ref{VHONLS_Stokes_spec}(b)-(c), instead of involving primarily the second mode as in  simulations with steep SPB initial data. 
The number and type of  spectral  transitions across values of $\Gamma$ is summarized in Figure~\ref{VHONLS_Stokes_vary_G}(b). While the overall number of transitions generally decreases with stronger damping, this trend is not strictly monotonic. All the spectral  transitions in V-HONLS occur at complex critical points, signaling increased interaction between the first two nonlinear modes.

A key difference in the long-term evolution lies in how viscosity modifies
the system’s phase-space dynamics. In V-HONLS, damping suppresses long-lived proximity to the unstable Stokes wave, thereby limiting the duration and recurrence of modulational instability.
This is not apparent from the band length evolution alone, but is evident in the spectral plots for later times.  As shown in the representative V-HONLS
spectral plot, Figure~\ref{VHONLS_Stokes_spec}(c), the spectrum
never returns to the initial perturbed Stokes wave
configuration, unlike the  HONLS
evolution which repeatedly revisits such a configuration (cf. Figure~\ref{HONLS_Stokes_spec}(a)).

As a result, the V-HONLS simulation exhibits a 
consistently lower phase variance (PVD), as shown in Figure~\ref{VHONLS_Stokes_str}(c).
In contrast, the  HONLS simulation (Figure~\ref{HONLS_Stokes_str}(c)) shows
higher PVD values, due to excursions near the modulationally  unstable Stokes state,  supporting the repeated growth of incoherent modes.

\vspace{6pt}
\noindent {\bf Summary of Results (V-HONLS, Moderately steep  wave initial data):}
For perturbed Stokes wave initial data, viscous damping partially suppresses incoherence by
 limiting excursions near the modulationally unstable Stokes wave,
 resulting in lower phase variance, fewer critical point crossings, 
 and less persistent multimodal disorder compared to HONLS.
Rogue waves are typically broad and non-soliton-like; at higher 
 values of $\Gamma$ some rogue waves manifest as one-mode SRWs.

\begin{figure}[htp!]
\centerline{
\includegraphics[width=0.3\textwidth]{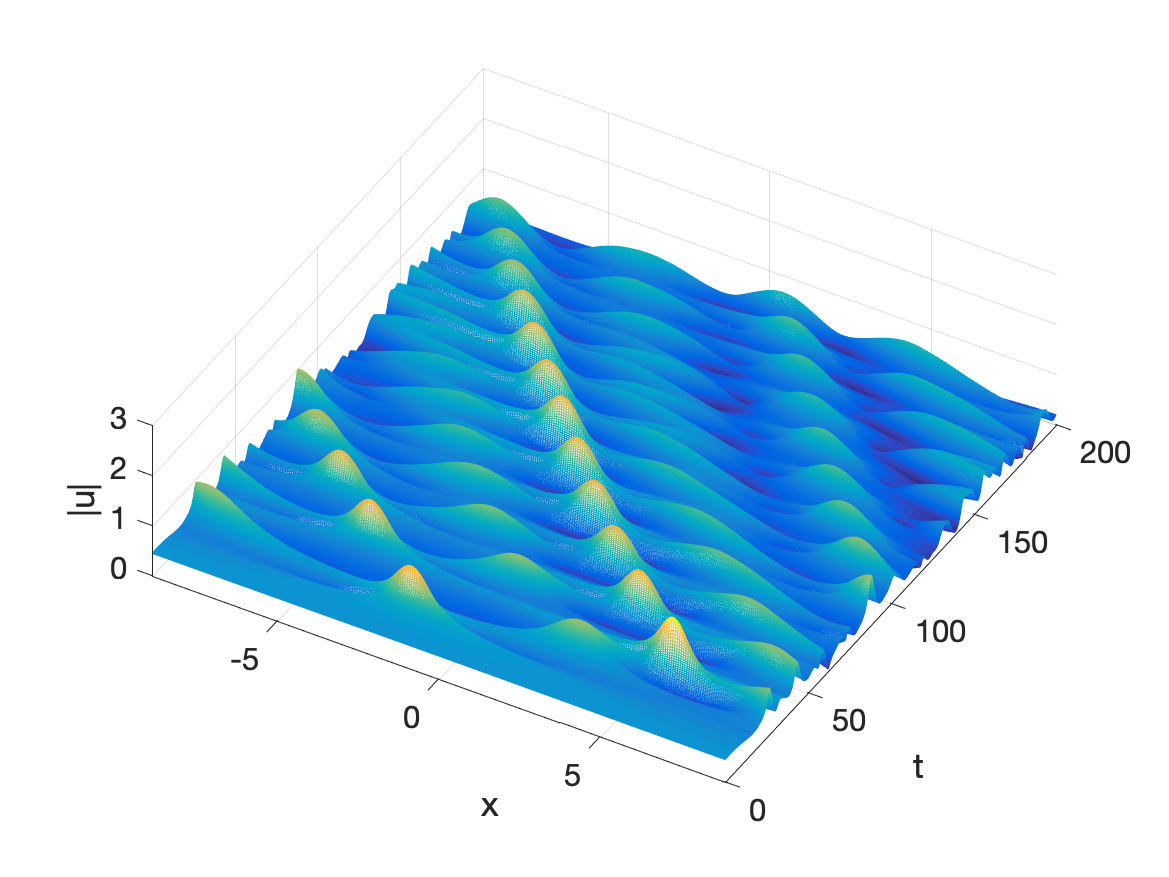}
\hspace{6pt}\includegraphics[width=0.25\textwidth]{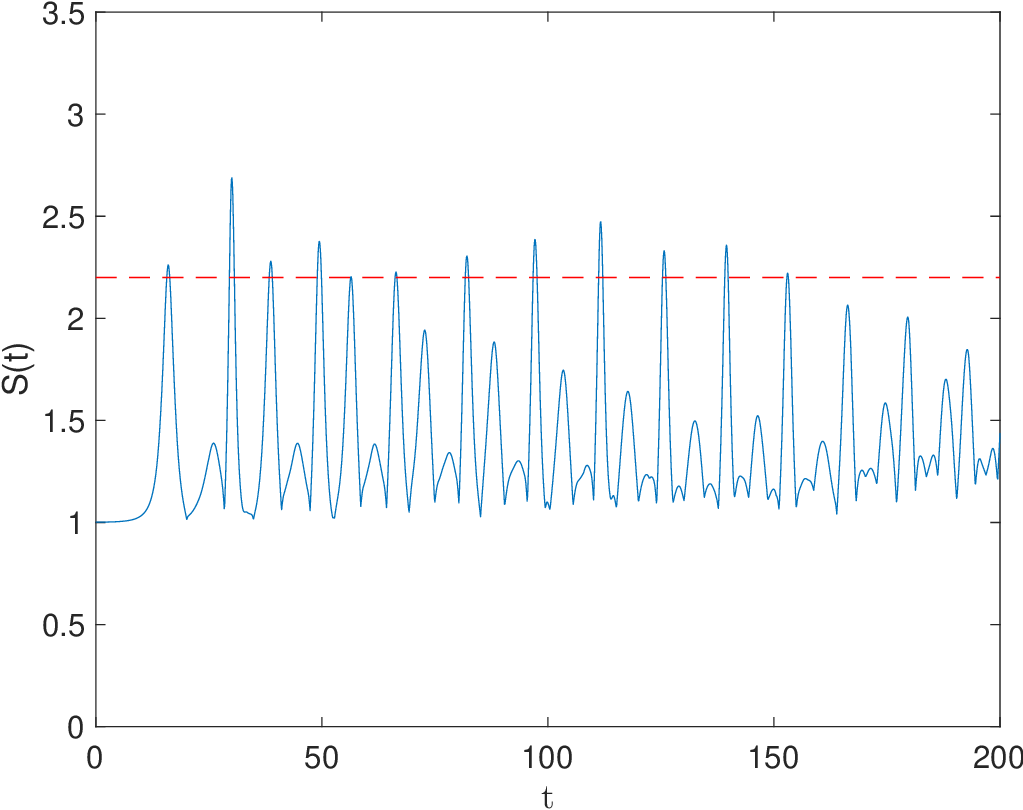}
\hspace{6pt}\includegraphics[width=0.25\textwidth]{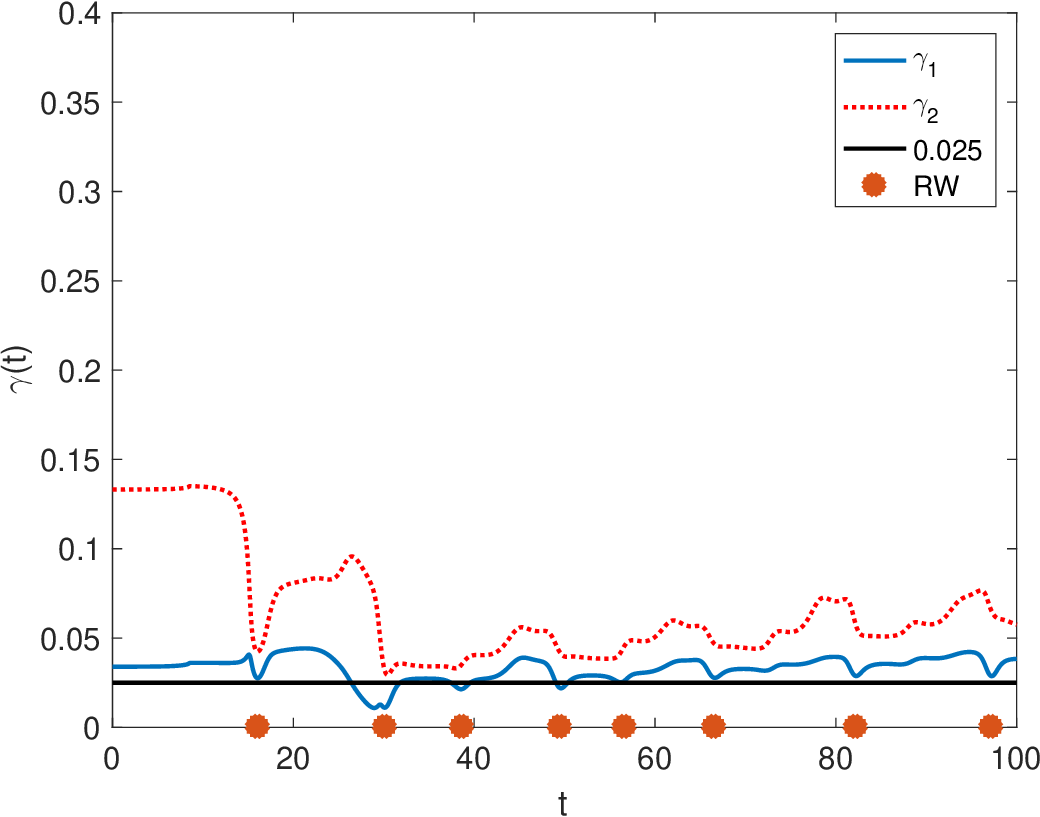}
}  
\vspace{6pt}
\centerline{
\includegraphics[width=0.25\textwidth]{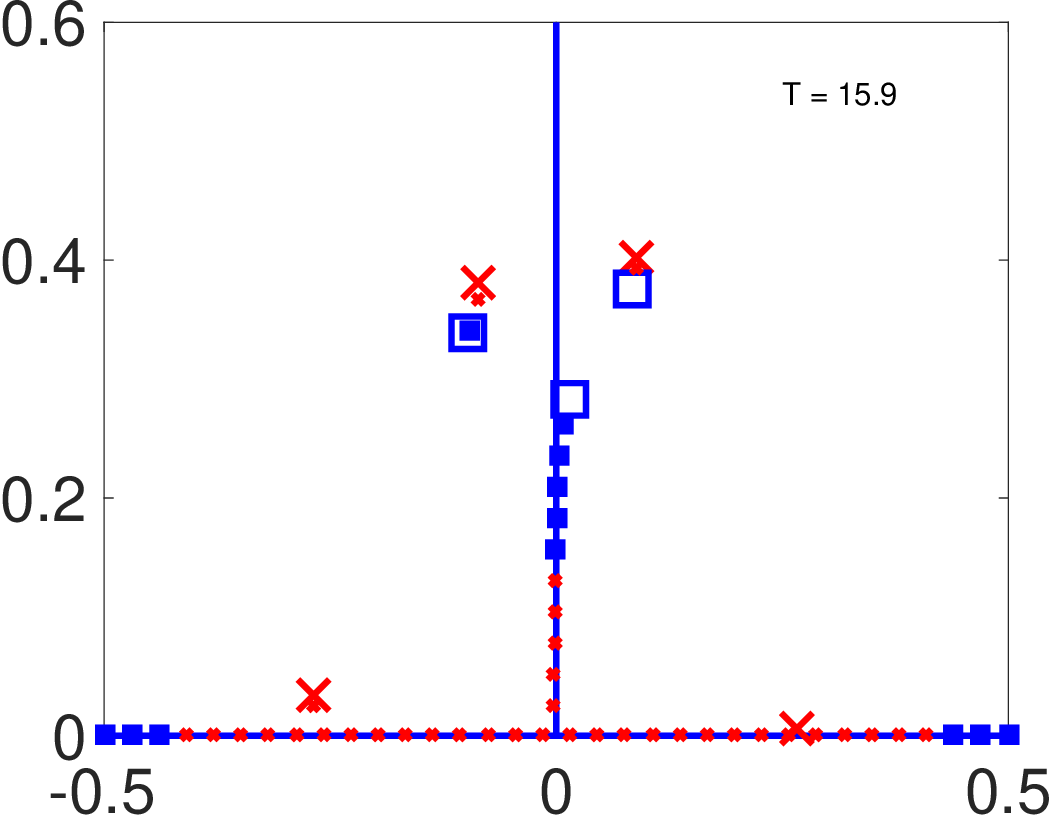}
\hspace{6pt}  \includegraphics[width=0.25\textwidth]{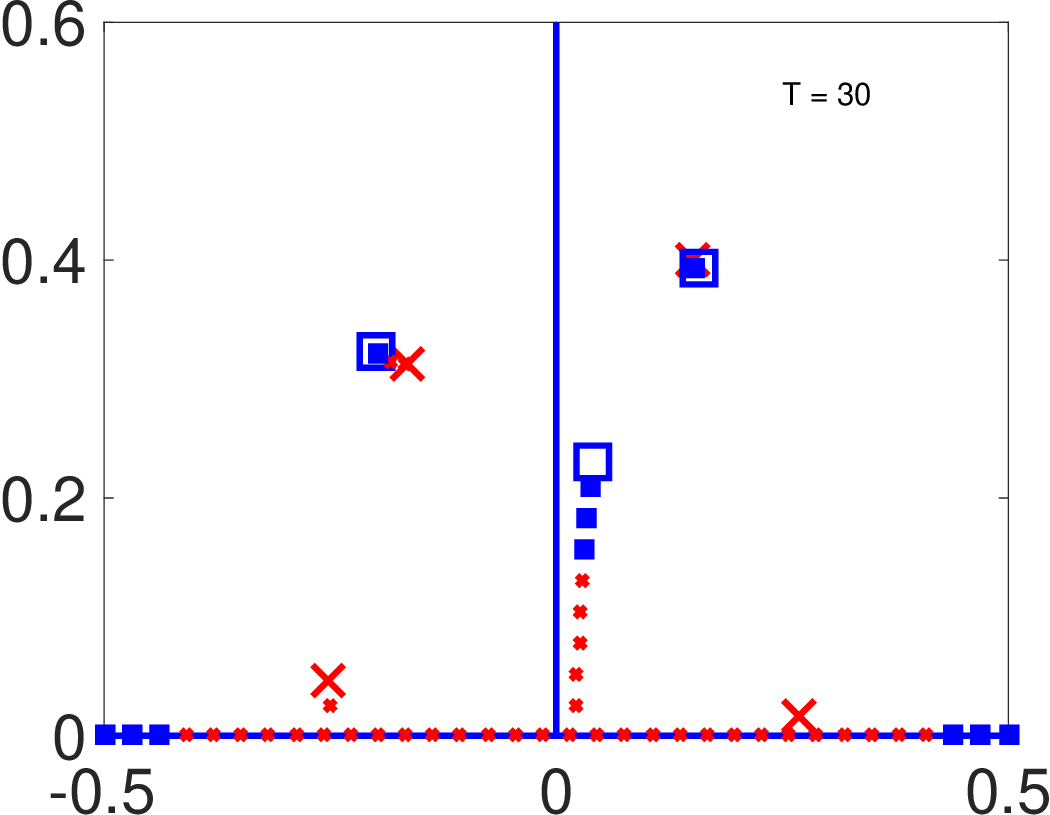}
\hspace{6pt}\includegraphics[width=0.25\textwidth]{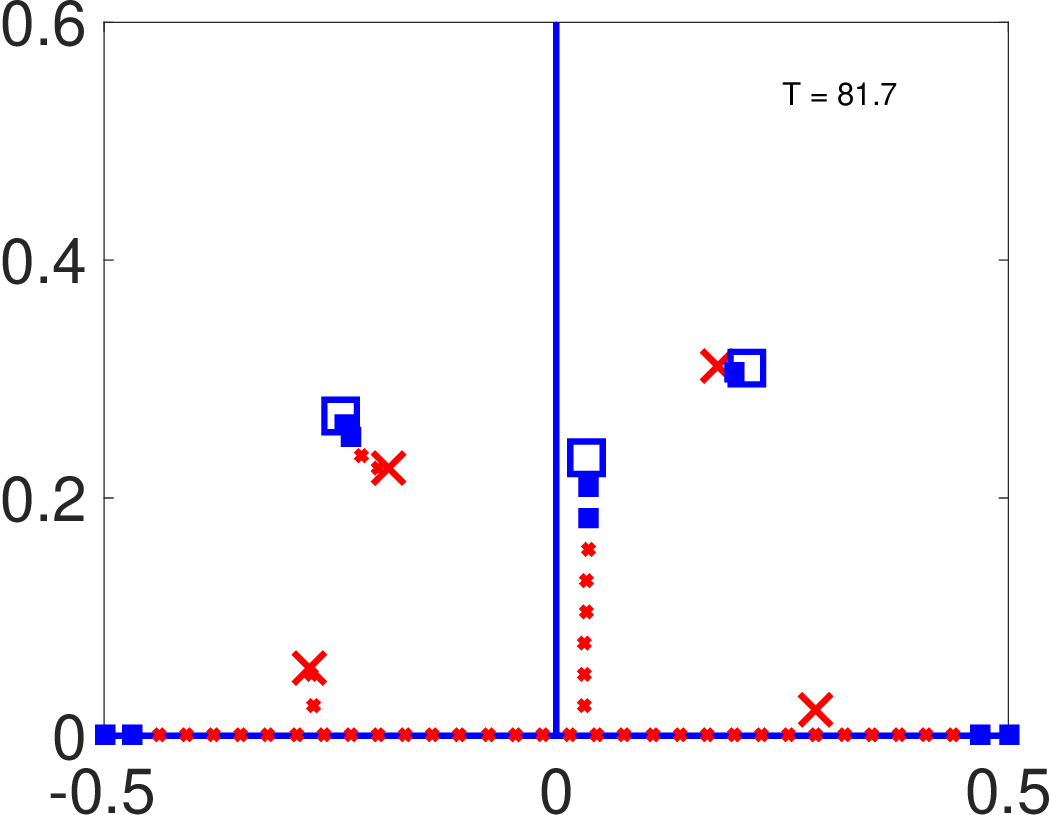}
}  
\caption{NLD-HONLS with $\beta = 0.1$ and perturbed Stokes initial data \rf{Stokes_ic} : (a) $|u(x,t)|$ for $0\leq t\leq 200$,
  (b) the strength $S(t)$, and (c) the band lengths $|\gamma_1(t)|$
  and $|\gamma_2(t)|$. The horizontal line marks the
  soliton-like threshold and the  red dots denote  rogue wave events. The Floquet spectra at: (d) $t = 15.9$, (e) $t = 30$, (f) $t = 81.7$. 
}
\label{NLD_Stokes_spec}
\end{figure}

\begin{figure}[htp!]
\centerline{
\includegraphics[width=0.25\textwidth]{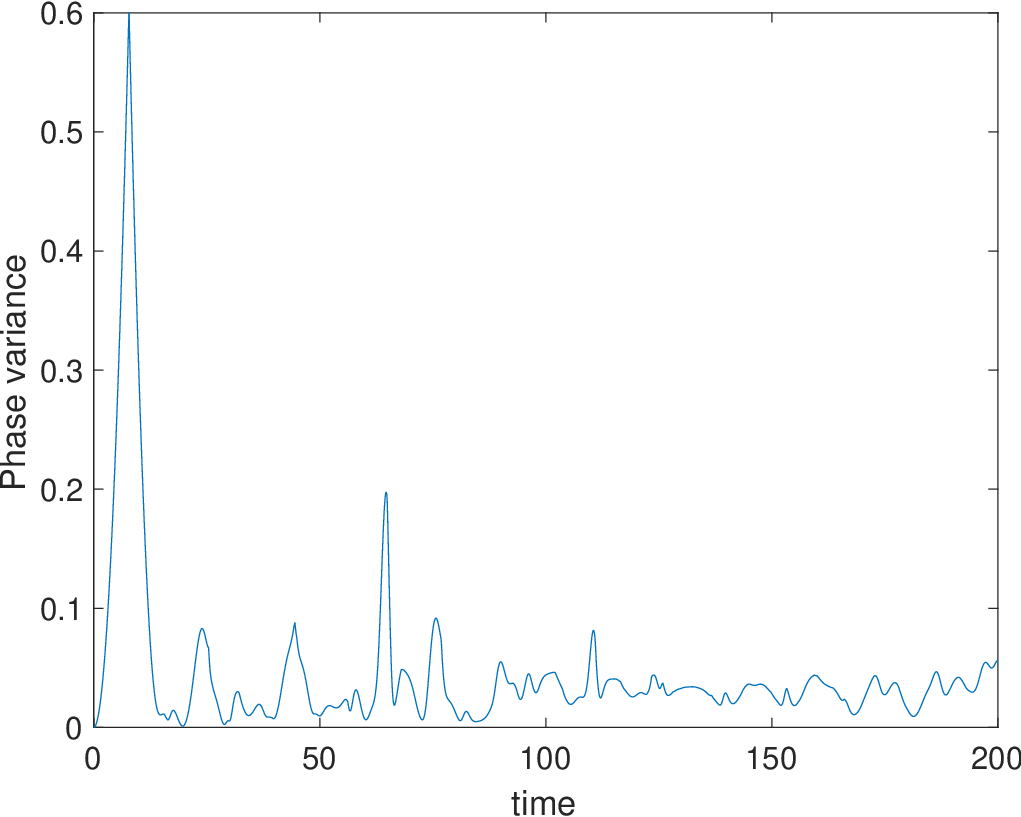}
\hspace{6pt}\includegraphics[width=0.25\textwidth]{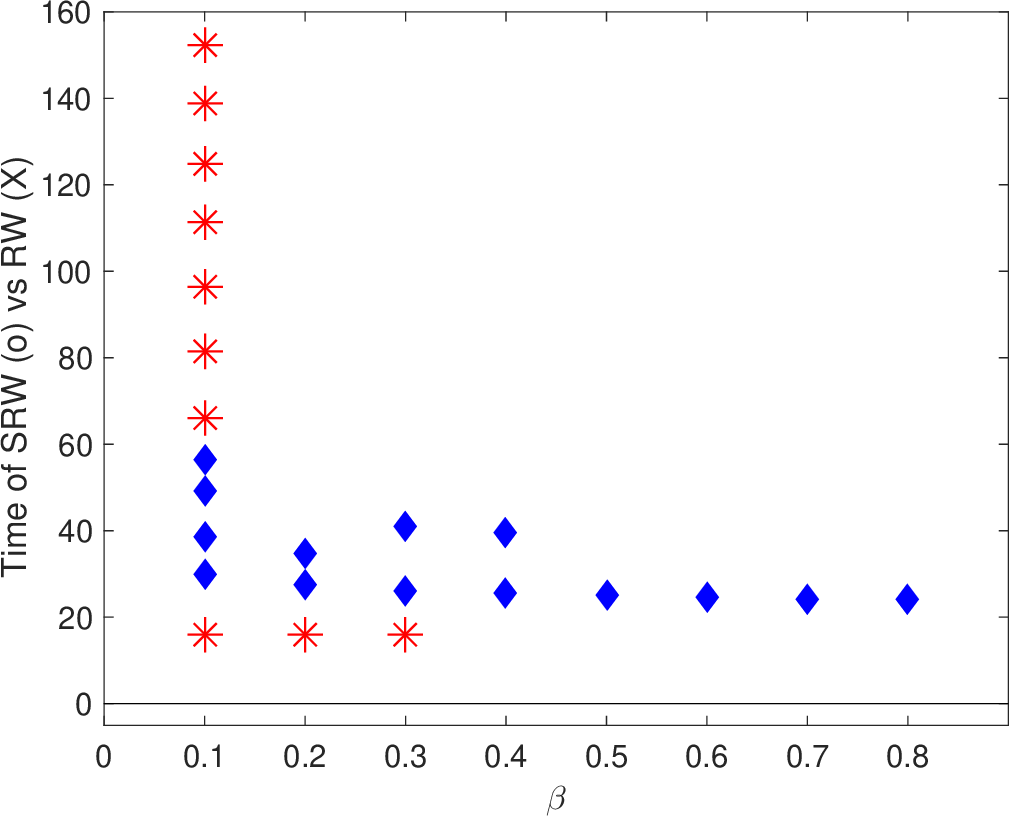} 
}
\caption{NLD-HONLS evolutions with  perturbed Stokes initial data
  \rf{Stokes_ic}:  (a) the phase variance PVD$(t)$ for  $\beta = 0.1$
  and (b) timeline showing  rogue wave events as damping parameter
  $\beta$ varies.
 Magenta dots indicate the occurrence of 2-mode SRWs, blue diamonds 1-mode SRWs,
    and stars mark generic rogue waves.}
\label{NLD_Stokes_PVD}
\end{figure}

\subsubsection{Moderately Steep Wave Initial Data: NLD-HONLS} 
Figures~\ref{NLD_Stokes_spec}(a) and (b)  provide the evolution of the surface $|u(x,t)|$ and the
strength $S(t)$ under the NLD-HONLS equation with $\beta = 0.1$,
using perturbed Stokes wave initial data \rf{Stokes_ic} over the interval $0 < t < 200$.  The strength plot reveals frequent rogue wave events up to approximately  $t = 152$, significantly more than those observed in the V-HONLS and HONLS models.

\vspace{6pt}
\noindent {\bf The Floquet spectral evolution up to the first rogue wave event:}
As previously discussed, for Stokes-type initial data, the spectral evolution under NLD-HONLS qualitatively mirrors that of the  HONLS and V-HONLS models up to the first rogue wave event at
$t = 15.9$.  This event is a  generic rogue wave (Figure~\ref{NLD_Stokes_spec}(d)),
and aligns closely with similar  events 
in the HONLS simulation
at $t = 15.9$ (Figure~\ref{ic_Stokes_spec}(c)) and the V-HONLS simulation $ t = 16.1$ 
(Figure~\ref{VHONLS_Stokes_spec}(a)).

\vspace{6pt}
\noindent {\bf Later stage NLD-HONLS dynamics:}
The length of the spectral  band in the right quadrant,
$\gamma_1$, remains generally small, slightly exceeding the
threshold for a one soliton-like state.
However, after the first rogue event, it repeatedly contracts enough
to satisfy the soliton-like criterion, indicating
recurrent
formation of one-mode SRWs. This behavior is
clearly visible in
the bandlength evolution plot Figure~\ref{NLD_Stokes_spec}(c).
In contrast, the length
of the band in the left quadrant, $\gamma_2$, consistently remains larger
than the threshold for a soliton-like mode. 
As with the  steep wave initial data. The Floquet spectrum is well-organized and critical point crossings do not occur
in the NLD-HOLNLS evolutions for perturbed Stokes data.

One-mode SRWs are obtained until approximately $t = 57.2$.
A representative example of this state is shown at $t = 30$  in Figure~\ref{NLD_Stokes_spec}(e).
After this time, rogue waves become broader and less localized, reflecting the  growing band lengths $|\gamma_1|$ and $|\gamma_2|$,
as exemplified by the rogue event at $t = 81.7$ in
Figure~\ref{NLD_Stokes_spec}(f). 

Once the NLD-HONLS evolution passes the initial development of the modulational instability (for $ t \gtrapprox 15.9$), the  Floquet spectral configurations become
robust,  reflecting the emergence of underlying  dynamical coherence.
This is  further supported by low values of the phase variance  PVD diagnostic for $t \gtrapprox 15$. 
Figure \ref{NLD_Stokes_PVD}(a) shows the evolution of the  PVD: it reaches a
maximum of $\mbox{PVD}_{max}\approx 0.6$  during the early stage of modulational instability,  but then drops rapidly as the effects of 
nonlinear damping effects are felt and coherent structures start to form.

Figure~\ref{NLD_Stokes_PVD}(b) shows a timeline of SRWs and generic rogue wave events in NLD-HONLS simulations with Stokes-type initial data for $\beta \in[0.1,0.8]$.
 As clearly seen in the plot, the total number of rogue wave events decreases as  $\beta$ increases, and only for $\beta = 0.1$ do  generic rogue waves occur after
  $t = 57.2$.
Notably,  when  $\beta \ge 0.4$, the first rogue waves that emerge are  localized coherent one-mode SRWs rather than broader generic events.
 highlighting the organizing effect of stronger damping.

Further as $\beta$ is varied, Figure~\ref{NLD_vary_Beta}(b) shows consistently low averaged PVD values are obtained, remaining below 0.08 in all cases, similar to those  observed in the SPB simulations. These low PVD values are consistent 
  with the formation of sharply localized SRWs.

  A clear contrast in behavior emerges with the prior SPB case: the system does not sustain a soliton-like state for any extended period. Instead, transient  one-mode soliton-like regimes arise, as indicated by  contractions in
  the band length $\gamma_1$.

\vspace{6pt}
\noindent {\bf Summary of Results (NLD-HONLS, Moderately steep wave initial data):}
The Floquet spectrum is well-organized, critical point crossings are absent, and repeated contractions of the spectral band length 
$\gamma_1$
support the emergence of  coherent one-mode SRWs until about 
$t = 57.2$
After this time, rogue waves become broader and less localized as the bands increase, yet the overall evolution remains notably more structured than in the HONLS and V-HONLS  models.
Furthermore, 
 the NLD-HONLS evolution maintains lower phase variance throughout.

\subsection{Spectral Downshifting in the NLD-HONLS and V-HONLS equations}
\subsubsection{Energy and Momentum Evolution under NLD-HONLS and V-HONLS}
In this section, we examine the evolution of key wave properties, the energy and momentum, which offer insight into the underlying dynamics, focusing on their behavior within the NLD-HONLS and V-HONLS frameworks. 

The total wave energy $E(t)$ and momentum $P(t)$ are defined by
\be
E(t) = \frac{1}{L} \int_0^L |u|^2\,\mathrm{d}x, \qquad
P(t) = \frac{i}{2L}\int_0^L (uu^*_x -u^*u_x)\,\mathrm{d}x.
\label{EandPdefn}
\ee
Using the Fourier series,
\begin{align*}
u(x,t) = \sum_{k = -\infty}^\infty \hat{u}_k(t) e^{\ri \mu_k x}, \qquad
|u(x,t)|^2 = \sum_{k=0}^\infty (\alpha_k(t) e^{\ri \mu_k x} + \alpha^*_k(t) e^{-\ri \mu_k x} ),
\end{align*}
one obtains the Fourier representations for the energy and momentum,
\begin{align}
 E(t)  = \sum_{k = -\infty}^\infty |\hat{u}_k|^2, \qquad  P(t)  = \sum_{k = 1}^\infty \mu_k(|\hat{u}_k|^2 -|\hat{u}_{-k}|^2).
\end{align}
where $\mu_k = 2\pi k/L$.
In particular, the momentum  provides a  measure of the spectral asymmetry, which is central to understanding the mechanism of frequency downshifting.

For equation \rf{DHONLS} the evolution of the wave energy and momentum are as follows:

\be
  \frac{dE}{dt} = -\left[ 2\Gamma E + 4\eps\beta B + 2\eps\Gamma P\right],\qquad
  \frac{dP}{dt} = -\left[2\Gamma P - 4\eps\beta S + 4\eps\Gamma Q\right]
\label{derivE&derivP}
\ee
where
\[B = \frac{1}{L}\int_0^L |u|^2 {\cal H}\left(|u|^2_x\right)\, dx,\quad
  Q = \frac{1}{L}\int_0^L |u_x|^2 \, dx,\quad
      S = \frac{1}{L}\int_0^L  \Im\left(u u_x^*\right){\cal H}\left(|u|^2_x\right) \, dx\]
and $\Im(z)$ is the imaginary part of $z$. Clearly when  $\beta >  0$ and $\Gamma = 0$ then $\frac{dE}{dt} < 0$ for all time indicating continuous energy loss due to nonlinear mean flow damping. Similarly when  $\Gamma >0$ and $\beta =  0$ the system exhibits energy dissipation via viscous damping. 
Note that $E(t)$ and $P(t)$  are conserved under NLS or HONLS dynamics ($\Gamma = \beta = 0$).

   \subsubsection{Measuring Downshifting}
   Modulational instability drives spectral downshifting by enabling energy exchange between the carrier wave and sidebands, as described by the conservative NLS and HONLS models. This interaction is typically reversible, leading to temporary downshifts during cycles of modulation and demodulation. However, Lake et al. \cite{LYF1978} observed that at higher steepness, the transfer to lower sidebands can become irreversible—an effect not accounted for in conservative formulations.

   Two distinct spectral diagnostics --  the spectral center and the
   spectral peak -- are used to characterize frequency downshifting,
   as they capture different aspects of the wave energy distribution.
\begin{itemize}
\item {\bf Frequency downshifting diagnostics:}
\begin{itemize}

      \item The spectral mean 
  \[
  k_m (t): =  \frac{\sum_{k = -\infty}^\infty \mu_k|\hat{u}_k|^2}{\sum_{k = -\infty}^\infty |\hat{u}_k|^2} = \frac{P(t)}{E(t)},
\]
also called the spectral center, is a weighted average of the spectral content of the wave.    %where $\hat{u}_k$(t)  are the Fourier coefficients of the wave field $u(x,t)$,
    
  \item The spectral peak,  $k_{peak}$,  represents the dominant wave frequency.
    It is the wave number of the Fourier mode of maximal amplitude at a given time t. In essence, $k_{peak}$ corresponds to the  highest energy mode.
  \end{itemize}
\end{itemize}

In terms  of the spectral mean, frequency downshifting is considered to occur when
$k_m$ is decreasing, while in the spectral peak sense downshifting  occurs  when
the spectral peak changes from its original value $k_{peak} = 0$ to a lower mode \cite{UK1994,KO1995}.

If the spectral energy is tightly localized around the peak, then
$k_m \approx k_{peak}$. 
However, the same value of $k_m$
can arise from different distributions of $|\hat{u}_k|^2$. 
When analyzing frequency downshifting, researchers typically track both the spectral mean $k_m$ and the spectral peak $k_{peak}$ over time. Accordingly, in our numerical experiments we apply the following criterion \cite{CESS2023}:

\noindent {\bf A criterion for frequency downshifting:}
\vspace{6pt}
\noindent
{\em Frequency downshifting occurs when both of the following conditions are met:
  1) $k_m$ is decreasing and 
  2) $k_{peak}$ moves to a lower mode. }

  For each of the V-HONLS and NLD-HONLS models,
$E(t)$ and $P(t)$  decrease at different rates allowing for the possibility of downshifting.
If both conditions are met, our focus turns to an aspect of downshifting which is related to $k_{peak}$.
  Since $k_{peak}$ may shift to a sideband mode and then
  back to the carrier mode
 repeatedly while $k_m$ is decreasing, we are interested in
 identifying the last  time at which the carrier mode is the dominant mode.  

\begin{dfn}
\label{tds}
    The {\em time of permanent downshift} $t_d$ is 
defined as the time at which the carrier wave ceases to be the dominant mode.  Note that $t_d$ is determined by $k_{peak}$.
\end{dfn}

\subsubsection{Timing of Downshift Relative to the Last Rogue Wave}
In this subsection we briefly  demonstrate that permanent downshift occurs in
both the V-HONLS and NLD-HONLS models. A more thorough discussion of the
underlying  downshifting mechanisms in  these models can be found in
\cite{CESS2023, SS2015}.
Here  we specifically examine the roles of SRWs and the type of
dissipation in driving permanent frequency downshifting.

\begin{figure}[htp!]
\hspace{6pt}\includegraphics[width=0.3\textwidth]{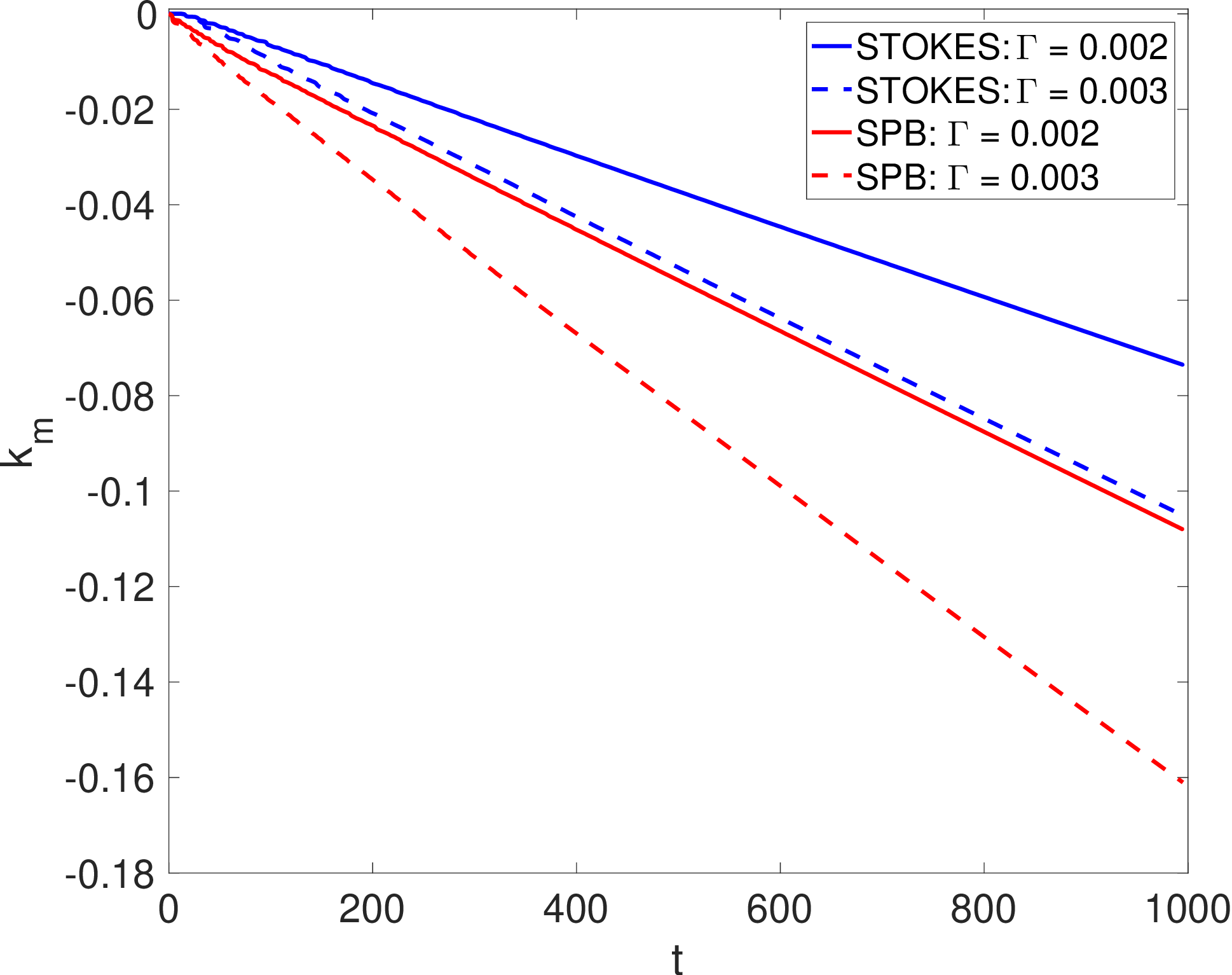}
\includegraphics[width=0.3\textwidth]{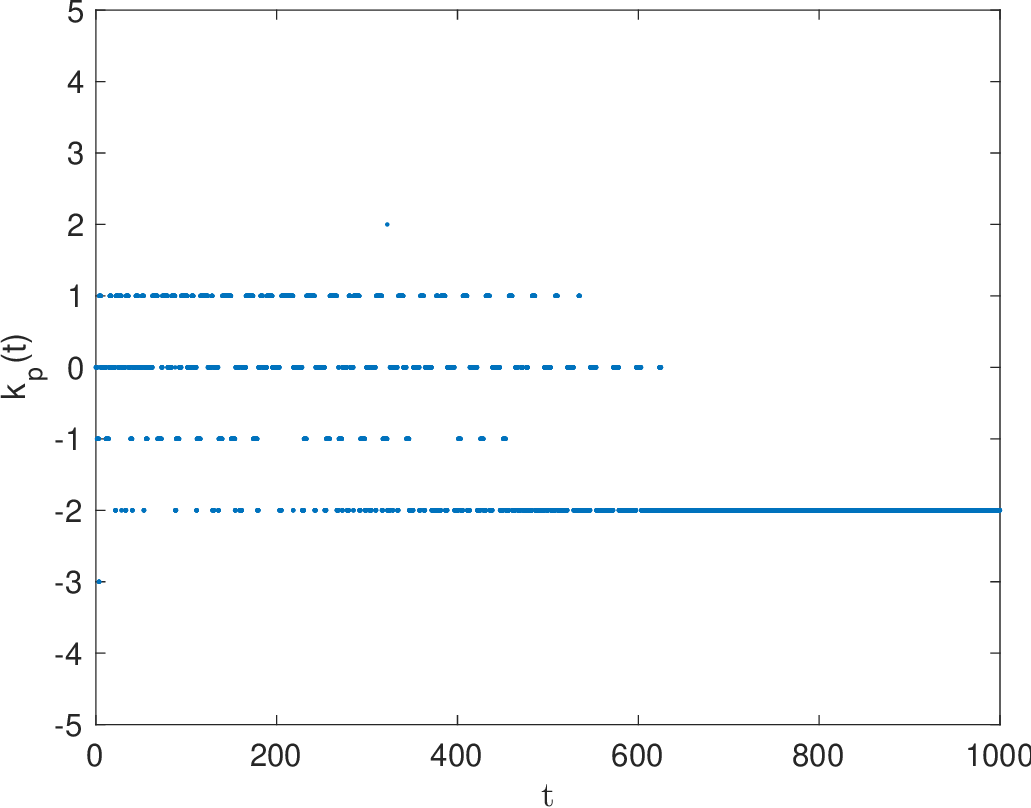}
\includegraphics[width=0.3\textwidth]{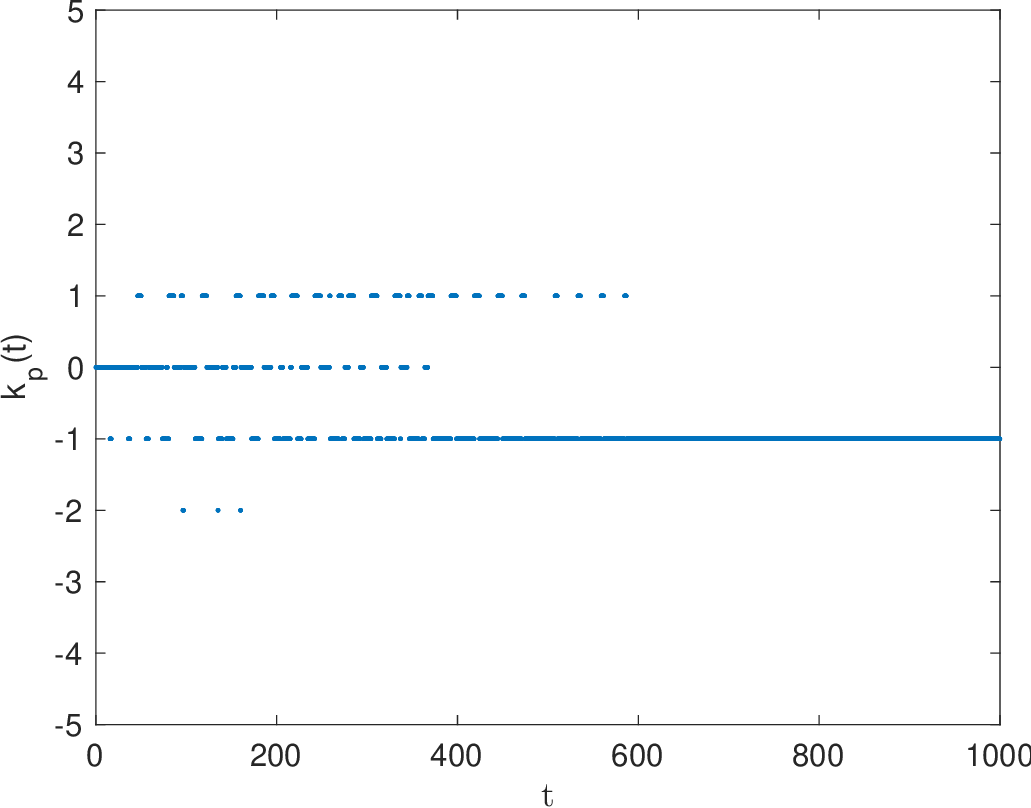}
\caption{Evolution under the V-HONLS model:  
  (a) Spectral  mean   $k_m$ for SPB initial data \rf{SPB_ic}
  (solid and dashed red curves) and perturbed Stokes
  initial data \rf{Stokes_ic} (solid and dashed blue curves) with $\Gamma = 0.002, 0.003$ respectively;
  (b) Spectral peak   $k_{peak}$ for
 SPB initial data with $\Gamma = 0.002$;  (c) Spectral peak $k_{peak}$ for perturbed Stokes
 initial data with $\Gamma = 0.002$.
}
\label{DS_VHONLS}
\end{figure}

We systematically
examine the  evolution of the
spectral peak $k_{peak}$  and the spectral mean $k_m$  
in the  V-HONLS and NLD-HONLS models for  
steep SPB and moderately steep Stokes-type initial conditions,
across damping parameter ranges $ \Gamma \in [0.001,0.008]$  
and $ \beta \in [0.1,0.8]$.

\begin{figure}[htp!]
\centerline{
\includegraphics[width=0.3\textwidth]{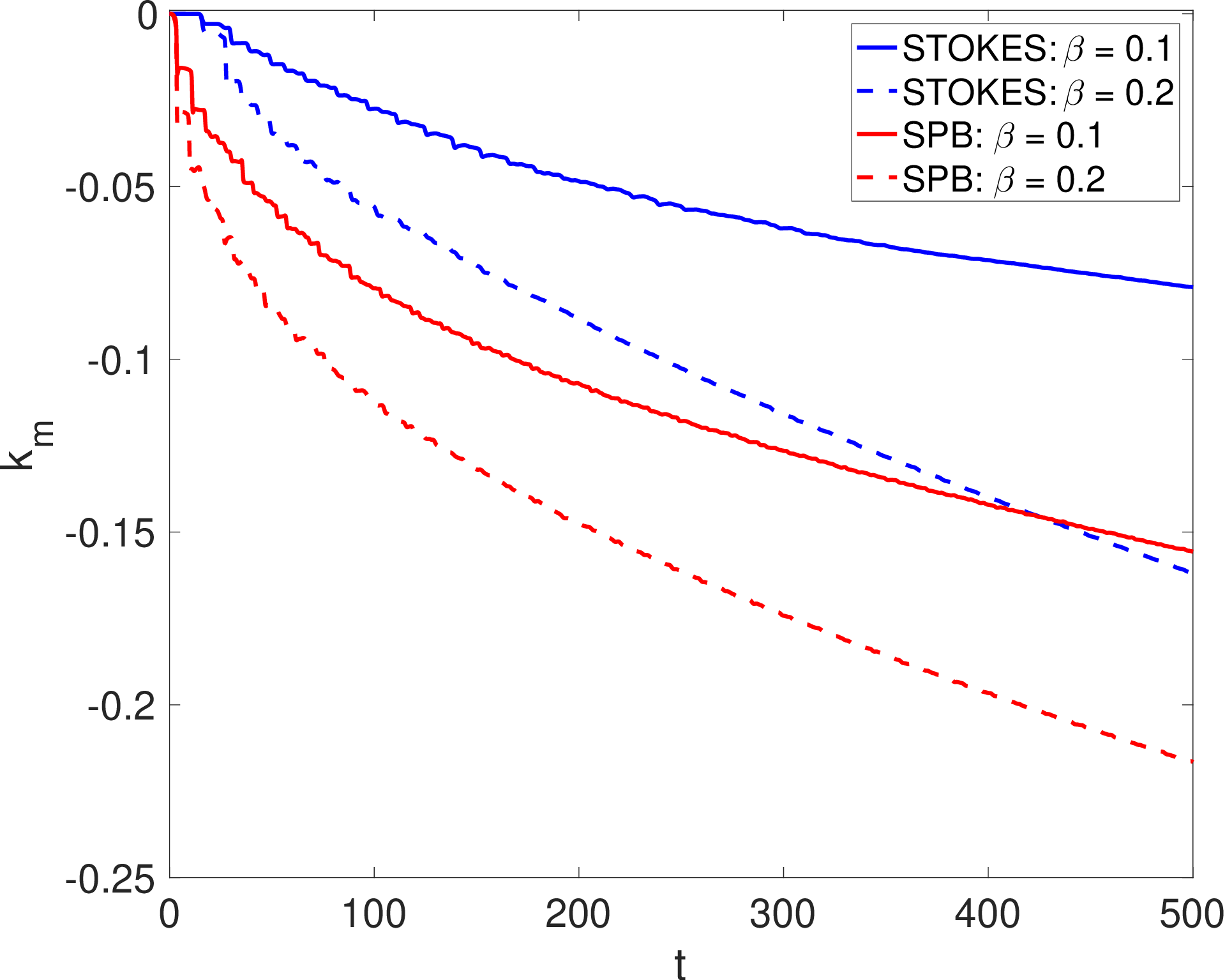}
\includegraphics[width=0.3\textwidth]{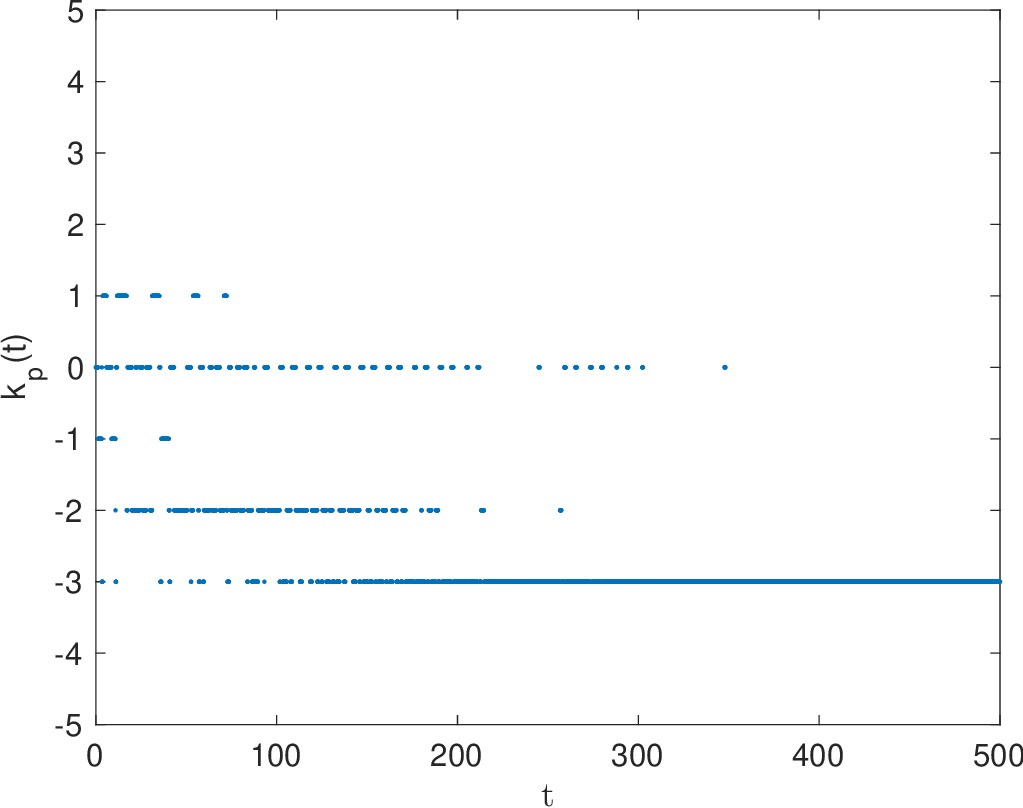}
\includegraphics[width=0.3\textwidth]{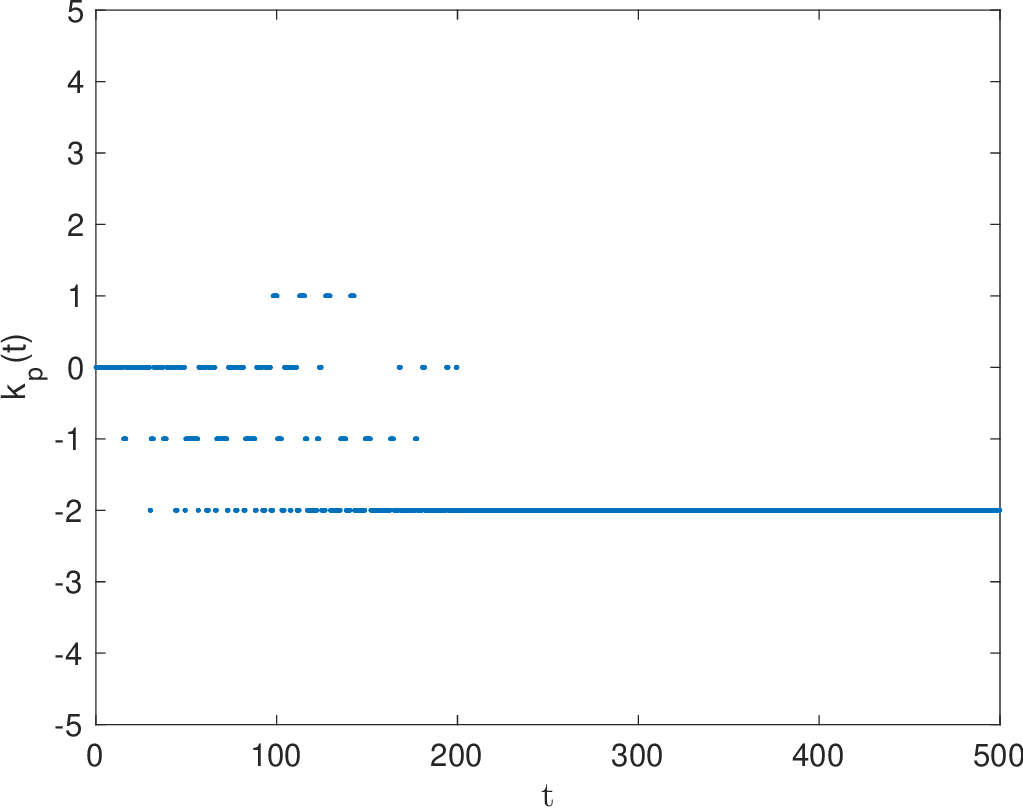}
}
\caption{Evolution under the NLD-HONLS model: (a)
  Spectral  mean   $k_m$ for SPB initial data \rf{SPB_ic}
  (solid/dashed red curves) and perturbed Stokes
  initial data \rf{Stokes_ic} (solid/dashed blue curves) with
  $\beta = 0.1, 0.2$.;
    (b) spectral peak   $k_{peak}$ for
 SPB initial data  with  $\beta = 0.1$;  (c) spectral peak $k_{peak}$ for perturbed Stokes
  initial data with $\beta = 0.1$.}
\label{DS_NLDHONLS}
\end{figure}

Representative results for the evolution of $k_m$
in  the V-HONLS model, for $\Gamma = 0.002$ and $\Gamma = 0.003$,
are shown in 
Figure~\ref{DS_VHONLS}(a). The red and dashed red curves  correspond to simulations with SPB
initial data \rf{SPB_ic}, with  $\Gamma = 0.002$ and  $\Gamma =  0.003$,
respectively.
The blue and dashed blue curves correspond to 
moderately steep Stokes type data \rf{Stokes_ic}
for the same values of $\Gamma$.
Analogous results for the NLD-HONLS model are presented in
Figure~\ref{DS_NLDHONLS}(a), showing the 
evolution of $k_m$ for 
$\beta = 0.1$ and $\beta = 0.2$.

Frequency downshifting in $k_m$ is consistently observed,
across all simulations of both  the V-HONLS and NLD-HONLS models,
regardless of  initial conditions or damping strength
($\Gamma$ and $\beta$).
In every case,
$k_m(t)$ decreases over time -- strictly monotonically in the V-HONLS model -- indicating  persistent downshifting in the spectral mean. Further, the
cumulative shift $\Delta k_m = k_m(0) - k_m(t)$
increases with the damping strength, with larger values of  $\Gamma$ or
$\beta$ producing correspondingly larger downshifts. 

Since $k_m$ is an energy-weighted average wavenumber, a sustained 
decrease in $k_m$ indicates a redistribution of spectral energy toward
longer wavelengths, i.e. lower wavenumbers. This process typically
drives a corresponding shift in $k_{Peak}$, leading to permanent frequency downshifting.
However, $k_m$ alone does not provide direct information about the time of permanent downshift, which is  determined from the evolution of
$k_{peak}$.

Figures~\ref{DS_VHONLS}(b)-(c) show  the evolution of $k_{peak}$ for the
V-HONLS model with $\Gamma = 0.002$, using SPB and perturbed Stokes initial data, respectively.
Permanent downshift in $k_{peak}$ occurs at 
$t_d = 622$ for the SPB case and $t_d = 584$ for the perturbed
Stokes case.
Transient upward shifts in $k_{peak}$
 between sideband modes occurs even after this permanent
 shift away from the carrier frequency.
 Analogously, Figures~\ref{DS_NLDHONLS}(b)-(c) present the corresponding
 results for the NLD-HONLS model with $\beta =0.1$ ,
 where  permanent downshift occurs at  $t_d = 346$ for SPB initial data and $t_d = 199$ for perturbed Stokes initial data.

 \vspace{6pt}
 
\noindent {\bf Time of downshift vs time of last rogue wave:}
\begin{figure}[htp!]
\centerline{
\includegraphics[width=0.3\textwidth]{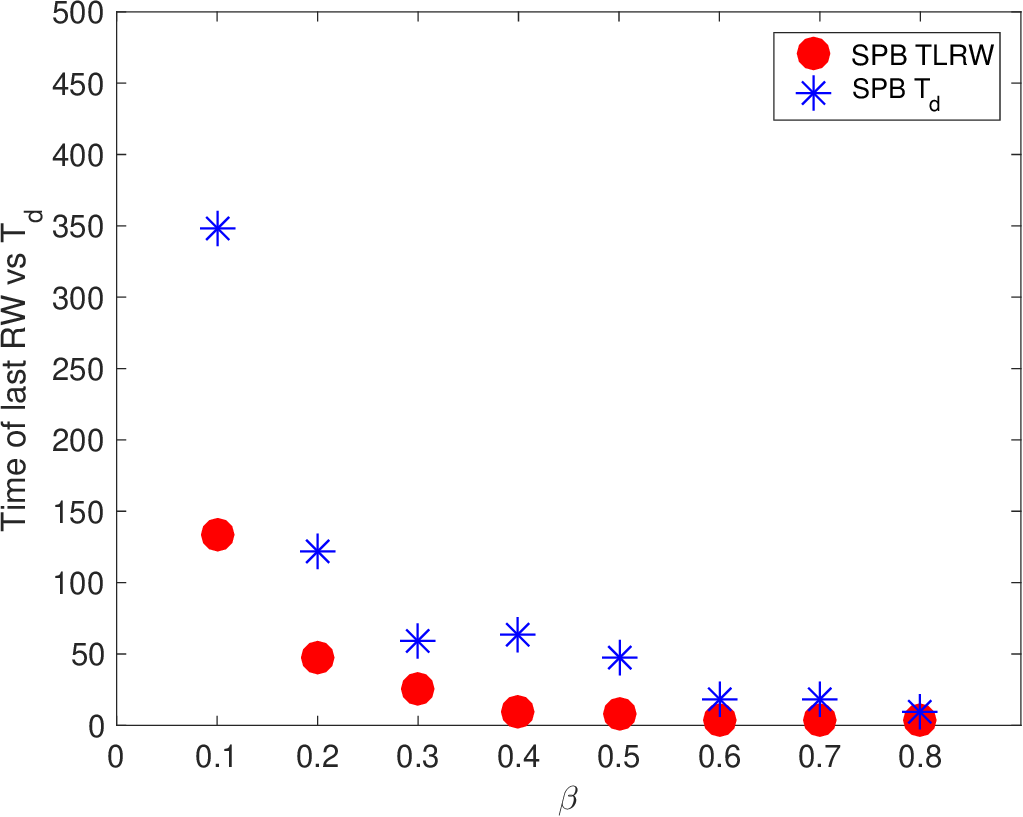}
\hspace{6pt}\includegraphics[width=0.3\textwidth]{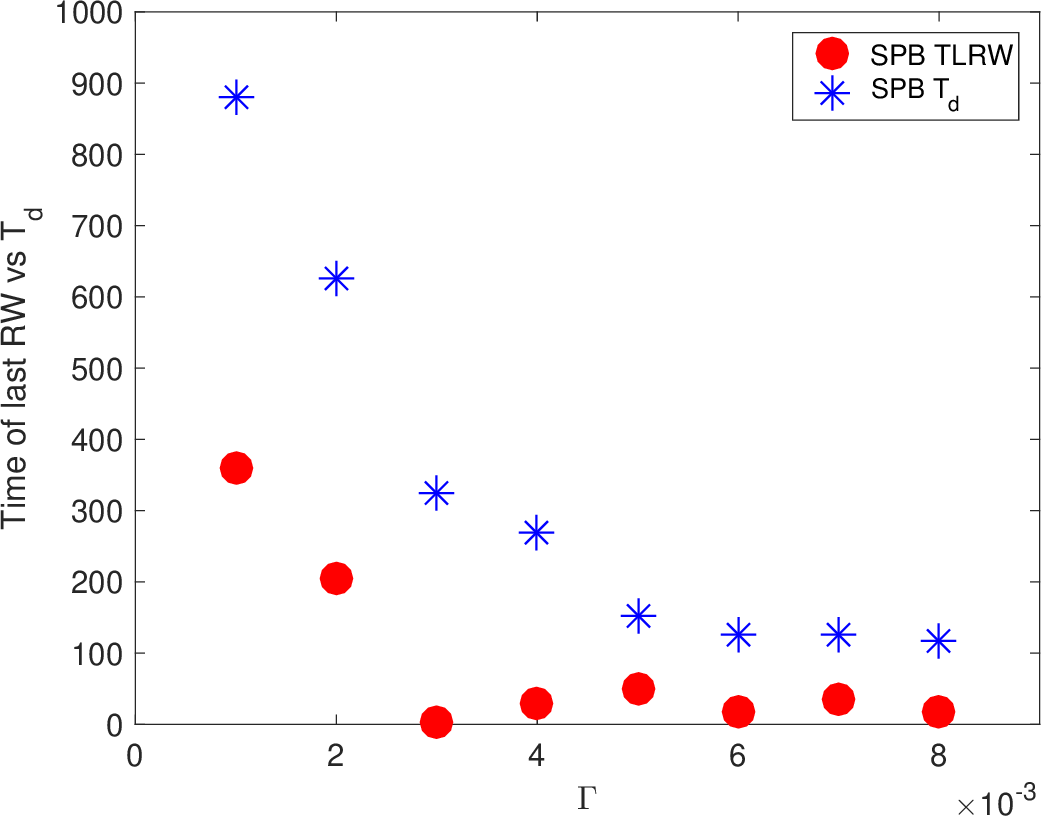}
}
\caption{Times of last rogue wave (red bullet) and permanent downshift $t_d$ (blue star) for steep SPB initial  data \rf{SPB_ic}:  
  a) NLD-HONLS model as
  $\beta$ varies and b) V-HONLS  model as $\Gamma$ varies.}
\label{TDS_RW1}
\end{figure}
 
\begin{figure}[htp!]
\vspace{12pt}
\centerline{
\includegraphics[width=0.3\textwidth]{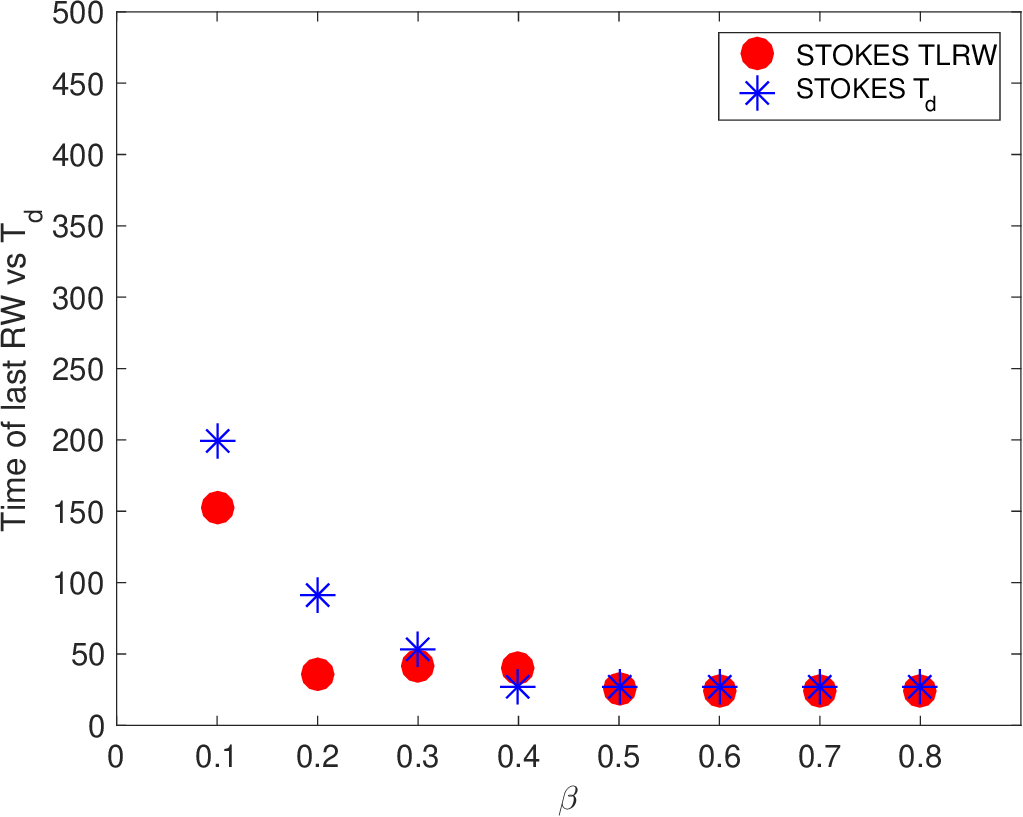}
\hspace{6pt}\includegraphics[width=0.3\textwidth]{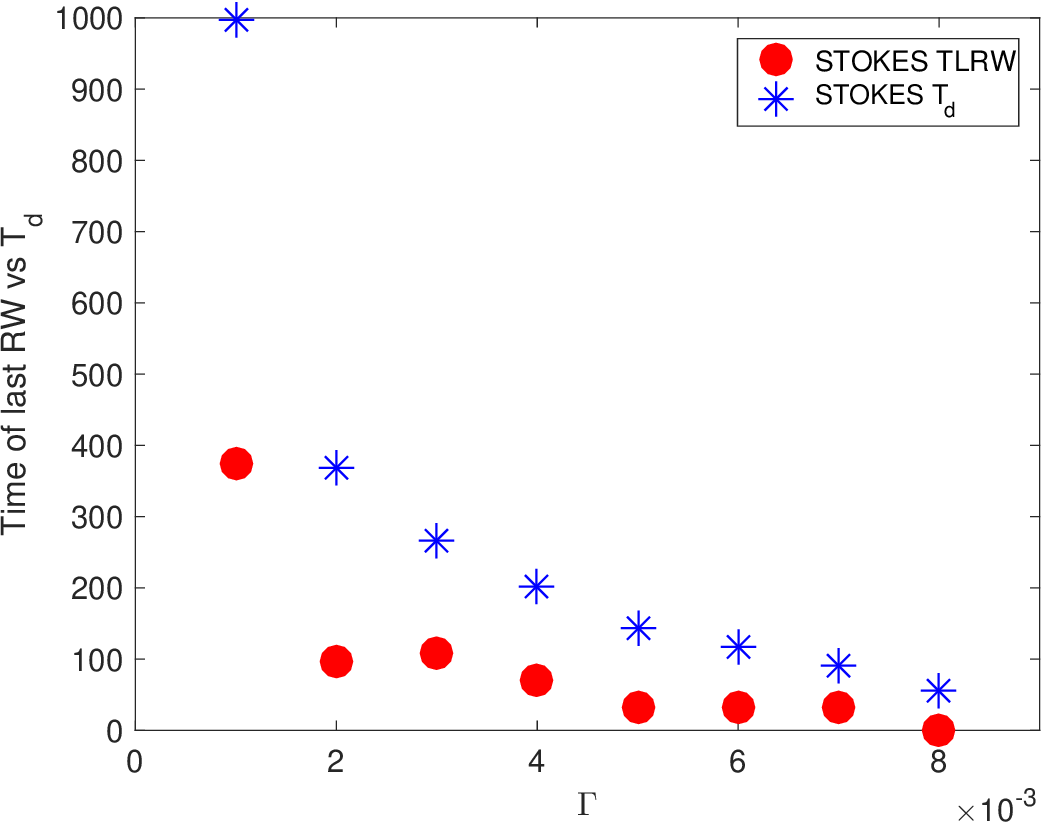}
}  
\caption{Times of last rogue wave (red bullet) and permanent downshift $t_d$ (blue star) for perturbed Stokes  initial  data \rf{Stokes_ic}:  
  a) NLD-HONLS model as
  $\beta$ varies and b) V-HONLS model as
  $\Gamma$ varies. }
\label{TDS_RW2}
\end{figure}

Although the evolution of $k_{peak}$ -- used to determine the time of permanent downshift $t_d$ --  
 is not explicitly shown for additional parameter choices, the corresponding values of $t_d$ for both the V-HONLS and NLD-HONLS models, across both types of initial data, are summarized in 
 Figures \ref{TDS_RW1} and \ref{TDS_RW2}. These results show a clear overall
 decrease in  $t_d$  as  $\Gamma$ or $\beta$ increases, with one exception for SPB initial data: a very slight
 increase in $t_d$ is
 observed at $\beta = 0.4$ before the downward trend resumes. 
We interpret this small
 deviation as a minor fluctuation associated with variability in the
 modulated background, rather than a change in the underlying decay trend.

The times of the last rogue waves, as $\Gamma$ and $\beta$ vary, are
extracted from the rogue wave timelines shown in 
Figures~\ref{VHONLS_vary_Gamma} and
Figures~\ref{VHONLS_Stokes_vary_G} for the V-HONLS model, and 
 Figures~\ref{NLD_vary_Beta} and
 Figures~\ref{NLD_Stokes_PVD} for the NLD-HONLS model, using
 both SPB and perturbed Stokes
initial  data.
These times are then compared with the time of permanent downshift in Figure~\ref{TDS_RW1} and Figure~\ref{TDS_RW2},  corresponding to
steep and moderately steep initial data, respectively.
Remarkably, across all experiments, the last rogue wave consistently precedes
the time of permanent frequency downshift.

 A key difference emerges between the models regarding  the time of permanent downshift.
In the NLD-HONLS model, permanent downshift typically follows soon
after the last rogue wave, and for large $\beta$, it occurs almost immediately for both types of initial data (Figure~\ref{TDS_RW1}(a) and Figure~\ref{TDS_RW2}(a)).  Since nearly all rogue waves in this model are one- or two-mode SRWs, 
this suggests that coherent SRW dynamics
may actively facilitate or accelerate the transition.
In contrast, the V-HONLS model generally shows a longer delay,
except under strong viscosity ((Figure~\ref{TDS_RW1}(b) and Figure~\ref{TDS_RW2}(b)). Here,  SRWs are largely replaced by broader, diffuse rogue waves as time evolves, indicating that the downshift develops  more gradually through cumulative  modal mixing.

Together, these results suggest that the two models follow
different pathways towards permanent downshifting:
i) In the NLD-HONLS model, coherence driven dynamics support
sustained soliton-like regimes and frequent
SRW formation, creating a close  link between rogue wave events and the time of  permanent downshift.
ii) In the V-HONLS model, disorganized Floquet spectral dynamics and only transient soliton-like regimes --  with fewer SRWs and instead  broader, generic rogue wave events -- lead to a
decoupling between rogue waves and the time of permanent downshift,
which then occurs more gradually.

\section{Conclusions}

This work presented a systematic comparison of rogue wave formation and spectral evolution in the HONLS equation and two of its dissipative extensions: the  nonlinear mean-flow damping model (NLD-HONLS) and linear viscous model (V-HONLS).
Using both steep  SPB initial data and moderately steep perturbed Stokes
wave data,  we examined how each model’s dissipation mechanism affects
four main aspects of the wave evolution: coherence, Floquet spectral organization, SRW formation, and frequency downshifting.
We found  that all three models can produce
soliton-like rogue waves,
but their frequency, persistence, and the nature of the
background wave field differ significantly depending on the damping mechanism. The differences in the background dynamics play a key role in shaping the conditions under which SRWs emerge. 

{\bf Conservative HONLS:}
For steep SPB initial data, the HONLS solution supports
a partially localized one-mode soliton-like state, with
intermittent emergence of short-lived two-mode SRWs.
The Floquet spectrum exhibits repeated complex critical point crossings and interaction with additional nonlinear modes, leading to recurrent transitions between localized and more
disordered  multi-mode states.
For moderately steep initial data, SRWs do not form;
instead rogue wave events are more diffuse and limited in number  due to
the
intrinsic chaotic background dynamics. 

{\bf Viscous Damping (V-HONLS):} 
The V-HONLS model applies uniform damping across all modes, leading to frequent critical point crossings, elevated phase disorder, and greater modal decoherence. As a result, rogue waves in this setting tend to be less localized and more diffuse. While SRWs may arise early in the evolution they are not sustained as in the NLD-HONLS model since viscosity drives modal mixing and suppresses sustained localization.  Frequency downshifting typically occurs
long after the last rogue wave, reflecting a more gradual transfer  of
energy  across modes rather than a sharply organized event.

{\bf Nonlinear Mean-Flow Damping (NLD-HONLS):}
For steep initial data, the NLD-HONLS model exhibits a distinct regime of sustained coherence, characterized by  the contraction of Floquet spectral
bands into  tiny, well-separated 
bands that remain localized over long times. This sustained spectral organization supports long-lived soliton-like states that enable the  repeated formation of  SRWs. In contrast to  the HONLS and V-HONLS models — where spectral localization is brief and quickly disrupted — the persistence of these tiny
Floquet bands in the  NLD-HONLS indicates a regulated 
nonlinear focusing process.
This behavior is consistent with a pre–wave-breaking
regime: the wave envelope evolves toward a maximally coherent,
steep structure representative of the final focusing stage immediately
preceding physical breaking in real water, {\sl without modeling the breaking itself}.

For moderately steep initial data, SRWs are less persistent
but still emerge against a stabilized background characterized by a well-organized Floquet spectrum and low phase variance.
Thus, nonlinear mean-flow damping suppresses disorder and maintains
coherence  across a range of initial conditions. 
Notably,
in this model, permanent frequency downshift typically follows
soon after the final SRW event, suggesting that coherent SRW dynamics may actively facilitate or trigger the downshift.

{\bf Implications for Rogue Wave Dynamics:}
Together, these results demonstrate that the type of dissipation fundamentally
alters  the structure and dynamics of rogue waves.
The formation and persistence of tiny Floquet spectral bands provide a
 spectral diagnostic for identifying coherent, pre–wave-breaking states. In particular, nonlinear mean-flow damping induces a low-dimensional, organized dynamical regime that supports SRWs, in contrast to the chaotic or decoherent regimes observed in the conservative and viscous cases. This study establishes a spectral framework for distinguishing coherent SRWs from diffuse rogue waves and offers insight into how different dissipation mechanisms govern
the development of pre-wave-breaking states in near-integrable wave systems.

\section{Appendix}
\noindent {\bf Floquet spectrum for the Stokes wave:}
The Floquet discriminant for the Stokes wave, $u_a(t)=a \e^{\ri (2 a^2 t+\phi)}$, $a \in \R^+$,
  is  given by $\Delta = 2 \cos(\sqrt{a^2 + \lambda^2} L)$.
The Floquet spectrum 
consists of the entire real axis and the band of complex spectrum
$[-\ri a,\ri a]$, ($\lambda_0^s =  \pm \ri a$ )
and infinitely many  double points 
\be
(\lambda_j^d)^2 = \left(\frac{j\pi}{L}\right)^2 - a^2,\quad j\in\mathbb{Z}, \quad j \ne 0.
\label{dps}
\ee
Complex double points $\lambda^d_j$ are obtained if  
$0 < (j\pi/L)^2 < a^2$.
The number of complex double points is the largest integer M such that
$0 < M < |a|L/\pi$. The remaining
$ \lambda_j^d$ for $|j| >M$ are real double points. 
The
Floquet spectrum for the Stokes wave with $N = 2$ unstable modes is 
shown in  Figure~\ref{ic_SPB_spec}(a).
The condition for  $\lambda_j^d$ to be  complex is
precisely the condition  for small perturbations of the Stokes wave 
$u_{\eps}(x,t) = u_a(t)(1 + \eps \e^{\ri\mu_j x + \sigma_j t})$, $\mu_j = 2\pi j/L$,
to be unstable. 
According to linear stability analysis all modes   $\mu_j$,
$0 < (j\pi/L)^2 < a^2$, will initially grow exponentially
with 
$\sigma_j^2 = \mu_j^2\left(4|a|^2 - \mu_j^2 \right)$ before
saturating due to the nonlinear terms.

\noindent {\bf Spatially periodic breather solutions of the NLS equation:}
 The spatially periodic breathers are homoclinic orbits of unstable Stokes waves $u_a(t)$ and can be explicitly constructed by means of B\"acklund transformations \cite{SZ1987}.
 In the case of a Stokes wave with two or more unstable
modes, the B\"acklund formula can be iterated algebraically to provide
a complete representation of the homoclinic manifold of $u_a(t)$. 

\begin{figure}[htp]
  \centerline{
 \includegraphics[width=0.3\textwidth]{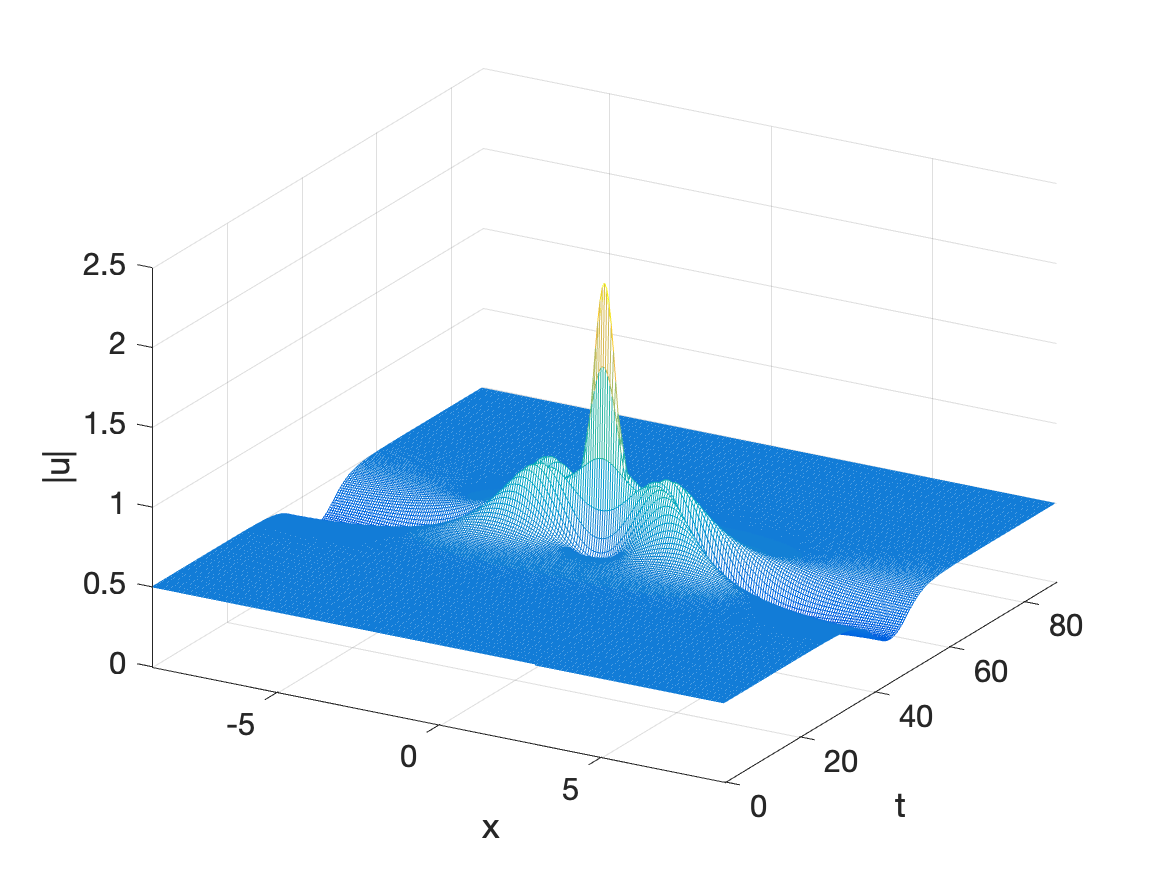}
   \includegraphics[width=0.3\textwidth]{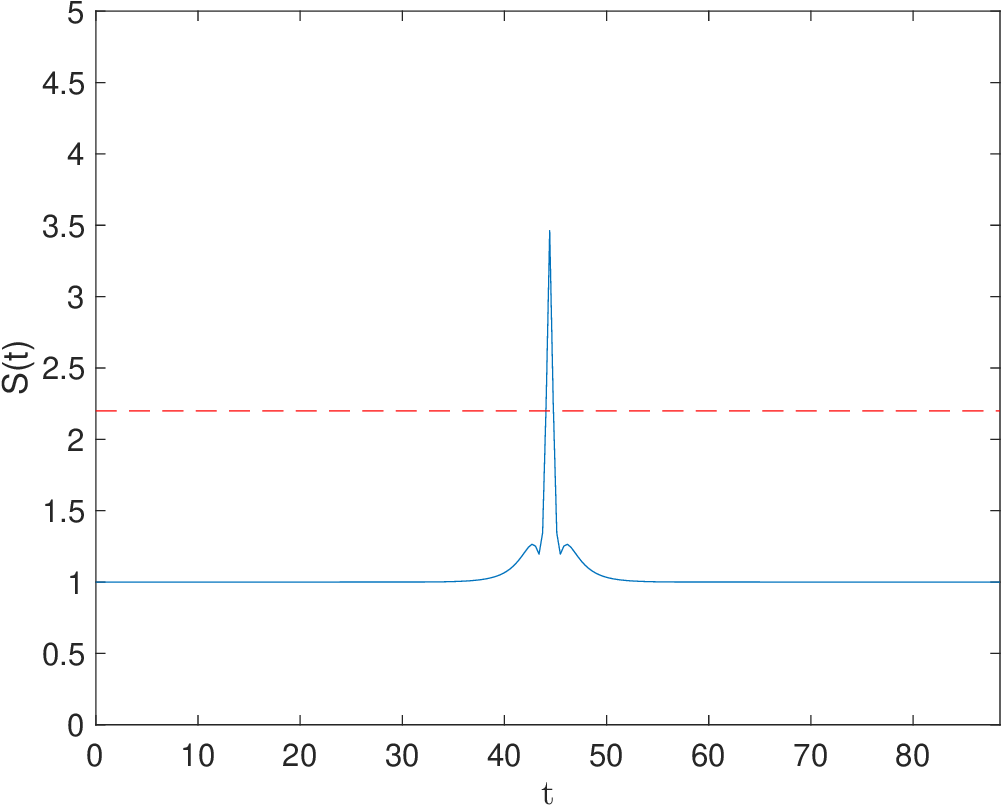}
  }
  \caption{Coalesced two-mode SPB for $\rho = 0$ and $\tau =0$: (a)  $|U(x,t)|$ and (b) S(t) for $0\le t \le 100$.}
  \label{Coal_U2}
  \end{figure}

The two-mode SPB over an unstable Stokes wave is given by: 
\begin{equation}
U(x,t;\rho,
\tau)=a\e^{2\ri a^2 t} \frac{N(x,t;\rho, \tau)}{D(x,t;\rho, \tau)},
\label{comboSPB}
\end{equation}
where

\[
\begin{array}{rl}
& N(x,t;\rho, \tau)  = 4 (\sin^2 p \cos 2p +\sin^2 q \cos 2q)  \\ & + 4\sin
  p(\sin^2 q -\sin^2 p) \cos 2q \cos\left({2k_1}x+\beta
  \right) \mbox{sech}(\rho -\sigma t) \\ & -2  \sin 2p \sin 2q \, \tanh (\tau -\delta t) \tanh
  (\rho -\sigma t) \\ &-4\sin q(\sin^2 q -\sin^2
  p)\cos\left({2k_2}x+\gamma \right) \mbox{sech}(\tau -\delta
  t)  \\
 & -4\sin p \sin q( \cos^2 p +  \cos^2 q) 
\cos \left(2k_1x+\beta \right) \cos \left(2k_2x+\gamma \right) 
 \mbox{sech}(\rho -\sigma t) \mbox{sech}(\tau -\delta t) \\\
&  - 4 \ri \sin 2p (\sin^2 q -\sin^2 p) \tanh (\rho -\sigma t) 
 4 \ri \sin 2q (\sin^2 q -\sin^2 p) \tanh (\tau -\delta t)  \\
&  + 4 \ri \sin p \sin 2q (\sin^2 q -\sin^2 p) \tanh (\tau -\delta t) \mbox{sech}(\rho -\sigma t)
\cos\left(2k_1x+\beta \right)  \\
&-  4 \ri \sin 2p \sin q (\sin^2 q -\sin^2 p) \tanh (\rho -\sigma t) \mbox{sech}(\tau -\delta t)
\cos\left(2k_2x+\gamma \right)  \\
& - 2 \sin 2p \sin 2q 
\sin \left(2k_1 x+\beta \right) \sin \left(2k_2 x+\gamma \right) 
 \mbox{sech}(\rho -\sigma t) \mbox{sech}(\tau -\delta t).
\end{array}
\]

and
\[
\begin{array}{rl}
 & D(x,t;\rho, \tau)  =  4(\sin^2p\cos^2q+\sin^2q\cos^2p) \\ 
 & + 4\sin q(\sin^2 q
  -\sin^2 p)\cos\left(2k_2x+\gamma \right) \mathrm{sech}(\tau
  -\delta t)\\  
 &-4\sin p(\sin^2 q - \sin^2 p) \cos\left(2k_1x+\beta \right) \mbox{sech}(\rho -\sigma t) \\
&- 4\sin p \sin q  (\cos^2 p + \cos^2 q)
\cos\left(2k_1x+\beta \right)
\cos\left(2k_2x+\gamma \right) \mbox{sech}(\rho -\sigma t)
\mathrm{sech}(\tau -\delta t)  \\
& -2 \sin 2p \sin 2q \, \tanh (\tau -\delta t) \tanh (\rho -\sigma t) \\ 
& -2 \sin 2p \sin 2q 
\sin \left(2k_1x+\beta \right) \sin \left(2k_2x+\gamma \right) 
 \mbox{sech}(\rho -\sigma t) \mbox{sech}(\tau -\delta t). 
\end{array}
\]
The above formula  contains two spatial
modes with wavenumbers $k_1=\pi/L$ and $k_2=2\pi/L$, and depends on two real parameters, $\rho$ and $\tau$ which determine the time at which the first and second modes are excited, respectively.
Figure~\ref{Coal_U2}(a)
shows 
the two-mode SPB with maximal amplitude, corresponding to the parameter choice
$\rho = 0$ and $\tau = 0$, where   both modes are simultaneously excited. In this case the solution exhibits a single localized focusing event and 
then asymptotically decays to the Stokes wave as $t\to\pm\infty$.
  The corresponding strength plot in Figure~\ref{Coal_U2}(b) confirms the
  occurence  of a single
  rogue wave event, consistent with this decay behavior.

%  similar to
% the first one observed in Figure~\rf{HONLS_SPB_str}(b).

 Since the  B\"acklund Transformation is isospectral, i.e. $\sigma(u_a(t)) = \sigma(U(x,t))$,
 the Floquet spectrum for the  two-mode SPB  
 is the same as for the Stokes wave with two unstable modes and is shown in
 Figure~\rf{ic_SPB_spec}(a).

 %This SPB serves as the NLS prototype for rogue-wave formation and provides the baseline behavior for comparison with the higher-order HONLS models studied in this work.

\noindent {\bf Soliton solutions of the NLS equation:}
Complete information about the soliton solutions of the
NLS equation, obtained using
the inverse scattering theory, is contained 
in the scattering data \cite{YUJI1995,AC1991}.
The discrete spectra in the upper half plane, $\lambda_n$, are 
the zeros of the first Jost coefficient.  In the case of two discrete eigenvalues,  $\lambda_n = \xi_n + \ri \eta_n$, $n=1,2$,  the
two-soliton solution is as follows:

Define the quantities
\[\ba{ll}
\phi_n &= \pi/2 - \arctan\left[\frac{(\xi_n - \xi_{\tilde n})^2 + \eta_n^2 - \eta_{\tilde n}^2}{2\eta_{\tilde n} (\xi_n - \xi_{\tilde n})}\right],\quad \tilde n=3-n,\\
\theta_n(t) &= \theta_n^0 - 4\xi_n\eta_n t,\qquad \sigma_n(t) = \sigma_n^0 + 2\left(\xi_n^2 - \eta_n^2\right) t,\\
\Delta(x,t) &= \cosh\left[2(\eta_1 + \eta_2) x - \theta_1(t) - \theta_2(t)\right] \\
&+\frac{4\eta_1\eta_2}{(\xi_1 - \xi_2)^2 + (\eta_1 - \eta_2)^2}\cos\left[2(\xi_1 -\xi_2) x + \sigma_1(t) - \sigma_2(t)\right]\\
&+\frac{(\xi_1 - \xi_2)^2 + (\eta_1 + \eta_2)^2}{(\xi_1 - \xi_2)^2 + (\eta_1 - \eta_2)^2}\cosh\left[2(\eta_1-\eta_2) x - \left(\theta_1(t) - \theta_2(t)\right)\right]
\ea
\]
where $\theta_n(t)$ and $\sigma_n(t)$ give the position and phase of the solitons. Then the associated two-soliton solution, with velocities given in terms of $\xi_n$
, is  given by \cite{YUJI1995}
\be
\ba{ll} u(x,t) =& -\frac{2}{\Delta}\left(\frac{(\xi_1 - \xi_2)^2 + (\eta_1 + \eta_2)^2}{(\xi_1 - \xi_2)^2 + (\eta_1 - \eta_2)^2}\right)^{1/2}\\
&\times\left[2\eta_1\cosh\left(2\eta_2 x-\theta_2(t)-\mbox{i}\phi_1\right)\mbox{e}^{-2\mbox{i}\xi_1 x - \mbox{i}\sigma_1(t)} +
  2\eta_2\cosh\left(2\eta_1 x - \theta_1(t) -\mbox{i}\phi_2\right)\mbox{e}^{-2\mbox{i}\xi_2 x - \mbox{i}\sigma_2(t)}\right].\ea
\label{two-soliton}
\ee

\section*{Acknowledgments}
We gratefully acknowledge support for this project by the Simons Foundation through award \#527565 (PI: C.M.~Schober).

\bibliographystyle{plain}

\end{document}